\documentclass[leqno]{amsart}

\usepackage[foot]{amsaddr}

\usepackage[normalem]{ulem}

\usepackage[numbers]{natbib}

\usepackage{color}
\usepackage{geometry}
\usepackage{subfiles}
\usepackage{graphicx}
\usepackage[section]{placeins}
\usepackage{hyperref}
\usepackage{stmaryrd}
\usepackage{amsmath,amssymb,amsfonts,amsthm,textcomp}
\usepackage{mathtools}
\usepackage{tensor}
\usepackage{multicol}
\usepackage{subfig}
\usepackage{subcaption}
\usepackage{comment}
\usepackage{enumitem}
\usepackage{rotating}
\graphicspath{
              {./Figures/}
             }

\newfont{\tenbbb}{msbm10}
\newfont{\svnbbb}{msbm8}

\newcommand{\bs}[1]{\boldsymbol{#1}}
\newcommand{\cl}[1]{\mathcal{#1}}

\newcommand{\bb}[1]{\mathbb{#1}}

\newcommand{\di}[1]{\,\mathrm{d}{#1}}

\newcommand{\dv}{\,\mathrm{d}v}

\newcommand{\Div}{\mathrm{div}\mskip2mu}

\newcommand{\Divs}{\Div_{\mskip-2mu\scriptscriptstyle\cl{S}}}

\newcommand{\pards}[2]{\frac{\partial{#1}}{\partial{#2}}}

\begin{document}

\title{A Phase-field Model for Apoptotic Cell Death}
\author{Daniel A. Vaughan$^{1,2}$, Anna M. Piccinini$^{2}$, Mischa Zelzer$^{2}$, Etienne Farcot$^{1}$, Bindi S. Brook$^{1}$, Kris van der Zee$^{1}$, and Luis Espath$^{1}$}
\address{1. School of Mathematical Sciences, University of Nottingham, Nottingham, United Kingdom, NG7 2RD, 2. School of Pharmacy, University of Nottingham, Nottingham, United Kingdom, NG7 2RD}
\email{luis.espath@nottingham.ac.uk}

\date{\today}

\begin{abstract}
\noindent
The process of programmed cell death, namely apoptosis, is a natural mechanism that regulates healthy tissue, multicellular structures, and homeostasis. An improved understanding of apoptosis can significantly enhance our knowledge of biological processes and systems. For instance, pathogens can manipulate the apoptotic process to either evade immune detection or to facilitate their spread. Furthermore, of particular clinical interest is the ability of cancer cells to evade apoptosis, hence allowing them to survive and proliferate uncontrollably. Thus, in this work, we propose a phase-field framework for simulating intrinsic or extrinsic apoptosis induced by a cytotoxic phase, including deriving the configurational mechanics underlying such phenomena. The homogeneous analysis identifies degradation and recovery regimes under a fixed and evolving cytotoxic phase. To showcase model capabilities, we present simulations exhibiting different types of cellular dynamics produced when varying the mechanisms underlying apoptosis. The model is subsequently applied to probe different morphological transitions, such as cell shrinkage, membrane blebbing, cavity formation and fragmentation. Lastly, we compare the characteristics observed in our simulations to electron microscopy images, providing additional support for the model.   
\\
\textbf{AMS subject classifications:}
$\cdot$
74N20 
$\cdot$
92B05 
$\cdot$
74F25 
$\cdot$
92C15 

\end{abstract}

\maketitle

\tableofcontents                        


\section{Introduction}

Apoptosis, the process of programmed cell death, is a key component in a range of processes including cellular turnover, robust immune response to pathogens, embryonic development and chemical-induced cell death~\cite{elmore2007apoptosis}. Irregularities in this process, resulting in either enhanced or reduced apoptosis, are a causative factor in many human diseases, such as autoimmune and neurodegenerative disorders (excessive cell loss precipitating conditions such as Alzheimer's disease) and many types of cancer (inadequate cell clearance)~\cite{bertheloot2021necroptosis,castillo2021apoptosis,vitale2023apoptotic}. Hence, the ability to modulate and understand the life or death of a cell is of high interest. Apoptotic cells can be identified by their distinct morphological features. These include separation from other cells, chromatin condensation and DNA fragmentation, irregular bulging in the plasma membrane (finger formation), cell shrinkage, and the formation of apoptotic bodies, which are then degraded in lysosomes~\cite{nossing202350, alberts2002programmed}. These morphological changes can be observed using microscopy techniques, with electron microscopy providing superior cellular details and reproducibility compared to light microscopy~\cite{kari2022programmed}. The highly complex and sophisticated mechanisms of apoptosis can be triggered by a range of factors, including components of the immune system and cytotoxic molecules administered as part of a chemotherapeutic regimen~\cite{elmore2007apoptosis}. Given the complexity of apoptosis, a thorough understanding of how environmental cues shape this process could provide a valuable platform for devising targeted therapies and refining diagnostic strategies for personalised medication. 

Apoptosis can be initiated by one of two canonical pathways, see for instance, Lossi~\cite{Lossi2022_apoptosis} and Mustafa et al.~\cite{Mustafa2024_ApoptosisOverview}.
\begin{itemize}
\item[\emph{(i)}] Intrinsic (mitochondrial) pathway: triggered by intracellular stresses such as DNA damage, oxidative imbalance, growth-factor withdrawal, or protein misfolding. These cues promote mitochondrial release of cytochrome \textit{c}, which activates a cascade of proteases (caspases) that orchestrate cellular disassembly.
\item[\emph{(ii)}] Extrinsic (death-receptor) pathway: initiated by extracellular ligands binding to surface receptors such as Fas or TNF, which in turn activate the same downstream caspase machinery.
\end{itemize}
In this work, we refer to these mechanisms as intrinsic and extrinsic apoptosis, respectively. Irrespective of the initiating route, apoptosis involves a spatially distributed cytotoxic phase that drives the progressive breakdown of cytoskeletal and organelle structures. Mathematically, this cytotoxic phase is modelled in an agnostic manner, without distinguishing whether its origin is intrinsic or extrinsic. The field representing the cell itself is denoted the cyto phase.

\begin{figure}[htbp]
    \centering
    \includegraphics[width=0.6\textwidth]{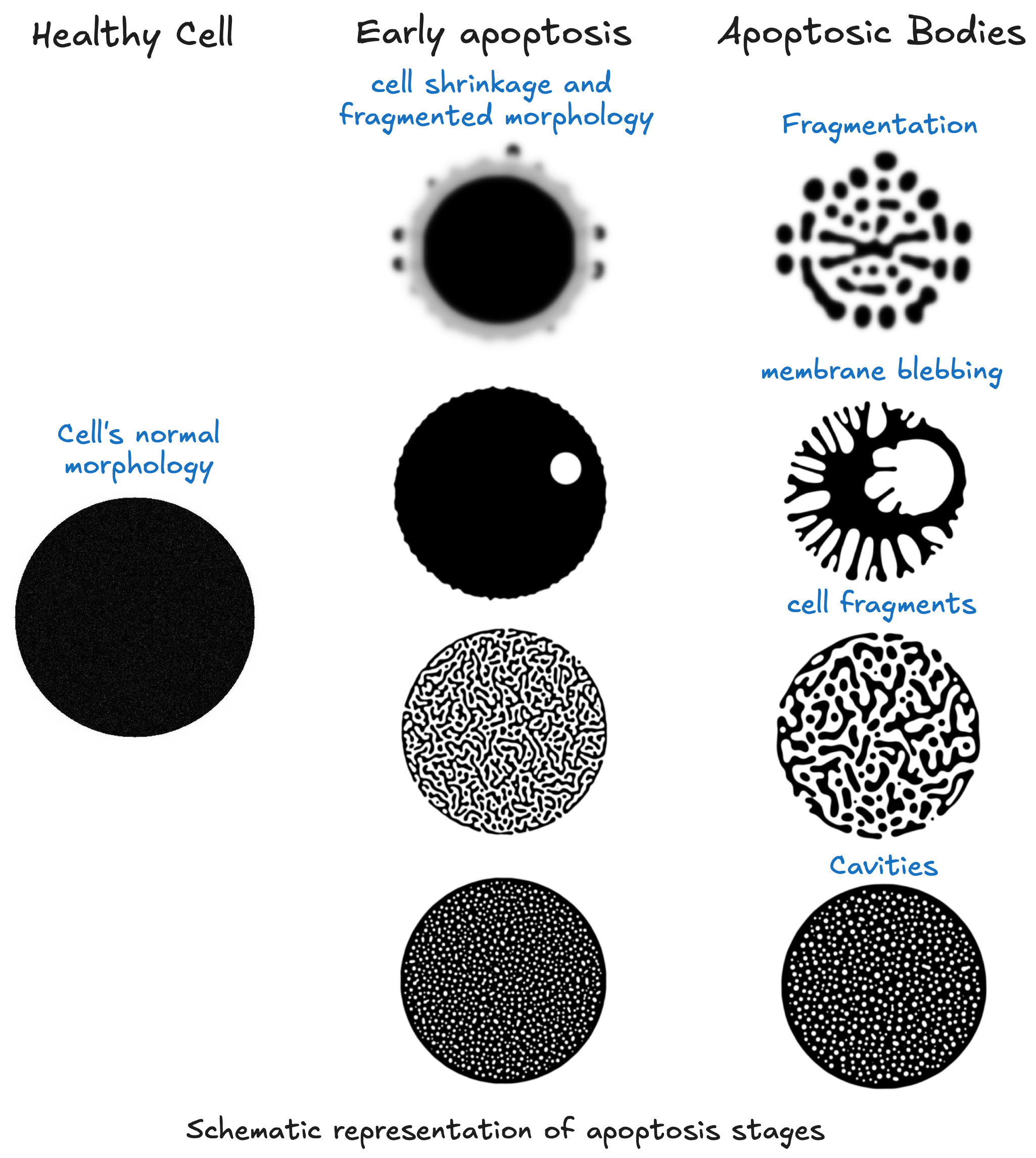}
    \caption{
        Schematic representation of apoptosis stages.
        The diagram illustrates the morphological progression from a healthy cell to apoptotic bodies.
        The stages include (left) normal morphology, (middle) early apoptosis characterised by cell shrinkage and membrane blebbing, 
        and (right) advanced apoptosis showing fragmentation and formation of apoptotic bodies.
    }
    \label{fig:apoptosis_schematic}
\end{figure}

In this study, we develop a multi-component non-conserved phase-field framework incorporating effective degenerative interactions between a cyto phase (cell) and a cytotoxic phase (triggered by the intrinsic or extrinsic apoptosis mechanism), to understand and model the physical characteristics exhibited by a cell undergoing apoptosis. This employs a coupled set of partial differential equations for the fields $\varphi$ (cyto phase) and $\sigma$ (cytotoxic phase). In our scenario, $\sigma$ is a non-conserved, evolving cytotoxic phase that triggers the degenerative reaction, which leads to apoptosis in the cell. It is not a concentration of a consumed reagent or therapeutic. The phase field $\varphi$ is also a non-conserved order parameter representing the cell. Phase-field modelling of apoptosis remains a relatively unexplored area of modelling for cellular dynamics and therefore provides an opportunity for this model to serve as a novel tool for investigating apoptosis. The use of multi-component phase-field models with reaction equations is well-established, with extensive literature not only in general contexts~\cite{bai2022chemo,gaziano2024phase,moure2018three}, but also in applications to biological phenomena, with published work to date focusing predominantly on modelling cell growth~\cite{wise2008three,hawkins2012numerical,colli2017optimal}.
While the chemical complexity of crystal formation and degradation is far simpler than in biological organisms, the analogy adopted here is geometric and thermodynamic. Both systems involve interface-driven evolution, anisotropic interfacial energy and morphological transitions. Here, anisotropy represents orientation-dependent interfacial energetics and motion. Even though apoptosis is underpinned by biochemical mechanisms, by adopting a thermodynamically driven model, we are able to focus on describing the result of the biochemical processes and linking them to system parameters.

To explore the model dynamics, numerical simulations illustrating finger formation (membrane blebs), nucleation and fragmentation are implemented in the Dedalus package~\cite{burns2020dedalus}. To demonstrate the model's capabilities, we provide a comparison of numerical simulations at selected time points and their similarity with real-world experimental data (electron microscopy images from You~et~al.~\cite{you2015kml001}) in two morphological categories: membrane blebbing (or finger formation) and size and number of cavity formations (throughout cavity formations will be referred to as nucleations). In this manuscript, we describe nucleations as a void ($\varphi$=0) surrounded by the phase field $\varphi=1$ and fragments as an area occupied by $\varphi=1$ and surrounded by a void ($\varphi$=0). While variability of biological systems does result in significant differences between images, we demonstrate that there are consistent similarities between the model output and cell imaging. We also quantify simulated shrinkage through the normalised cyto-phase area fraction.

In Section \S\ref{sc:continuum.theory}, we develop a phase-field framework to describe the dynamics of the cyto phase $\varphi$ and the cytotoxic phase $\sigma$, including the system that produces features observed in biological cells undergoing apoptosis. In Section \S\ref{sc:numerics}, we present computational simulations, used to illustrate finger formation, nucleation, and cell fragmentation during the degradation of the cyto phase field $\varphi$. Furthermore, in Section \S\ref{sc:experiments}, we compare finger formation, nucleation, and cell fragmentation with electron microscopy images from You~et~al.~\cite{you2015kml001}. As mentioned, continuum models for apoptosis remain largely unexplored, presenting a unique opportunity not only to extend the utility of these models to a new biological context, but also to gain mechanistic insight into the interplay between chemical signalling and mechanical deformation during programmed cell death. In Appendix \S\ref{sc:configurational.mechanics}, we present the mathematical derivation of the underlying configurational mechanics, including configurational forces and the configurational balance law. 
    
\subsection{Scope and modelling perspective.}

In this work, we model the mechanical morphology of apoptosis, independent of biochemical detail. The aim is to isolate the mechanical and interfacial consequences of apoptotic activation. To this end, we adopt a coarse-grained thermodynamic framework in which the cytotoxic phase is treated in a pathway-agnostic manner.

This perspective is motivated by the fact that thermodynamic principles and conservation laws are independent of the specific microscopic origin of the process, whether biological, chemical, or physical. This is precisely the role of coarse-grained phase-field modelling: to encode the energetic and kinetic structure governing interface evolution without resolving the full microscopic detail. While the biochemical cascade of apoptosis is highly complex, the resulting morphological transformations are governed by generic interfacial and energetic principles. The present model isolates these principles and provides a minimal framework for analysing the resulting dynamics.

Beyond reproducing morphological features, the framework admits a configurational-mechanical interpretation, allowing the motion of interfaces to be analysed in terms of configurational stresses and configurational forces that drive blebbing, nucleation, and fragmentation.

\section{A continuum theory for apoptotic processes}\label{sc:continuum.theory}

We model a single cell on a periodic domain $\Omega\subset\bb{R}^2$, with position $\bs{x}=(x,y)$ and time $t$. The dimensionless cyto field $\varphi$ distinguishes cellular material ($\varphi=1$) from its absence ($\varphi=0$), with intermediate values across diffuse interfaces. The dimensionless cytotoxic phase $\sigma$ represents effective apoptotic activation.

We represent the coupling between the cyto and cytotoxic phases using an effective reaction notation, following the continuum reaction frameworks of Clavijo et al.~\cite{Cla21} and Krambeck et al.~\cite{krambeck1970mathematical}. The symbols $\cl{A}_\varphi$ and $\cl{A}_\sigma$ denote cyto material and cytotoxic phase, respectively. The schematic interaction is
{\begin{equation}\label{equ:reaction}
 \varsigma_\varphi^r\cl{A}_\varphi+\cl{A}_\sigma
 \stackrel{k_1}{\rightharpoonup}
 \varsigma_\varphi^p\cl{A}_\varphi+(1+\varsigma_\sigma)\cl{A}_\sigma,
\end{equation}}
where $\varsigma_\varphi^r>\varsigma_\varphi^p\ge0$ specifies cyto-phase loss and $\varsigma_\sigma>0$ is an effective cytotoxic-phase gain. This notation describes coarse-grained field kinetics, (not a molecular stoichiometry for a particular apoptotic agent). The free-energy density and reaction driving terms use a common energy-density normalisation.

\subsection{Phase-field framework formulation}
The bounded reaction kernel is
\begin{equation}\label{equ:reac_equ}
 r=\frac{k_1}{\pi}g(\varphi)\arctan(k_2\sigma),
 \qquad g(\varphi)=\varphi(1-\varphi),
\end{equation}
where $k_1>0$ sets the amplitude and $k_2>0$ the sensitivity to the cytotoxic phase. The function $g$ localises the reaction to intermediate cyto-phase values. The reaction contributions to the two field balances are
\begin{equation}\label{equ:internal_mass_supply_rates}
(\varsigma_\varphi^p-\varsigma_\varphi^r)r \qquad\text{and}\qquad \varsigma_\sigma r.
\end{equation}

Next, consider the following energy functional
\begin{equation}\label{equ:Free_energy}
    \Psi[\varphi, \sigma] \coloneqq \int_{\Omega} \psi(\varphi,\nabla\varphi,\nabla\sigma) \dv,
\end{equation}
where the free-energy density $\psi$ is given as a function of the cyto phase field and regularized by the gradients of both phase fields, namely
\begin{equation}\label{equ:psi}
    \psi(\varphi,\nabla\varphi,\nabla\sigma) = f(\varphi) + \begin{bmatrix}\nabla\varphi & \nabla\sigma \end{bmatrix} \cdot \bs{\Sigma} \begin{bmatrix} \nabla\varphi \\ \nabla\sigma \end{bmatrix},
\end{equation}
Here $f$ is the bulk potential and $\bs{\Sigma}$ is a symmetric positive-definite matrix of gradient-energy coefficients. The energy depends on $\sigma$ through $\nabla\sigma$, with no bulk potential for the cytotoxic phase. The variational derivative of the free-energy density therefore supplies a diffusion term in the cytotoxic-phase equation.

The coupled kinetic equations are
\begin{equation}
    \begin{bmatrix}
        \tau & 0 \\[4pt]
        -\beta & \alpha
    \end{bmatrix}
    \begin{bmatrix}
        \dot{\varphi} \\[4pt] \dot{\sigma}
    \end{bmatrix}
        -
    \begin{bmatrix}
        (\varsigma_\varphi^p - \varsigma_\varphi^r)r \\[4pt]
        + \varsigma_\sigma r
    \end{bmatrix}
        =
    \begin{bmatrix}
        -\dfrac{\delta \Psi}{\delta \varphi} \\[7pt] -\dfrac{\delta \Psi}{\delta \sigma}
    \end{bmatrix}
\end{equation}
this can be written as, 
\begin{equation}\label{equ:variational_form}
    \tau \dot{\varphi} - (\varsigma_\varphi^p - \varsigma_\varphi^r)r = -\frac{\delta \Psi}{\delta \varphi} \quad \text{and} \quad \alpha \dot{\sigma} -\beta\dot{\varphi} - \varsigma_\sigma r = -\frac{\delta \Psi}{\delta \sigma},
\end{equation}
Superposed dots denote time derivatives. The positive kinetic coefficients $\tau$ and $\alpha$ govern relaxation of $\varphi$ and $\sigma$, respectively. The coefficient $\beta>0$ couples the instantaneous cytotoxic-phase rate to the cyto-phase rate. The non-conserved dynamics allow both local relaxation and reaction-driven loss of cyto phase. The variational derivatives are

\begin{equation}\label{equ:first_var}
    \dfrac{\delta \Psi}{\delta \varphi} = \dfrac{\partial \psi}{\partial \varphi} - \Div\dfrac{\partial \psi}{\partial\nabla\varphi}, \qquad\text{and}\qquad \dfrac{\delta \Psi}{\delta \sigma} = \dfrac{\partial \psi}{\partial \sigma} - \Div \dfrac{\partial \psi}{\partial\nabla\sigma},
\end{equation}
leading us to the following system
\begin{equation}\label{equ:system_pde}
    \left\{
    \begin{aligned}
        \tau\Dot{\varphi} - (\varsigma_\varphi^p - \varsigma_\varphi^r)r &= -\left(\dfrac{\partial \psi}{\partial \varphi} - \Div\dfrac{\partial \psi}{\partial\nabla\varphi}\right),
    \\[4pt]
        \alpha\Dot{\sigma} - \beta\Dot{\varphi} - \varsigma_\sigma r &= -\left(\dfrac{\partial \psi}{\partial \sigma} - \Div \dfrac{\partial \psi}{\partial\nabla\sigma}\right),
    \end{aligned} 
    \right.
\end{equation}
for which the phase fields $\varphi$ and $\sigma$ are periodic in $\Omega$. 

\subsection{Thermodynamic potential}
The thermodynamic potential, $f(\varphi)$, reflects a bistable state, promoting nontrivial equilibrium states for the phase field $\varphi$. We define it as
\begin{equation}\label{equ:double_well_potential}
    f(\varphi) = \frac{1}{4}\varphi^4 - \frac{1}{2}\varphi^3 + \frac{1}{4}\varphi^2.
\end{equation}
The polynomial structure is chosen to create a double-well potential. This setup drives phase separation, shaping an energy landscape that enables transitions between stable states.

\subsection{Specialisation to directional apoptosis}
We introduce directional interfacial energy through
\begin{equation}\label{equ:parameter_matrix}
    \bs{\Sigma}=
    \begin{bmatrix}
        \dfrac{\epsilon^2(\theta)}{2} & 0 \\
        0 & \dfrac{\gamma^2}{2} 
    \end{bmatrix},
\end{equation}
where $\gamma>0$ is the gradient-energy coefficient for the cytotoxic phase and $\epsilon$ is a function of the orientation $\theta$ of the interface, that is, the orientation of the gradient of the phase field $\varphi$. The dependency of $\epsilon$ on the orientation of the interface induces anisotropy during shrinkage, see for instance, the work by Kobayashi~\cite{kobayashi1993modeling}. Thus, accounting for this dependency of $\epsilon$ yields the following system of equations
\begin{equation}\label{equ:system_pde_direc}
    \left\{
    \begin{aligned}
        \tau\Dot{\varphi} - (\varsigma_\varphi^p - \varsigma_\varphi^r)r &= -\frac{\partial f}{\partial\varphi}+\Div\left(\epsilon\frac{\partial\epsilon}{\partial\theta}\frac{\partial\theta}{\partial\nabla\varphi}|\nabla\varphi|^{2}+\epsilon^{2}\nabla\varphi\right),
    \\[4pt]
        \alpha\Dot{\sigma} - \beta\Dot{\varphi} - \varsigma_\sigma r &= \Div(\gamma^{2}\nabla\sigma),
    \end{aligned} 
    \right.
\end{equation}
where the variable $\theta$ is the orientation of the interface with respect to the horizontal axis and defined as
\begin{equation}\label{equ:theta_def}
    \theta=\arctan\left(\frac{\partial\varphi/\partial y}{\partial\varphi/\partial x}\right)
\end{equation}
On a smooth angular branch with $|\nabla\varphi|>0$,
\begin{equation}
    \frac{\partial\theta}{\partial\nabla\varphi}=\frac{1}{|\nabla\varphi|^{2}}\left(-\dfrac{\partial\varphi}{\partial y},\dfrac{\partial\varphi}{\partial x}\right)
\end{equation}
and noting that $-\nabla\times(\varphi\bs{e}_z)=(-\partial\varphi/\partial y,\partial\varphi/\partial x)$, with $\bs{e}_{z}=[0,0,1]$, we arrive at
\begin{equation}
    \left\{
    \begin{aligned}\label{equ:direct_phi_pde}
        \tau \dot{\varphi} &= -\left(\frac{\partial f}{\partial \varphi} - \Div \left( \epsilon^2 \nabla \varphi - \epsilon \frac{\partial \epsilon}{\partial \theta} \nabla\times (\varphi \bs{e}_z) \right)\right) + (\varsigma_\varphi^p - \varsigma_\varphi^r) r, 
        \\[4pt]
        \alpha \dot{\sigma} &= \Div \left( \gamma^2 \nabla \sigma \right) + \beta \dot{\varphi} + \varsigma_\sigma r, 
        \\[4pt]
        \epsilon &= \Tilde{\epsilon}(\vartheta+\cos(\theta)), 
        \\[4pt]
        {\dfrac{\partial\epsilon}{\partial\theta}} &{= -\Tilde{\epsilon}\sin(\theta)}, 
        \\[4pt]
        \theta &= \arctan\left(\frac{\partial\varphi/\partial y}{\partial\varphi/\partial x}\right), 
        \\[4pt]
        r &= \frac{k_1}{\pi}g(\varphi)\arctan(k_2\sigma).
    \end{aligned}
    \right.
\end{equation}
The parameter $\tilde{\epsilon}$ sets the interfacial scale, while $\vartheta$ is the ratio of the constant contribution to the cosine amplitude in $\epsilon(\theta)$. Thus the relative angular modulation is $1/\vartheta$ ($\vartheta=25$ corresponds to a relative cosine amplitude of $0.04$). We use the bounded response to the cytotoxic phase $\arctan(k_2\sigma)$, where $k_2>0$ is dimensionless. It is linear to leading order for $|k_2\sigma|\ll1$ and saturates at large positive $\sigma$.

\subsection{Dimensionless Form}
Set $t=T\bar t$ and $\bs{x}=L\bar{\bs{x}}$, where $T$ and $L$ are reference time and length scales. We use positive degradation coefficients for $\varsigma_\varphi^r>\varsigma_\varphi^p$. Suppressing bars on the dimensionless fields and operators gives
\begin{align}\label{equ:varphi_pde_reac_nondim}
 \dot\varphi&=-\ell_f f'(\varphi)
 +\ell_\varphi\Div\left[(\vartheta+\cos\theta)^2\nabla\varphi
 +(\vartheta+\cos\theta)\sin\theta\,\nabla\times(\varphi\bs{e}_z)\right]
 -\ell_r^\varphi\varphi(1-\varphi)\arctan(k_2\sigma),
\end{align}
\begin{equation}\label{equ:sigma_pde_reac_nondim}
 \dot\sigma=\ell_1\Delta\sigma+\ell_2\dot\varphi
 +\ell_r^\sigma\varphi(1-\varphi)\arctan(k_2\sigma).
\end{equation}
Here $\Dot{\varphi}$ and $\Dot{\sigma}$ denote derivatives of $\varphi$ and $\sigma$ with respect to $\bar t$, and
\begin{equation}
\begin{aligned}
 \ell_f&=\frac{T}{\tau},&
 \ell_\varphi&=\frac{T\tilde\epsilon^2}{L^2\tau},&
 \ell_r^\varphi&=\frac{T(\varsigma_\varphi^r-\varsigma_\varphi^p)k_1}{\pi\tau},\\
 \ell_1&=\frac{T\gamma^2}{\alpha L^2},&
 \ell_2&=\frac{\beta}{\alpha},&
 \ell_r^\sigma&=\frac{T\varsigma_\sigma k_1}{\pi\alpha}.
\end{aligned}
\end{equation}
Substituting the cyto-phase rate into the cytotoxic-phase balance gives the net reaction contribution
\begin{equation}
 -\bigl(\ell_2\ell_r^\varphi-\ell_r^\sigma\bigr)
 \varphi(1-\varphi)\arctan(k_2\sigma).
\end{equation}
It depletes the cytotoxic phase when $\ell_2\ell_r^\varphi>\ell_r^\sigma$ and $0<\varphi<1$, $\sigma>0$. At the baseline, $1.5(960)-0.28=1439.72>0$. The full cytotoxic-phase rate also contains diffusion and the contribution of potential and interfacial relaxation through $\dot\varphi$.

We use $T=24$~h and $L=6$~$\mu$m as reference scales. To make the anisotropic scaling explicit, let $a(\theta)=\vartheta+\cos\theta$. The interfacial flux derived from $\frac12\epsilon(\theta)^2|\nabla\varphi|^2$ is
\begin{equation}
    \frac{\partial}{\partial\nabla\varphi}\left(\frac12\epsilon(\theta)^2|\nabla\varphi|^2\right)
    =\tilde{\epsilon}^{2}\left[a(\theta)^2\nabla\varphi+a(\theta)\sin\theta\,\nabla\times(\varphi\bs{e}_z)\right].
\end{equation}
The dimensionless interfacial term therefore has the prefactor $\ell_{\varphi}=T\tilde{\epsilon}^{2}/(L^{2}\tau)$ and agrees with the implemented spatial operator. An equivalent normalised convention is $\epsilon(\theta)=\epsilon_0(1+\delta\cos\theta)$, with $\epsilon_0=\vartheta\tilde{\epsilon}$ and $\delta=1/\vartheta$. Appendix~\ref{sc:local_threshold} derives homogeneous recovery thresholds and compares sustained with depleting cytotoxic phase. Reaction saturation gives an exposure-independent sufficient recovery condition for the first reported reaction parameter set.

\subsection{Interpretation of model parameters}

The parameter $k_1$ sets the reaction amplitude and $k_2$ the sensitivity to the cytotoxic phase. The ratio $\ell_2=\beta/\alpha$ couples the cytotoxic phase to the rate of cyto-phase change. The coefficient $\epsilon(\theta)$ assigns an orientation-dependent energetic cost to cyto-phase gradients. These are effective constitutive parameters for the morphological model.

These interpretations are not tied to specific biochemical pathways, but provide a consistent mapping between model parameters and effective biological processes at the level of interfacial mechanics.

\section{Numerical Simulations}\label{sc:numerics}
The coupled phase-field model simulates different types of topological transitions underlying the apoptotic phenomenon occurring in cells when induced by the presence of a cytotoxic phase field. Here, we simulate specific characteristics of apoptotic processes, namely finger formation, fragmentation, and nucleation as seen in the work by You~et~al.~\cite{you2015kml001}, with the aim of understanding parameter values triggering different transitions.

\subsection{Physical and Numeric Parameter list}
Parameter values required to replicate the cellular dynamics shown throughout this manuscript are listed in Table~\ref{tab:Parameter_List}. The dynamic moduli $\alpha$ and $\tau$ are chosen such that the dynamics of the phase field $\varphi$ evolve more slowly than those of the phase field $\sigma$. This reflects the intracellular process in which the cell’s response to the apoptotic agent occurs at a slower rate than the change in the cytotoxic phase.

\begin{table}[h!]
\caption{
    Physical and chemical parameters required for simulating apoptosis. The parameters $\ell_2$, $\ell_{\varphi}$, $\ell_{r}^{\varphi}$, $\ell_{r}^{\sigma}$ and $k_{2}$ are explored, and example values have been proposed in this table with the corresponding simulation for illustration. Remaining parameters have the listed values for the examples in this section (Appendix~\ref{sc:controlled_simulations} specifies the controlled comparisons). 
}
\begin{tabular}{c|c|c}
    \hline
    Parameter & Value & Physical description \\
    \hline
    $\ell_{f}$ & $3.33\times10^{3}$ & Bulk relaxation coefficient \\
    $\ell_{\varphi}$ & $1.33\times10^{-4}$ & Interfacial coefficient $T\tilde{\epsilon}^{2}/(L^{2}\tau)$ \\
    $\ell^{\varphi}_{r}$ & 960 (Figure~\ref{fig:finger_formation}), 4240 (Figure~\ref{fig:higher_reaction_rate}) & Cyto degradation magnitude \\
    $\vartheta$ & 25 & Inverse relative cosine amplitude \\
    $\ell_{1}$ & 1.0 & Cytotoxic-phase diffusion coefficient  \\
    $\ell_{2}$ & 1.5 (Figure~\ref{fig:apoptotic_simulation}) & Rate-coupling coefficient $\beta/\alpha$ \\
    $\ell^{\sigma}_{r}$ & 0.28 (Figure~\ref{fig:finger_formation}), 1.27 (Figure~\ref{fig:higher_reaction_rate}) & Cytotoxic-phase reaction magnitude \\
    $k_{2}$ & 8.0 (Figure~\ref{fig:finger_formation}), 1.0 (Figure~\ref{fig:higher_reaction_rate}) & Sensitivity to the cytotoxic phase \\
    $\varsigma_\varphi^p$ & 0 & Stoichiometric coefficient for the production of $\varphi$ \\
    $\varsigma_\varphi^r$ & 1 & Stoichiometric coefficient for the reactant required for $\varphi$ \\
    $\varsigma_\sigma$ & 1 & Effective cytotoxic-phase gain \\
\end{tabular}
\label{tab:Parameter_List}
\end{table} 

The numerical cases vary the interfacial coefficient $\ell_\varphi$, the reaction coefficients $\ell_r^\varphi$ and $\ell_r^\sigma$, the sensitivity $k_2$, and the coupling $\ell_2$. A larger $\ell_2$ strengthens the contribution of cyto-phase loss to cytotoxic-phase depletion when $\dot\varphi<0$.

\subsection{Initial conditions}
The main-text simulations use a $512 \times 512$ square grid, a Real Fourier basis and the RK443 scheme for time integration. A dealiasing factor of $3/2$ was applied to prevent aliasing errors. Preliminary convergence tests were performed to ensure mesh independence. The time step was fixed at $\Delta\bar t=10^{-5}$, selected after analysing step sizes in the range $\Delta\bar t\in[10^{-7},10^{-3}]$. This value was found to be sufficiently small to avoid spurious numerical artefacts. Similarly, simulations performed with resolutions of $128 \times 128$, $256 \times 256$, $512 \times 512$, and $1024 \times 1024$ demonstrated that the $512 \times 512$ grid provides an optimal balance between numerical accuracy and computational efficiency.

While cell shape can vary depending on cell type and environment~\cite{zeng2022cell}, we restricted the cyto and cytotoxic phase fields to circular shapes, with the addition of noise to mimic cellular inhomogeneities. The centres $\bs{x}_0,\bs{x}_1$ and radii $R_0,R_1$ specify the two initial circular regions, with $S_0$ and $S_1$ being the profiles before clipping. Coordinates and radii are dimensionless. The Gaussian noise has standard deviation $\sqrt{0.1}$, with $\eta(x,y)\sim\cl{N}(0,0.1)$.
\begin{equation}\label{equ:initial_cond}
\left\{
\begin{aligned}
    S_{0}(\bs{x}) &= \frac{1}{2}\left[1-\tanh\!\left(\frac{|\bs{x}-\bs{x}_0|^2-R_{0}^2}{\tilde{\epsilon}}\right)\right] + \eta, \\
    S_{1}(\bs{x}) &= \frac{1}{2}\left[1-\tanh\!\left(\frac{|\bs{x}-\bs{x}_1|^2-R_{1}^2}{\tilde{\epsilon}}\right)\right] + \eta, \\
    \varphi(\bs{x},0) &= \min\{1,\max\{0,S_{0}(\bs{x})\}\},\\
    \sigma(\bs{x},0) &= \min\{1,\max\{0,S_{1}(\bs{x})\}\}, \\
    \theta_0 &=0, \\
    {\epsilon(\theta_{0})} &{= \tilde{\epsilon}a(\theta_0)}, \\
    &\text{with periodic boundary conditions on }\partial\Omega\text{ for }\bar t>0.
\end{aligned}
\right.
\end{equation} 

The initial perturbations represent unresolved cellular heterogeneity. Since $g(\varphi)=\varphi(1-\varphi)$ vanishes at $\varphi=1$, downward perturbations activate interior reactions. Their subsequent growth is governed by the coupled equations. Appendix~\ref{sc:local_threshold} shows that sufficiently small homogeneous perturbations decay when $\ell_r^\varphi\arctan(k_2\sigma)<\ell_f/2$ and grow when the inequality is reversed. Thus, seeding does not itself guarantee degradation. The spatial simulations describe morphological evolution from the specified heterogeneous initial states. The cavity position, number and onset may depend on that initial heterogeneity.

\subsection{Finger formation and nucleation during apoptosis}

\begin{figure}[htbp]
    \centering
    \begin{tabular}{cc}
        \Large $\varphi(\bs{x},t)$ \hspace{0.12\textwidth}  $\sigma(\bs{x},t)$ & \Large $\varphi(\bs{x},t)$ \hspace{0.12\textwidth}  $\sigma(\bs{x},t)$ \\
        \includegraphics[width=0.375\textwidth]{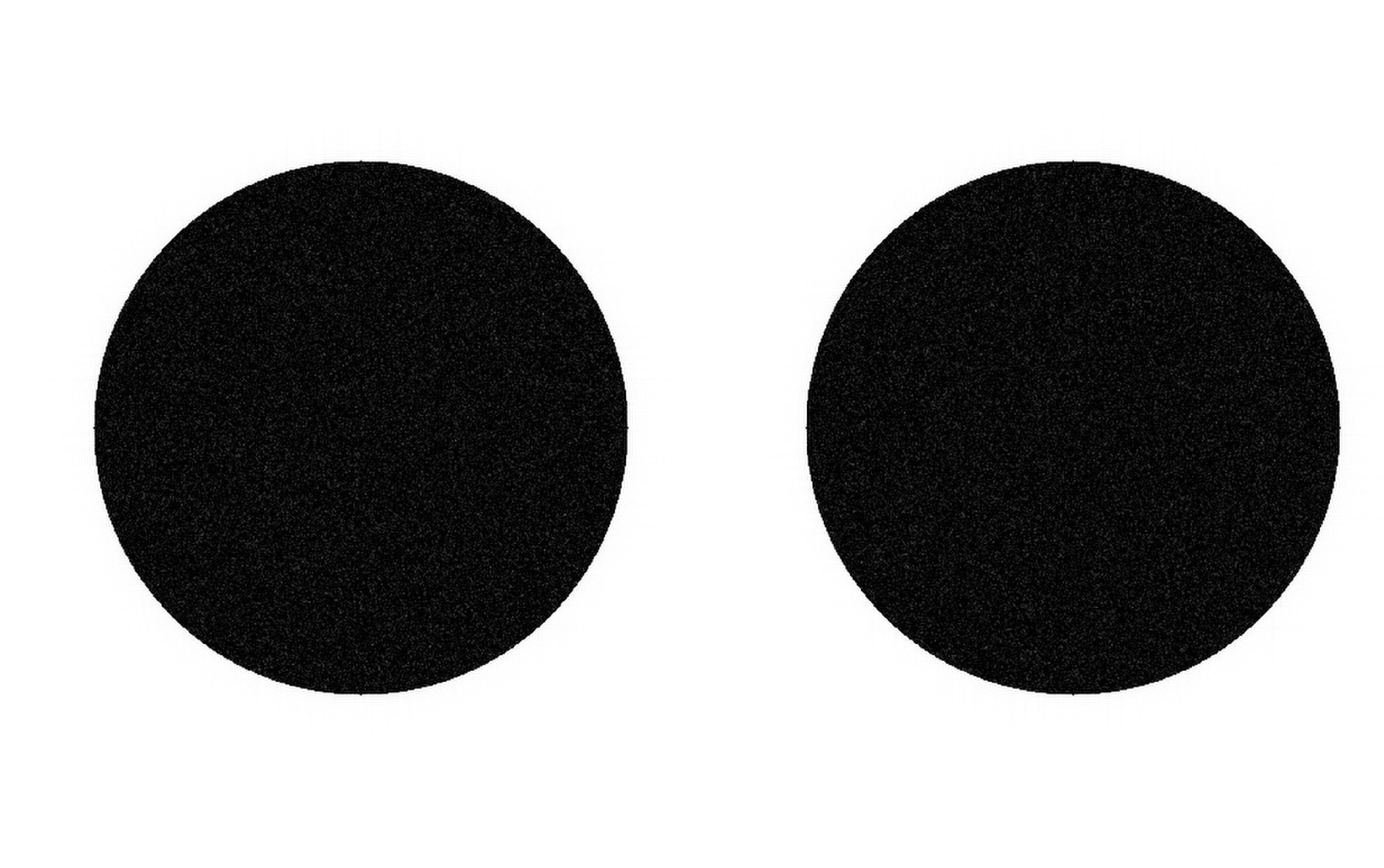} &
        \includegraphics[width=0.375\textwidth]{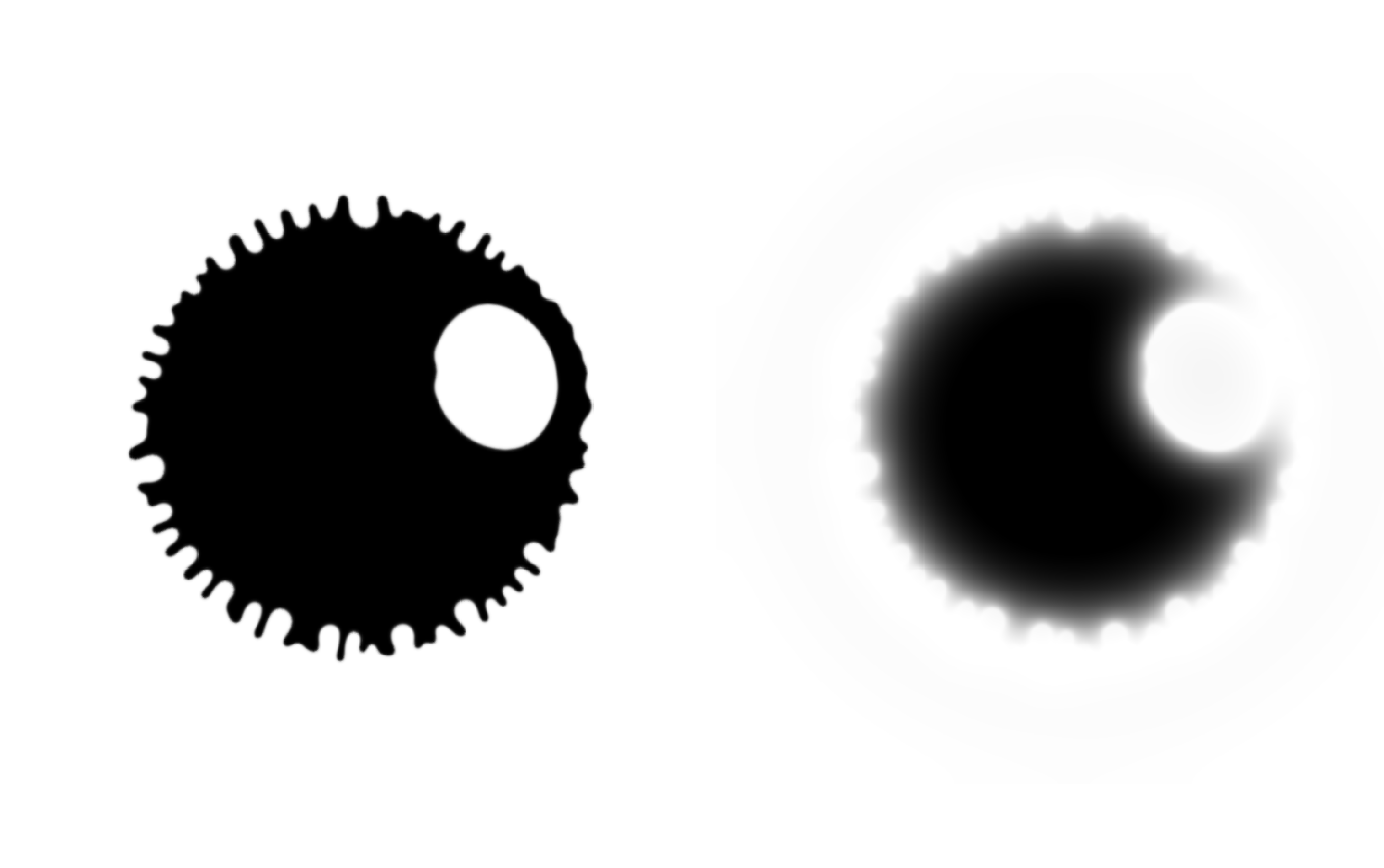} \\
         \Large (A) $\Bar{t}=0$ & \Large (B) $\Bar{t}=3.98$ \vspace{0.7cm} \\
         \Large $\varphi(\bs{x},t)$ \hspace{0.12\textwidth}  $\sigma(\bs{x},t)$ & \Large $\varphi(\bs{x},t)$ \hspace{0.12\textwidth}  $\sigma(\bs{x},t)$ \\ 
        \includegraphics[width=0.375\textwidth]{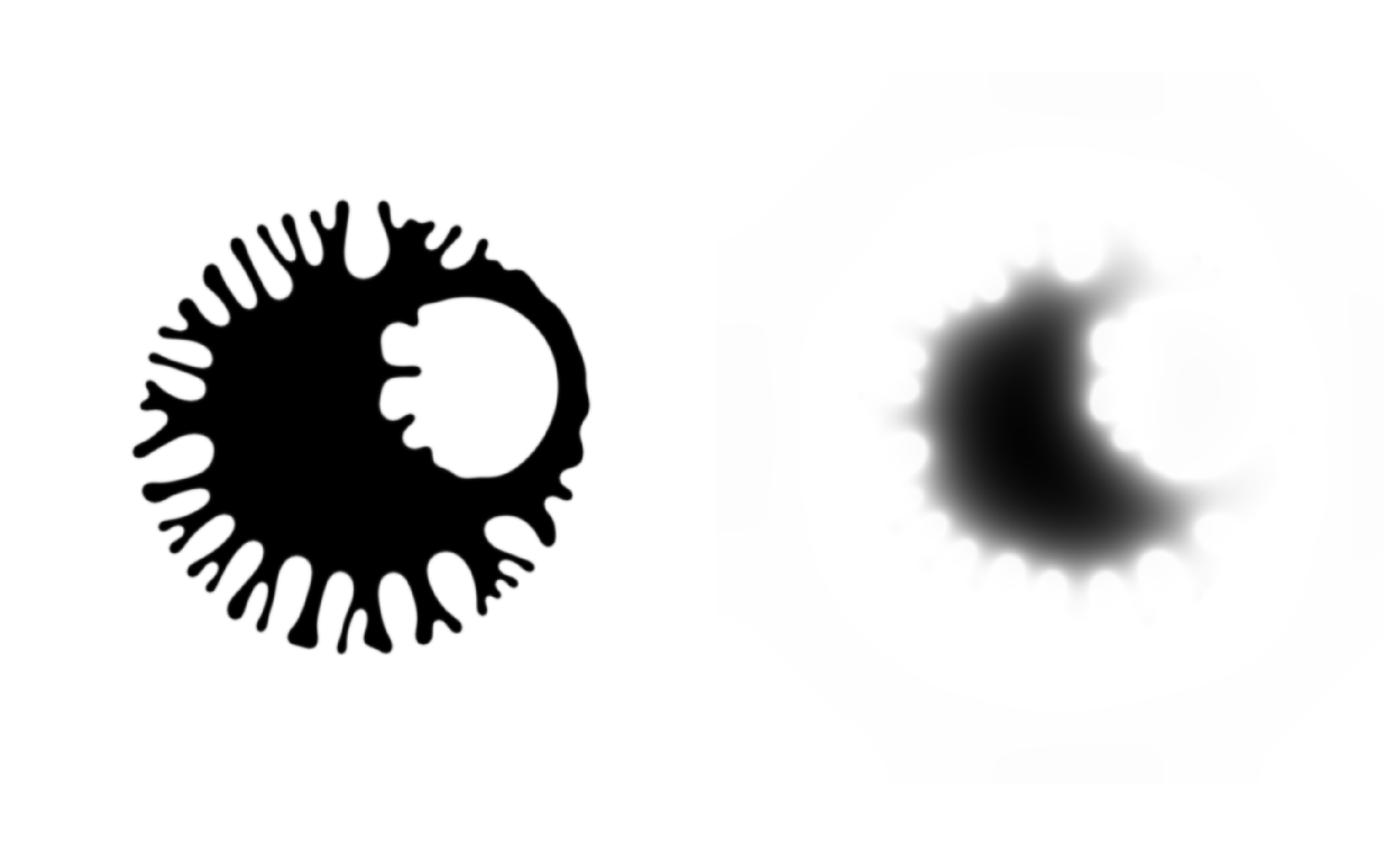} &
        \includegraphics[width=0.375\textwidth]{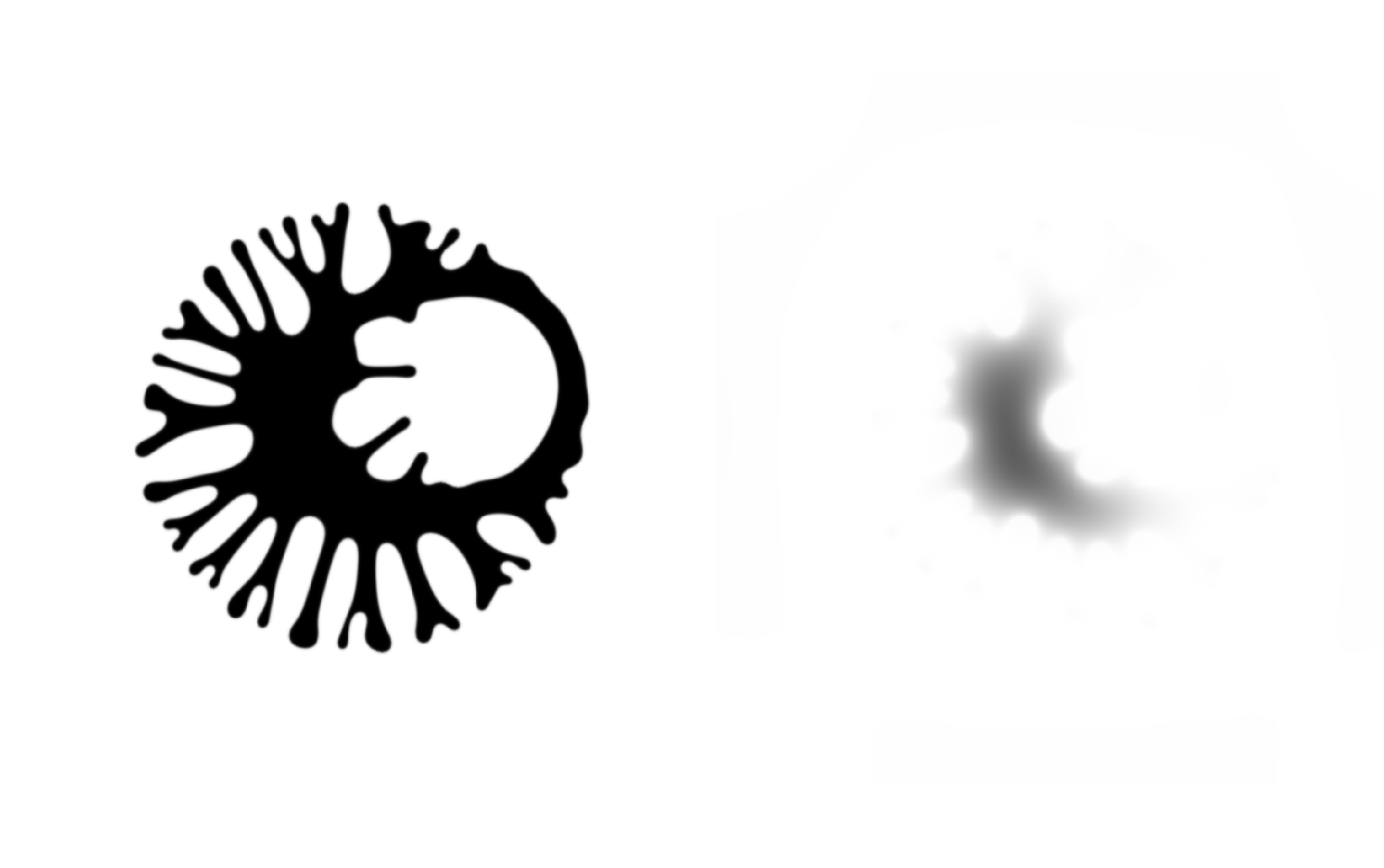} \\
        \Large (C) $\Bar{t}=7.99$ & \Large (D) $\Bar{t}=11.76$ \\
    \end{tabular}
    \caption{
        Progression of the $\varphi$ and $\sigma$ phase fields at the following nondimensionalised time points at $\Bar{t}=0.00 \ \text{(A)}, \ 3.98 \ \text{(B)}, \ 7.99 \ \text{(C)}, \ 11.76 \ \text{(D)}$. Using parameter values of:  $\ell_{\varphi}=1.333\times10^{-4}$, $\ell_{2}=1.5$, $\ell^{\varphi}_{r}=960$, $\ell^{\sigma}_{r}=0.28$ and $k_{2}=10.0$. The remaining parameters in the simulation are listed in Table~\ref{tab:Parameter_List}. 
    }
    \label{fig:apoptotic_simulation}
\end{figure}

The cytotoxic phase $\sigma$ drives degradation of the cyto phase $\varphi$. Figure~\ref{fig:apoptotic_simulation} provides simulations illustrating the cyto phase field degradation when the interface is selected to be $\ell_{\varphi}=1.33\times10^{-4}$ and the remaining parameters are $\ell_2=1.5$, $\ell_{r}^{\varphi}=960$, $\ell_{r}^{\sigma}=0.28$ and $k_{2}=10.0$. This interfacial coefficient is chosen to produce a sharp interface based on the length scales for the size of the cyto phase field. In the simulation output, the cyto phase field $\varphi$ diminishes at a high rate, producing the formation of distinct fingers, with concurrent depletion of the cytotoxic phase $\sigma$. It can also be seen that nucleations can form in the cyto phase field. The factor $\varphi(1-\varphi)$ localises the reaction to regions with $0<\varphi<1$. Therefore, inhomogeneities in the initial conditions allow for reactions to occur in the interior of the cyto phase field, hence enabling nucleations to form. However, these nucleations happen at a much lower rate. For this reason, it can be seen that the cytotoxic phase is depleted at the cyto phase field interface (membrane), starting at $\bar{t}$=3.98 and continuing in $\bar{t}$=7.99 and $\bar{t}$=11.76.

\subsection{Interface width influence}
At fixed bulk coefficient and orientation, the planar interface thickness scales as $a(\theta)\sqrt{\ell_\varphi/\ell_f}$, as derived in Appendix~\ref{sc:interface_motion}. Increasing $\ell_\varphi$ therefore broadens the interface and the region where the reaction is active.
\begin{figure}[htbp]
    \centering
    \begin{tabular}{cccc}
        \Large $\varphi(\bs{x},t)$ & \Large $\varphi(\bs{x},t)$ & \Large $\varphi(\bs{x},t)$ & \Large $\varphi(\bs{x},t)$ \\
        \includegraphics[width=0.23\textwidth]{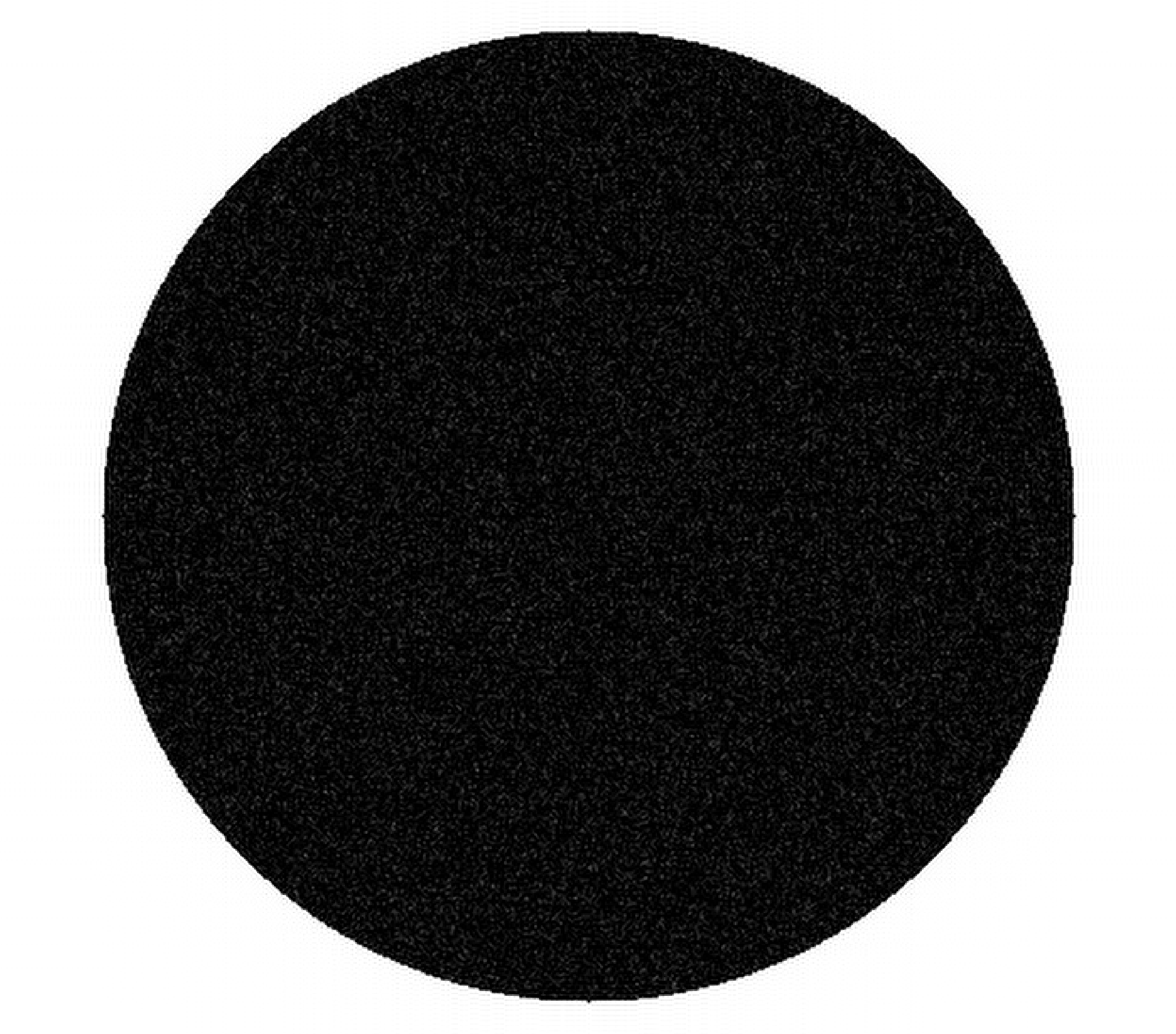} &
        \includegraphics[width=0.23\textwidth]{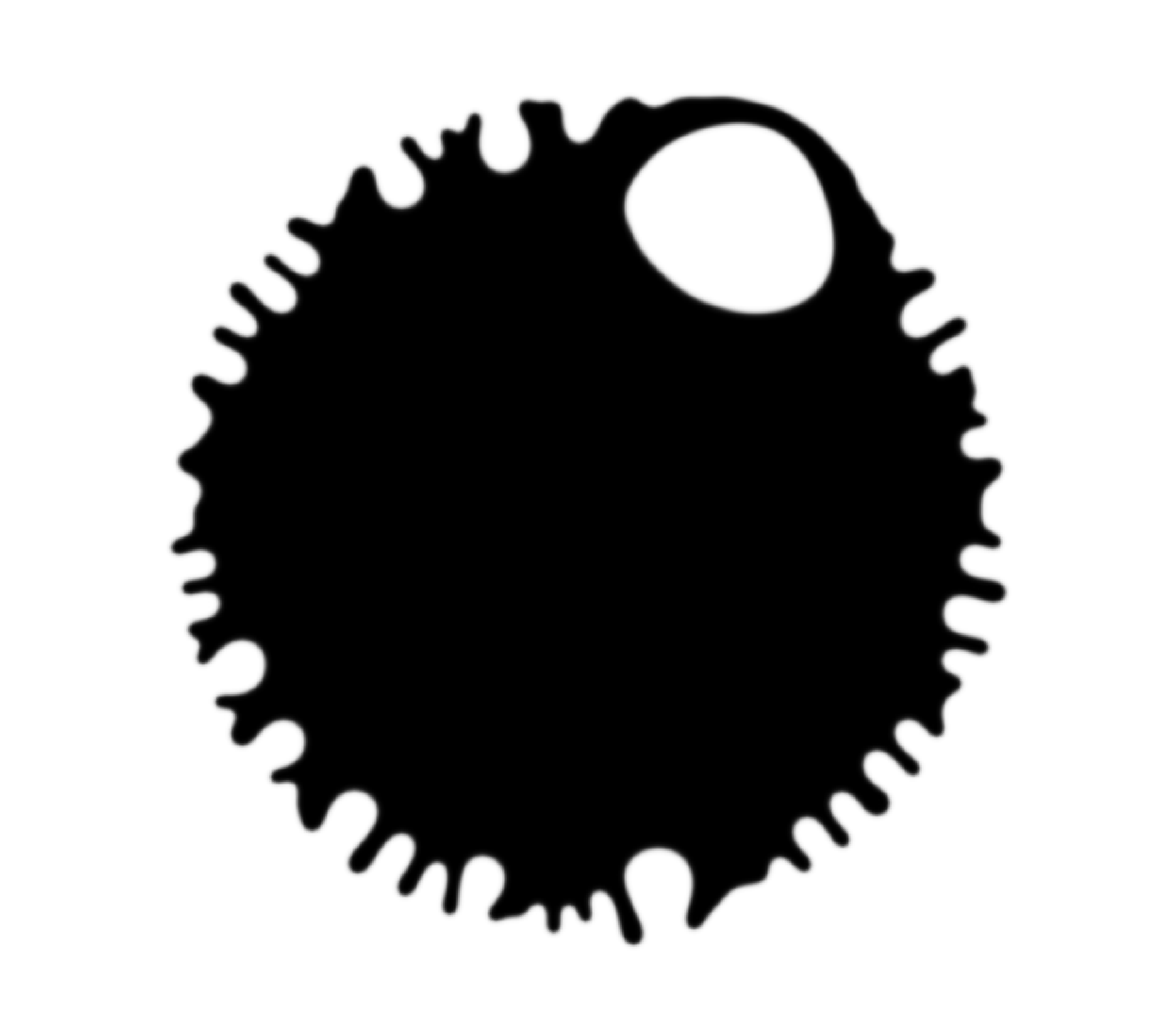} &
        \includegraphics[width=0.23\textwidth]{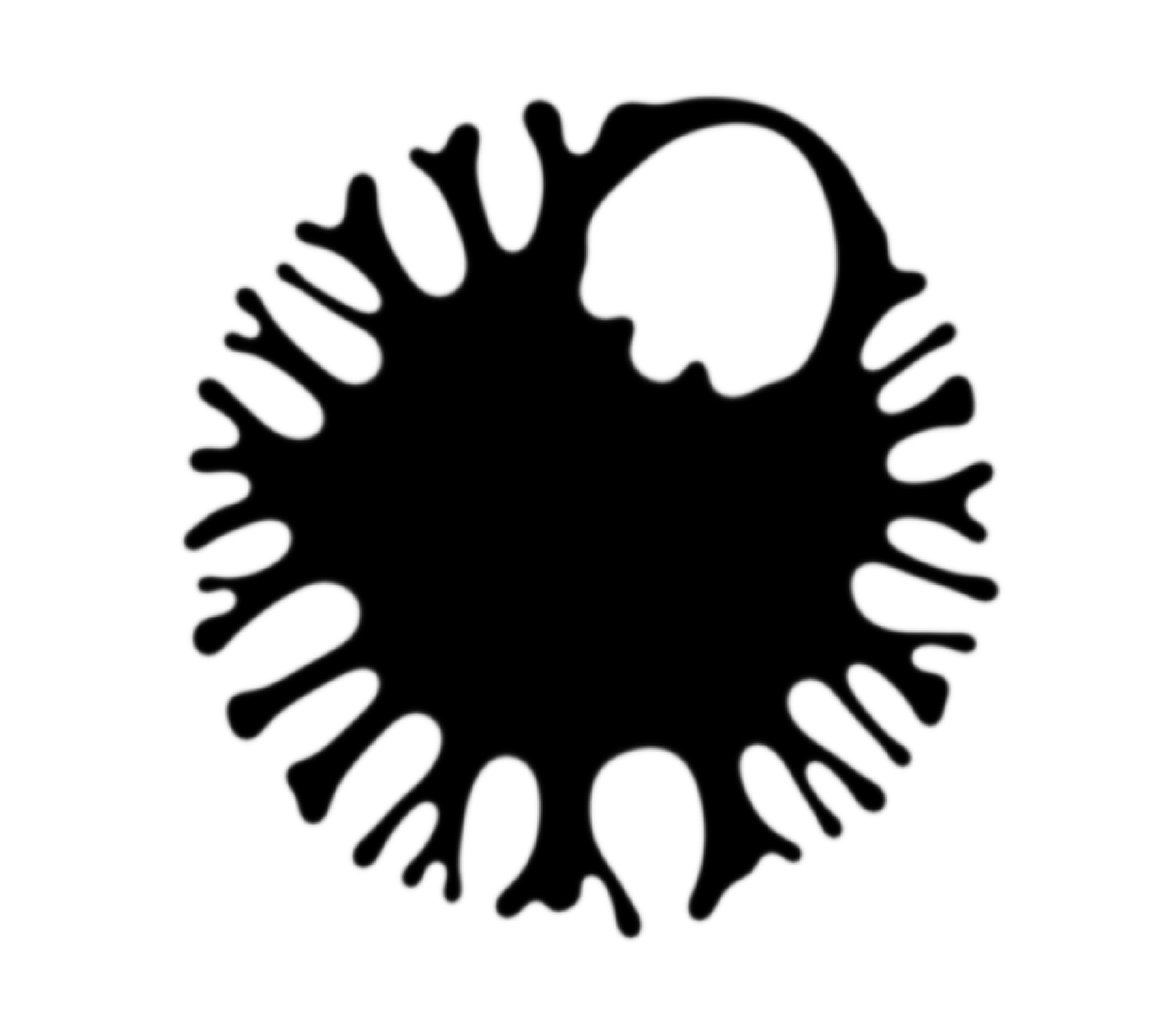} &
        \includegraphics[width=0.23\textwidth]{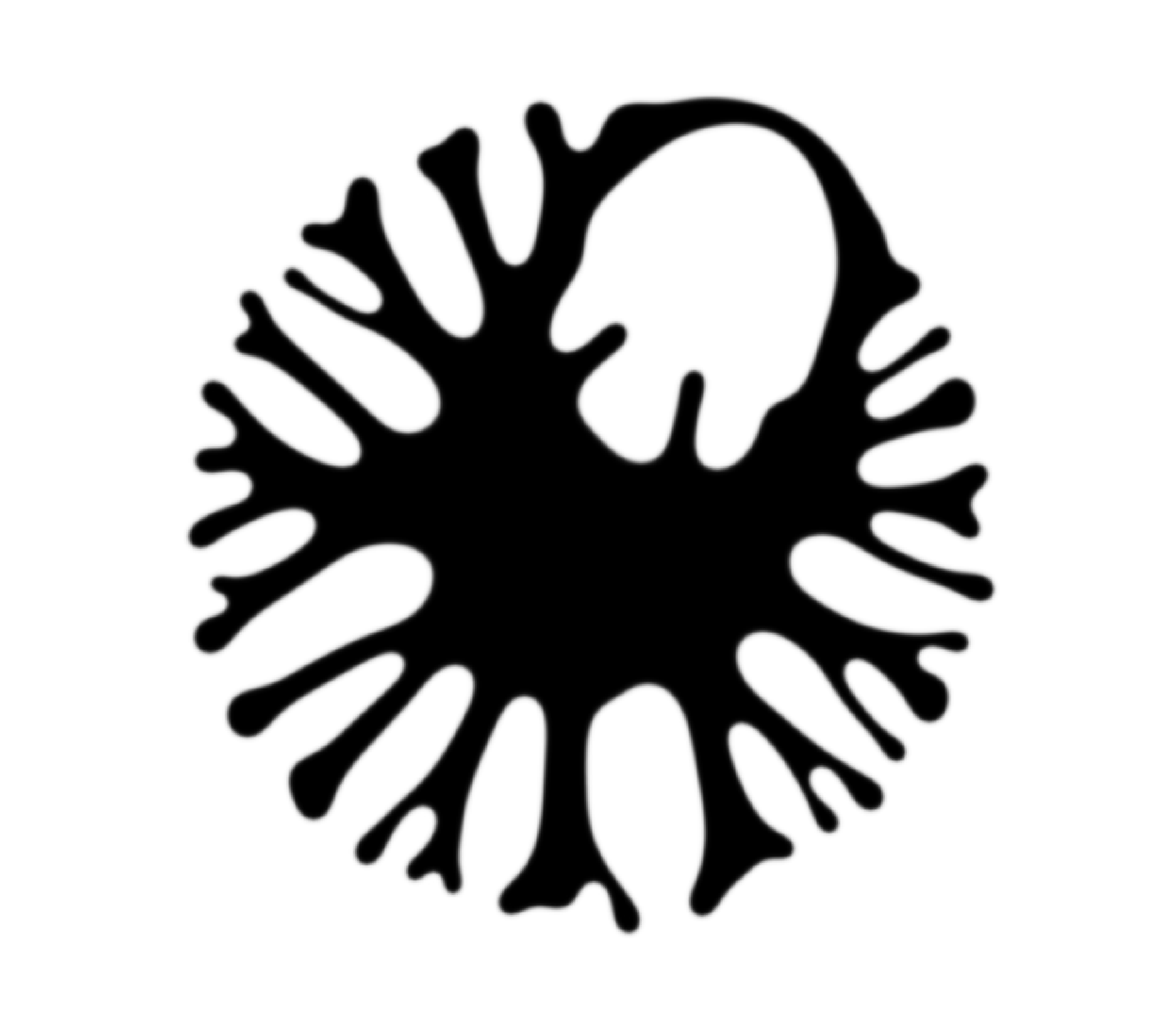} \\
        \Large (A) $\Bar{t}=0.00$ & \Large (B) $\Bar{t}=4.14$ & \Large (C) $\Bar{t}=8.35$ & \Large (D) $\Bar{t}=11.98$ 
    \end{tabular}
    \caption{
        Evolution of the phase field $\varphi$ at the nondimensionalised time points $\bar{t}=0.00$, (A) 4.14 (B), 8.35 (C), 11.98 (D) with $\ell_{\varphi}=1.333\times10^{-4}$, $\ell_{2}=1.5$, $\ell^{\varphi}_{r}=960$, $\ell^{\sigma}_{r}=0.28$, and $k_{2}=8.0$. The remaining parameters required to recreate the simulation are as listed in Table~\ref{tab:Parameter_List}. Distinct finger formation can be observed.
    }
    \label{fig:finger_formation}
\end{figure}  

\begin{figure}[htbp]
    \centering
    \begin{tabular}{cccc}
        \Large $\varphi(\bs{x},t)$ & \Large $\varphi(\bs{x},t)$ & \Large $\varphi(\bs{x},t)$ & \Large $\varphi(\bs{x},t)$ \\
        \includegraphics[width=0.23\textwidth]{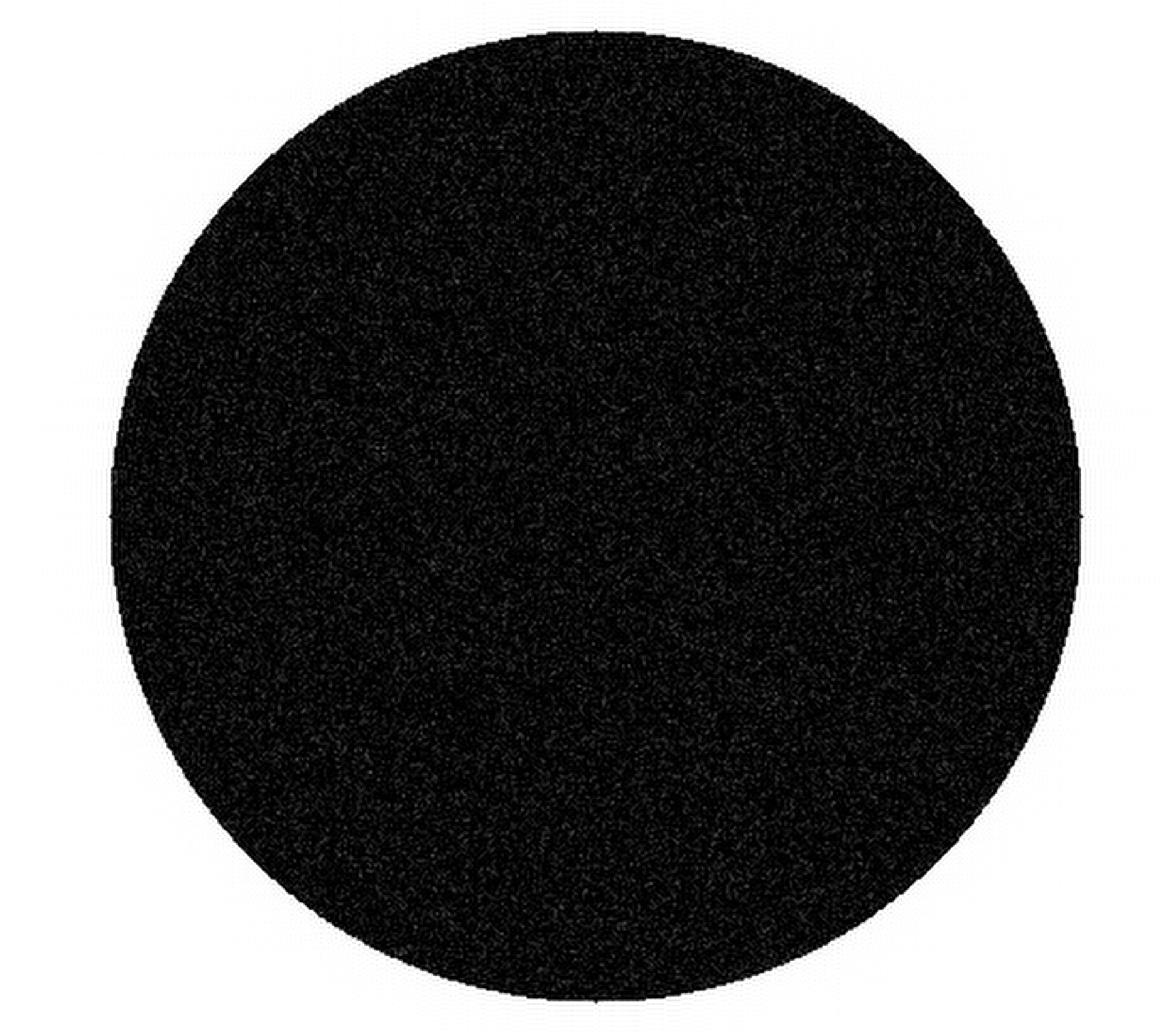} &
        \includegraphics[width=0.23\textwidth]{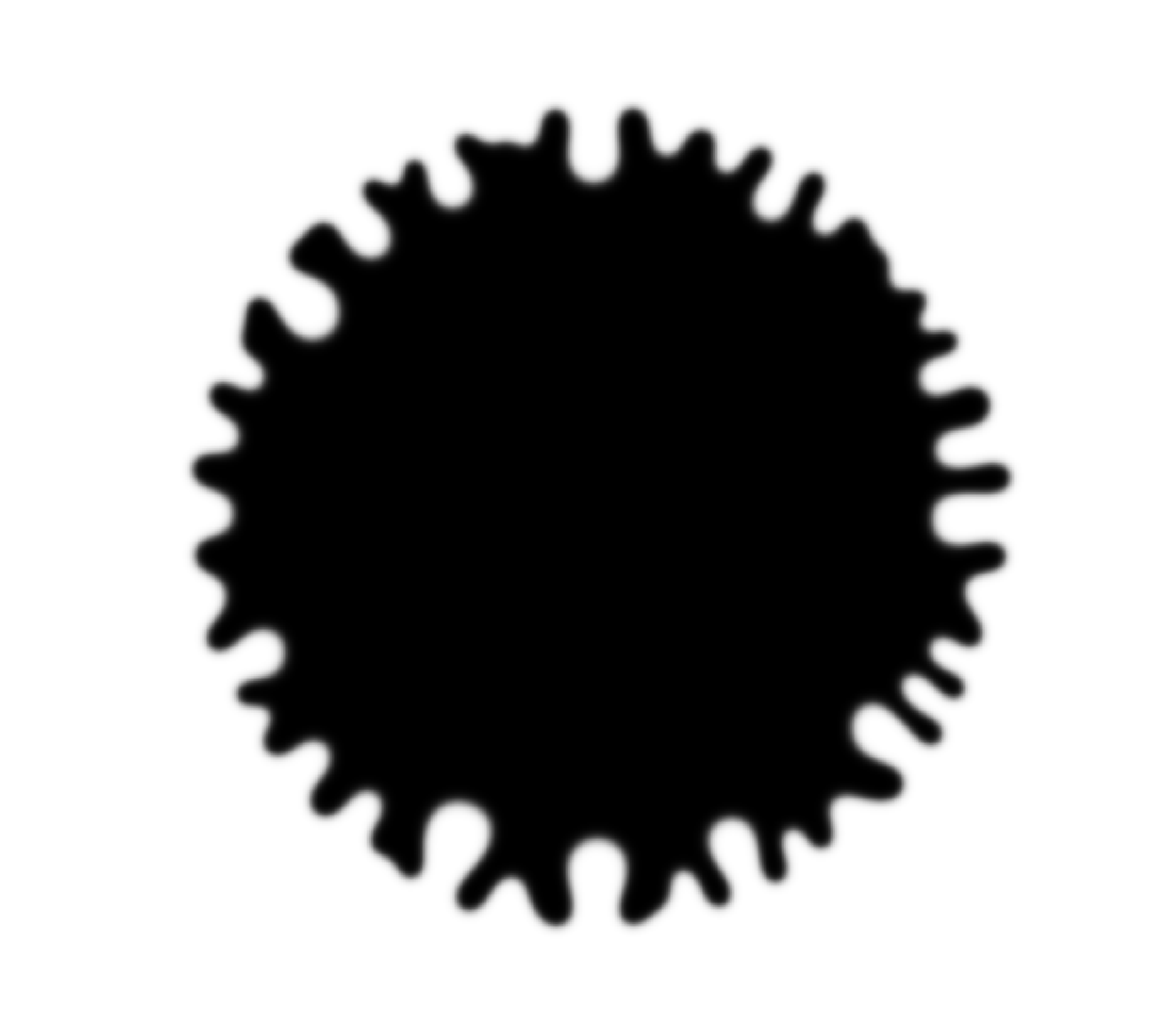} &
        \includegraphics[width=0.23\textwidth]{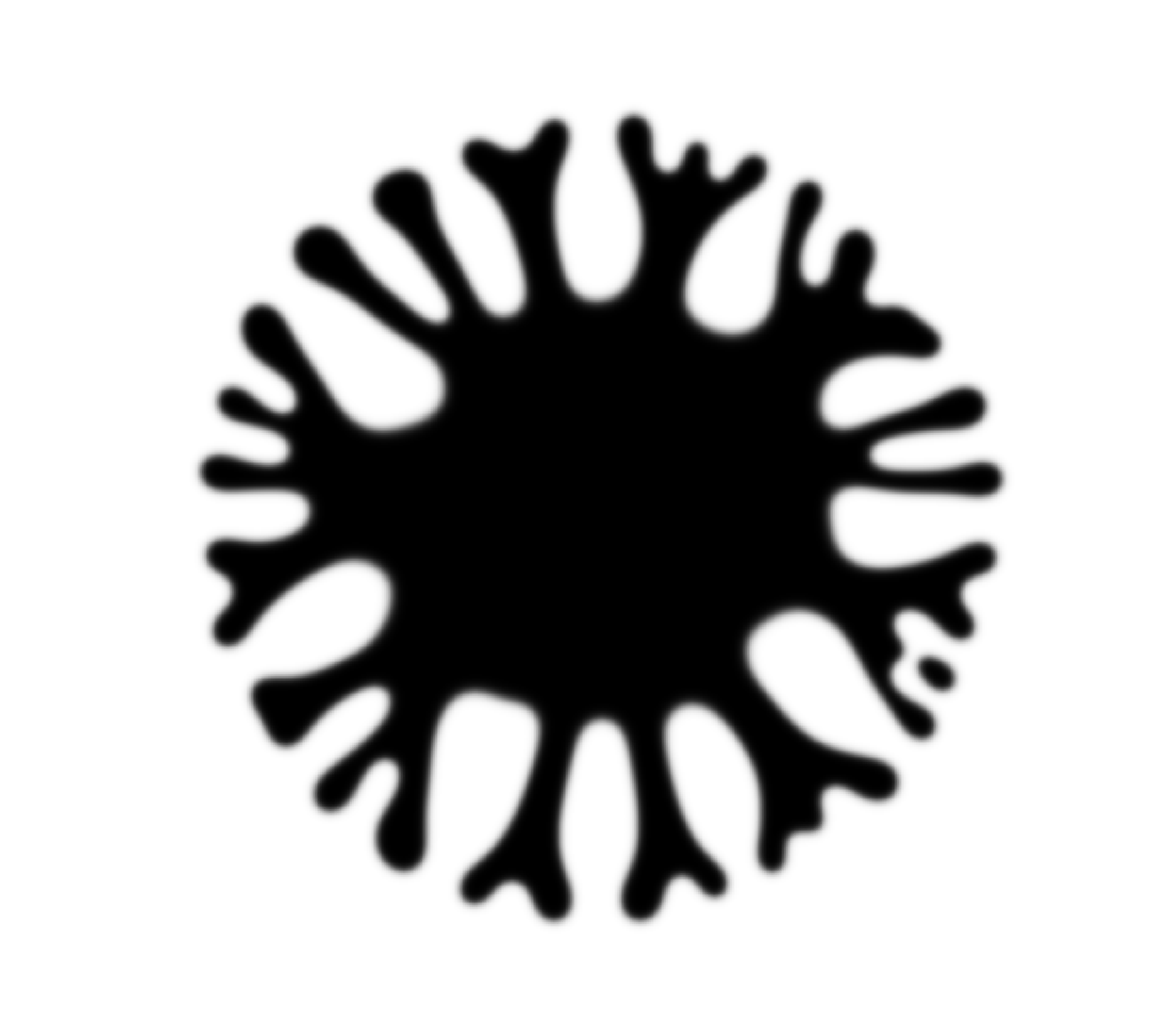} &
        \includegraphics[width=0.23\textwidth]{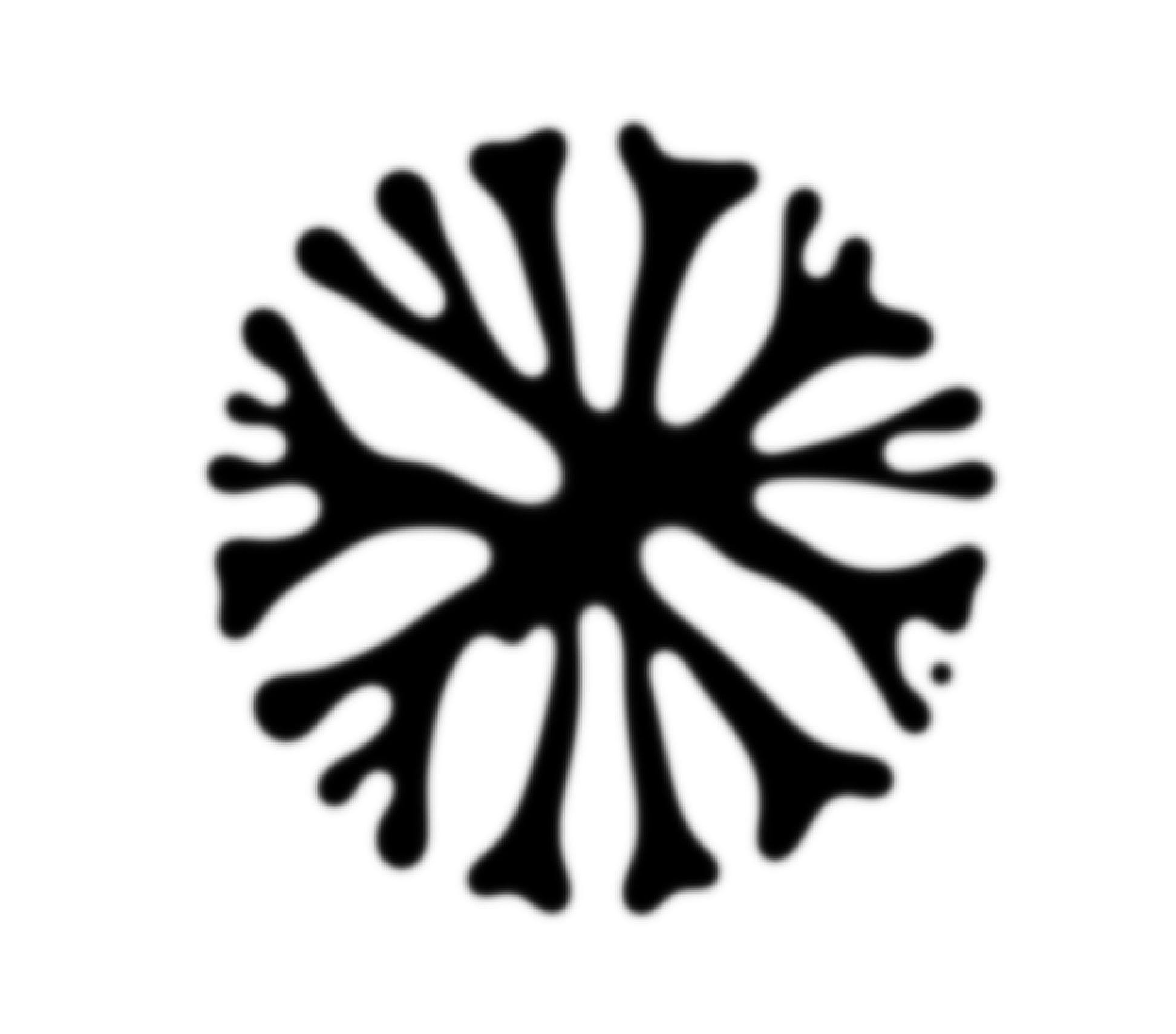} \\
        \Large (A) $\Bar{t}=0.00$ & \Large (B) $\Bar{t}=4.08$ & \Large (C) $\Bar{t}=7.51$ & \Large (D) $\Bar{t}=11.95$ 
    \end{tabular}
    \caption{
        Evolution of the phase field $\varphi$ at the time points at $\Bar{t}=0$ (A), 4.08 (B), 7.51 (C), 11.95 (D) of $\varphi$ with interface width increased to $\ell_{\varphi}=5.333\times10^{-4}$ in comparison to the simulation presented in Figure~\ref{fig:finger_formation}. The following parameters $\ell_{2}=1.5$, $\ell^{\varphi}_{r}=960$, $\ell^{\sigma}_{r}=0.28$ and $k_{2}=8.0$. The remaining parameters required to recreate the simulation are as listed in Table~\ref{tab:Parameter_List}. Finger formation in $\varphi(\bs{x},t)$ can be observed as forming during the degradation of the phase field. A fragment can be seen to form from one finger.
    }
    \label{fig:finger_formation_diffuse}
\end{figure} 

\begin{figure}[htbp]
    \centering
    \begin{tabular}{cccc}
        \Large $\varphi(\bs{x},t)$ & \Large $\varphi(\bs{x},t)$ & \Large $\varphi(\bs{x},t)$ & \Large $\varphi(\bs{x},t)$ \\
        \includegraphics[width=0.23\textwidth]{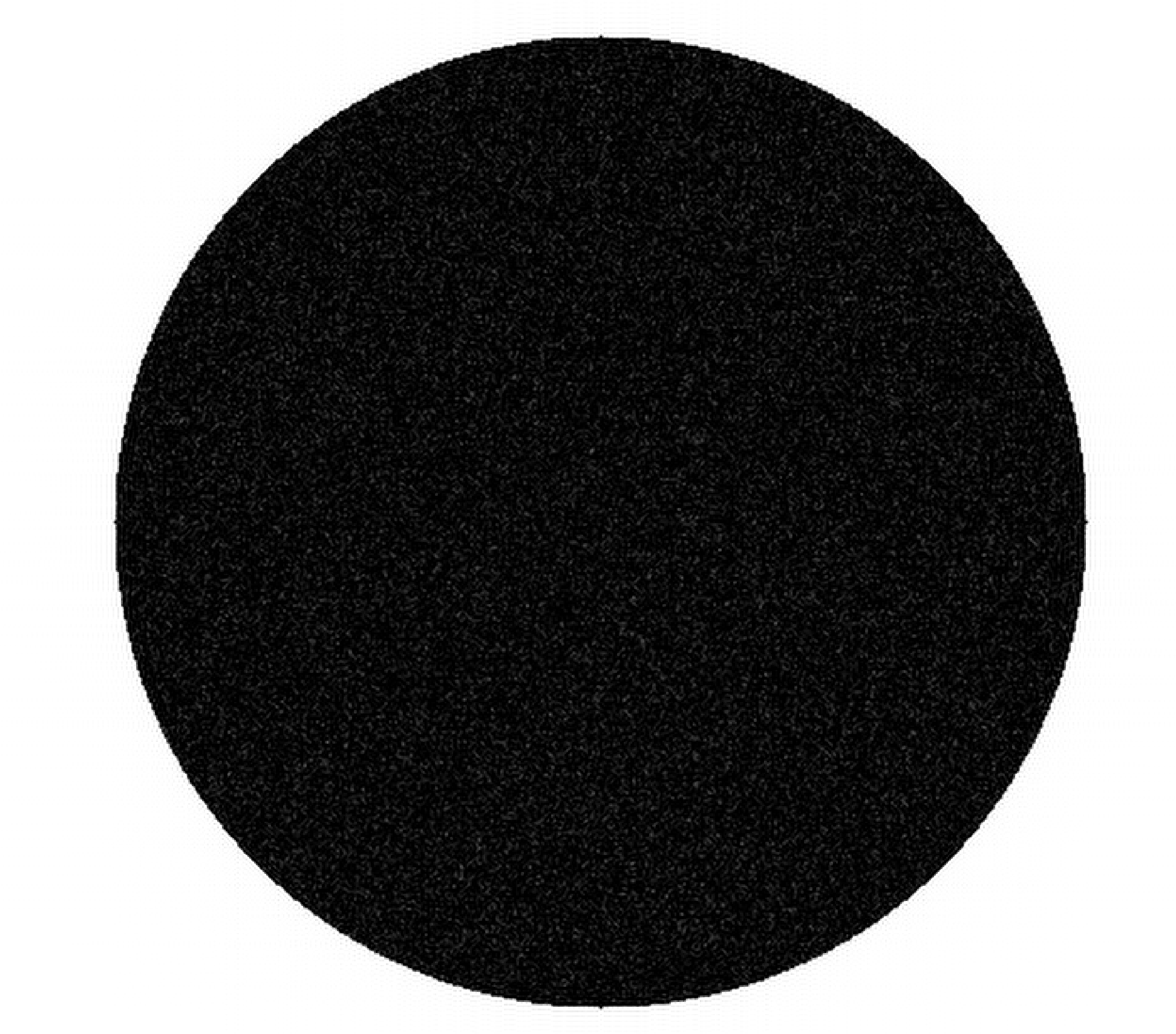} &
        \includegraphics[width=0.23\textwidth]{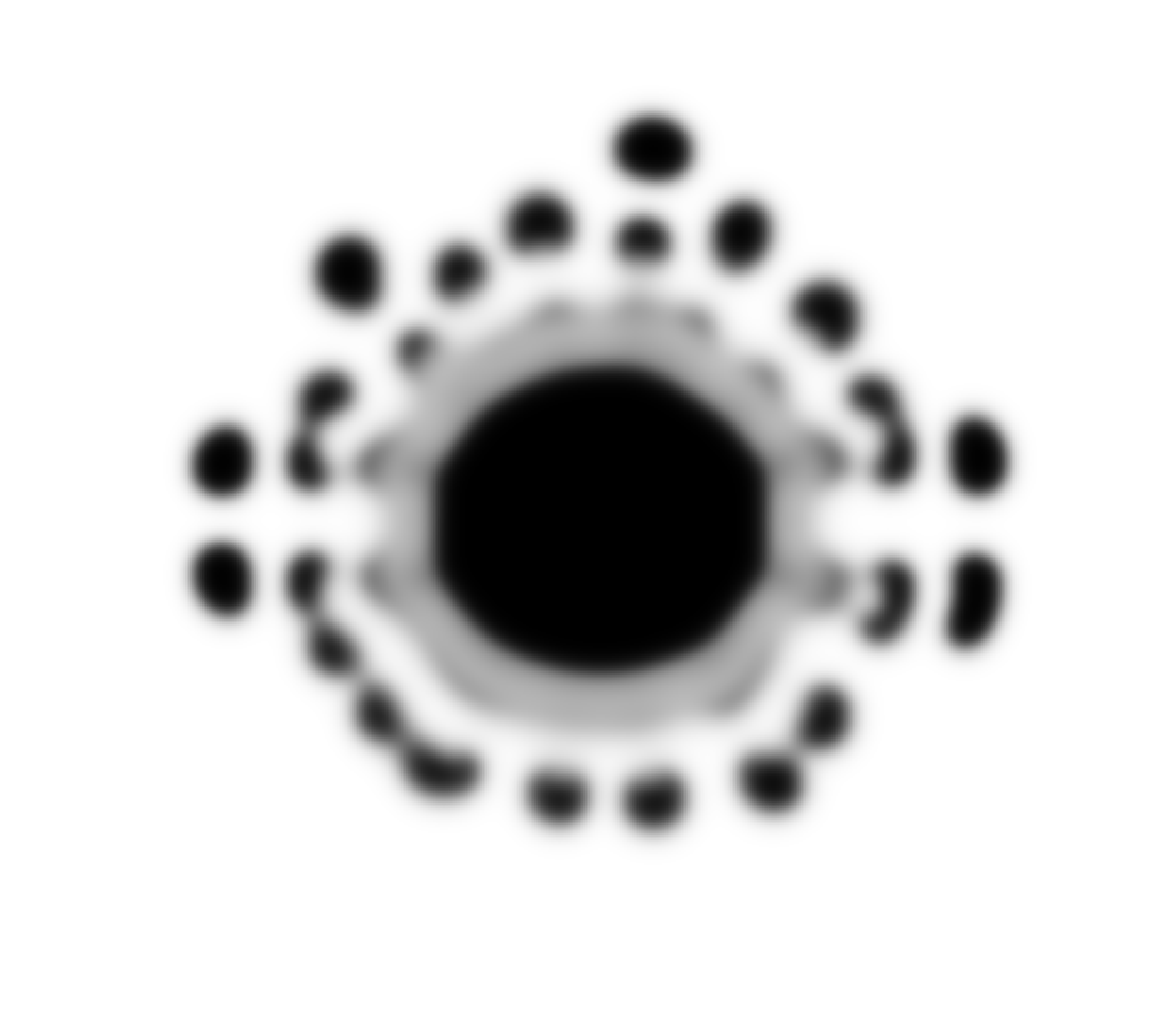} &
        \includegraphics[width=0.23\textwidth]{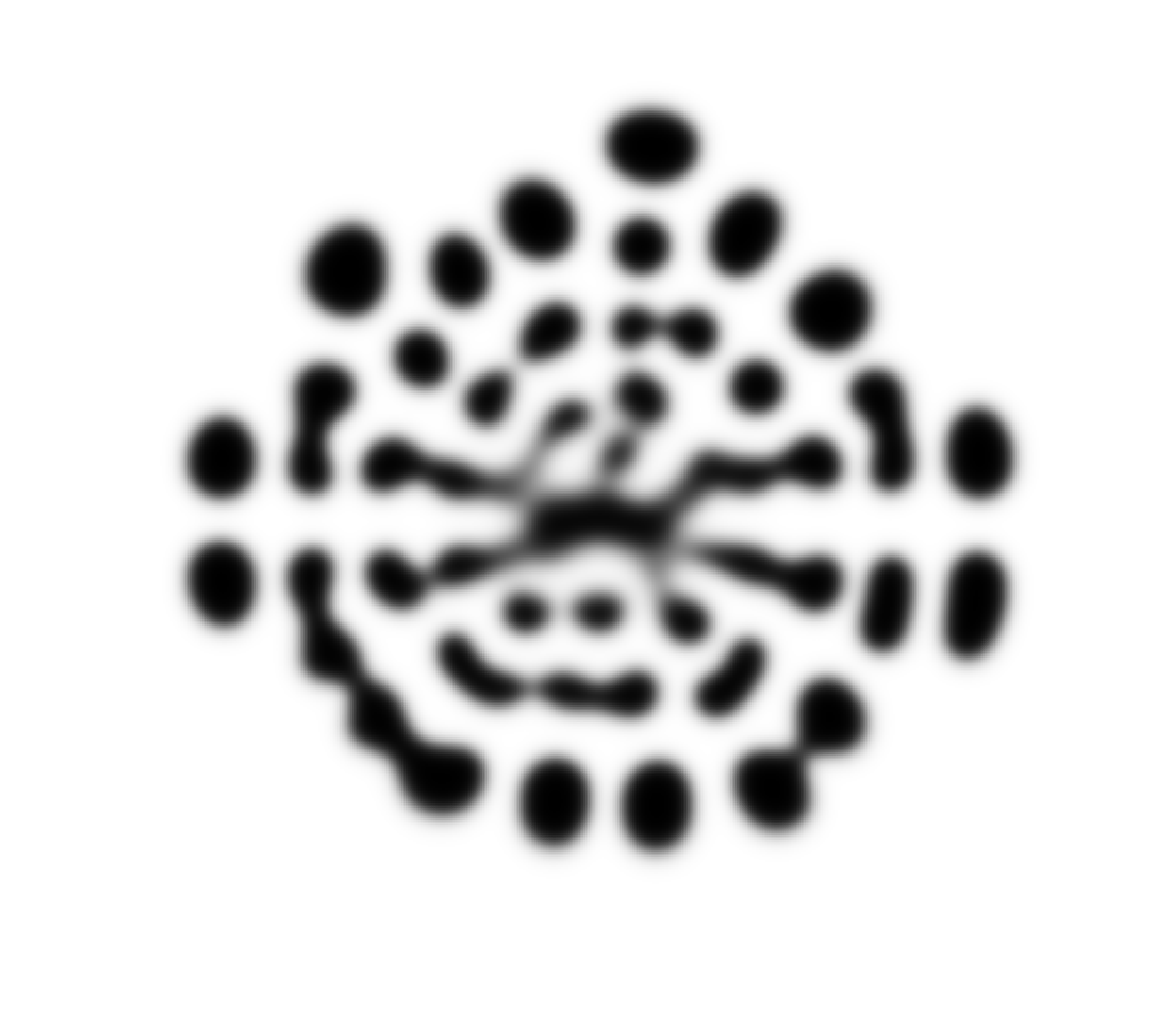} &
        \includegraphics[width=0.23\textwidth]{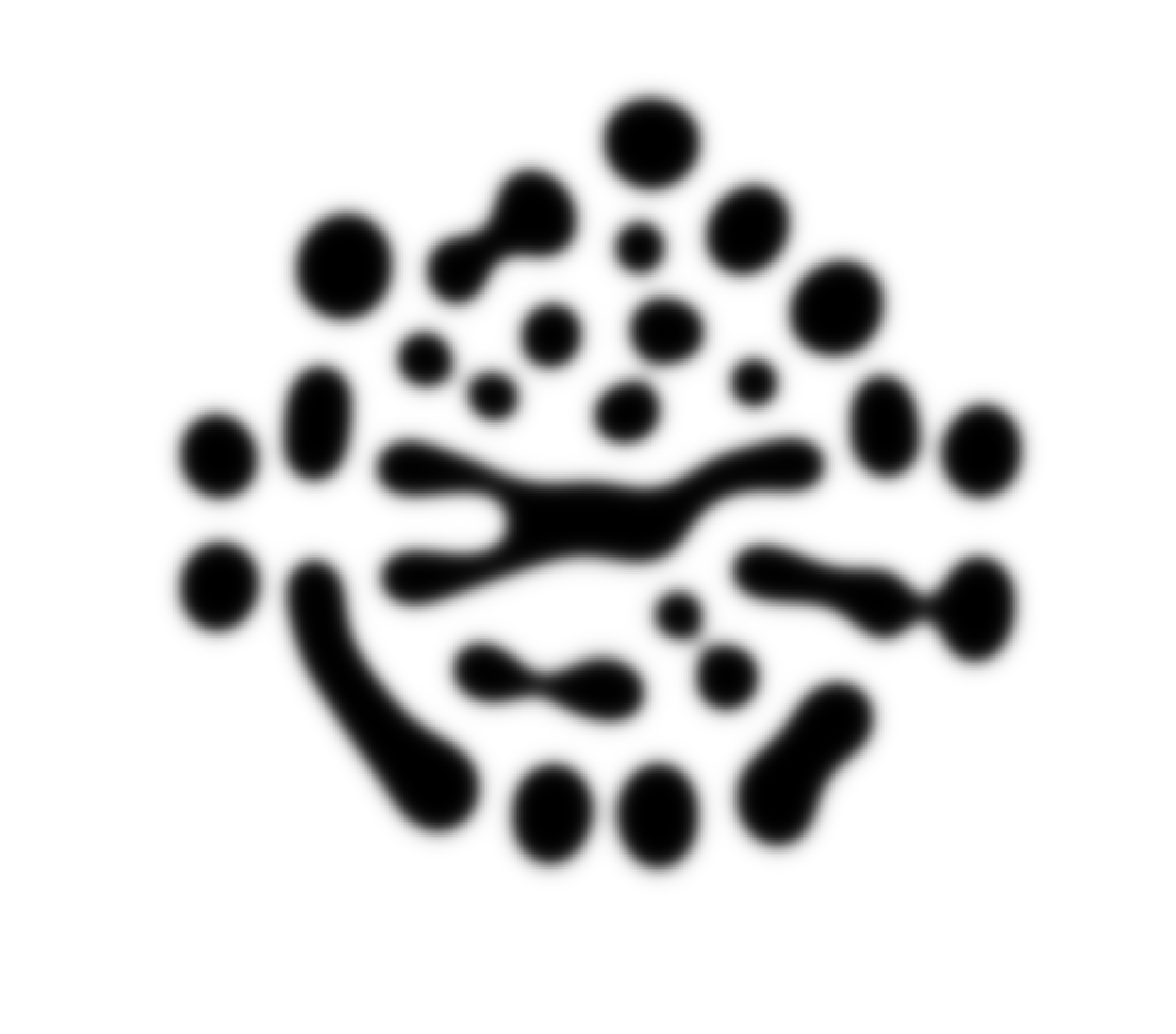} \\
         \Large (A) $\Bar{t}=0.00$ & \Large (B) $\Bar{t}=2.57$ & \Large (C) $\Bar{t}=4.42$ & \Large (D) $\Bar{t}=11.42$
    \end{tabular}
    \caption{
        Evolution of the phase field $\varphi$ at nondimensionalised time $\Bar{t}=0.00$ (A), 2.57 (B), 4.42 (C), 11.42 (D) of $\varphi$ with an interface width $\ell_{\varphi}=2.133\times10^{-3}$, $\ell_{2}=1.5$, $\ell^{\varphi}_{r}=960$, $\ell^{\sigma}_{r}=0.28$, and $k_{2}=8.0$. The remaining parameters required to recreate the simulation are as listed in Table~\ref{tab:Parameter_List}. No finger formation observed around the interface and fragmentation of the phase field $\varphi(\bs{x},t)$ is present during this simulation of apoptosis.
    }
    \label{fig:finger_formation_diffuse_max}
\end{figure}

It is known in phase-field equations that changes to the interface width can have a range of effects on the resulting dynamics, as a broader interface leads to a loss of detail in the simulation, as stated by Qin and Bhadeshia~\cite{qin2010phase}. As can be seen in Figures~\ref{fig:finger_formation}, \ref{fig:finger_formation_diffuse}, and \ref{fig:finger_formation_diffuse_max}, a lower $\tilde{\epsilon}$ value sharpens the interface, while at larger values, the phase field has a diffuse interface. Figure~\ref{fig:finger_formation_diffuse_max} uses the interfacial coefficient $\ell_{\varphi}=2.133\times10^{-3}$, and here, there is no finger formation during the degradation of the phase field $\varphi$. However, by reducing $\ell_{\varphi}=5.333\times10^{-4}$ (Figure~\ref{fig:finger_formation_diffuse}), fingers begin to form during the degradation of the phase field $\varphi$. At $\ell_{\varphi}=1.333\times10^{-4}$, finger formation in $\varphi$ is more distinct compared to simulations obtained with higher values for $\ell_{\varphi}$, as can be seen in Figures~\ref{fig:finger_formation_diffuse} and \ref{fig:finger_formation_diffuse_max}.

As seen in the bottom right-hand corner of Figure~\ref{fig:finger_formation_diffuse} (C) and in Figure~\ref{fig:finger_formation_diffuse_max} (B), (C), and (D), there is a pinch-off (fragmentation) from the finger during phase field $\varphi$ degradation. In these examples, pinch-off appears at the larger interfacial coefficients. A wider diffuse interface extends the region with $0<\varphi<1$, where the reaction is active, and also changes the interfacial contribution to the dynamics.

\subsection{Varying reaction strength and cytotoxic-phase coupling}
The parameters $\beta$, $k_{1}$, and $k_{2}$ are essential for controlling the rate at which the phase fields $\varphi$ and $\sigma$ evolve through reaction and rate coupling. The examples below compare the resulting transient morphologies. In this section, the interface width is kept as $\ell_{\varphi}=1.33\times10^{-4}$ to evaluate the effect of the other parameters. 

\subsubsection{Rate of apoptosis when varying of reaction parameters}
Varying $k_{1}$ or $k_{2}$ can greatly affect the rate of degradation exhibited by the cyto phase field $\varphi$, thus, exhibiting different dynamics in terms of nucleation and fragmentation.  

\begin{figure}[htbp]
    \centering
    \begin{tabular}{cccc}
        \Large $\varphi(\bs{x},t)$ & \Large $\varphi(\bs{x},t)$ & \Large $\varphi(\bs{x},t)$ & \Large $\varphi(\bs{x},t)$ \\
        \includegraphics[width=0.23\textwidth]{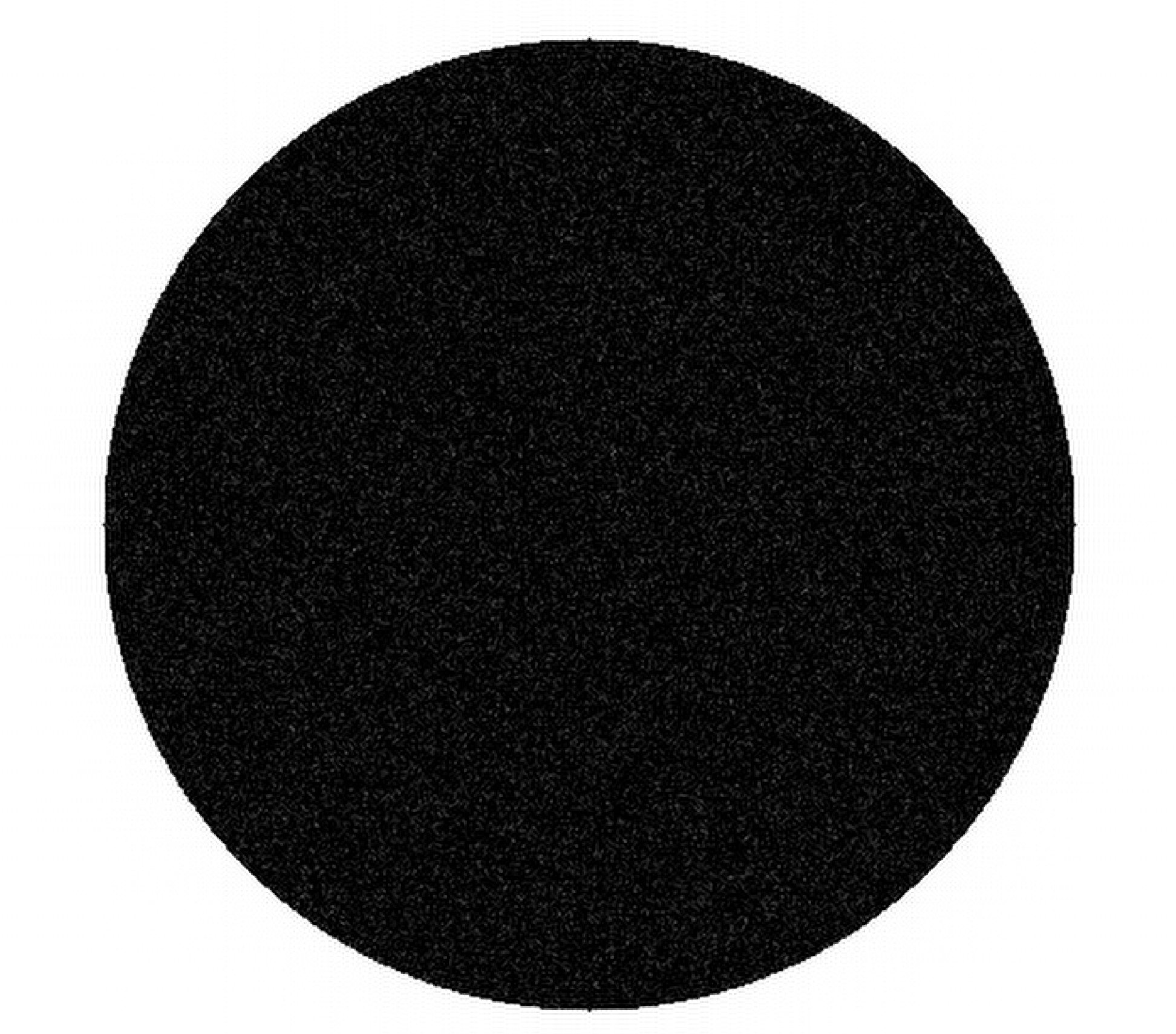} &
        \includegraphics[width=0.23\textwidth]{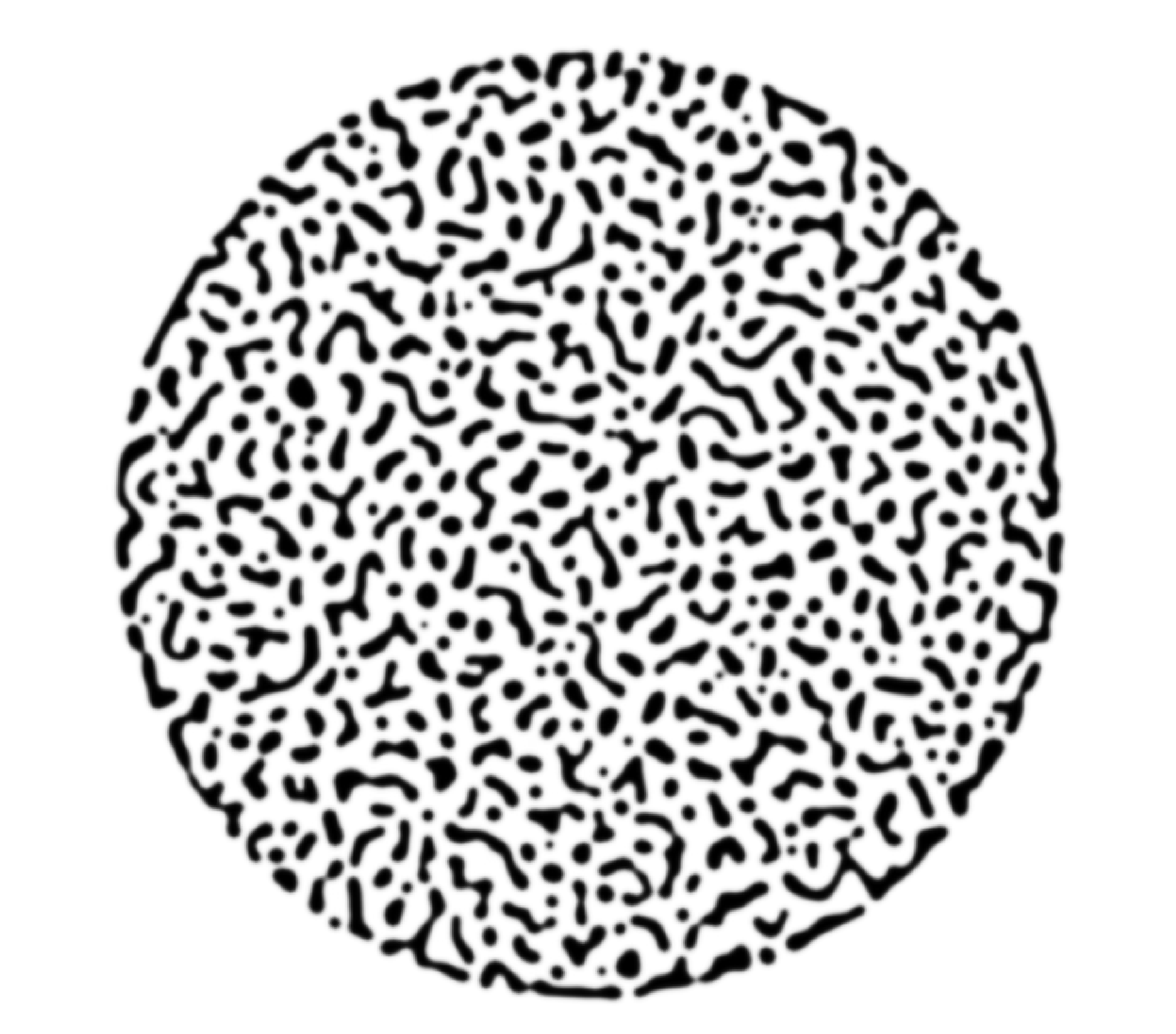} &
        \includegraphics[width=0.23\textwidth]{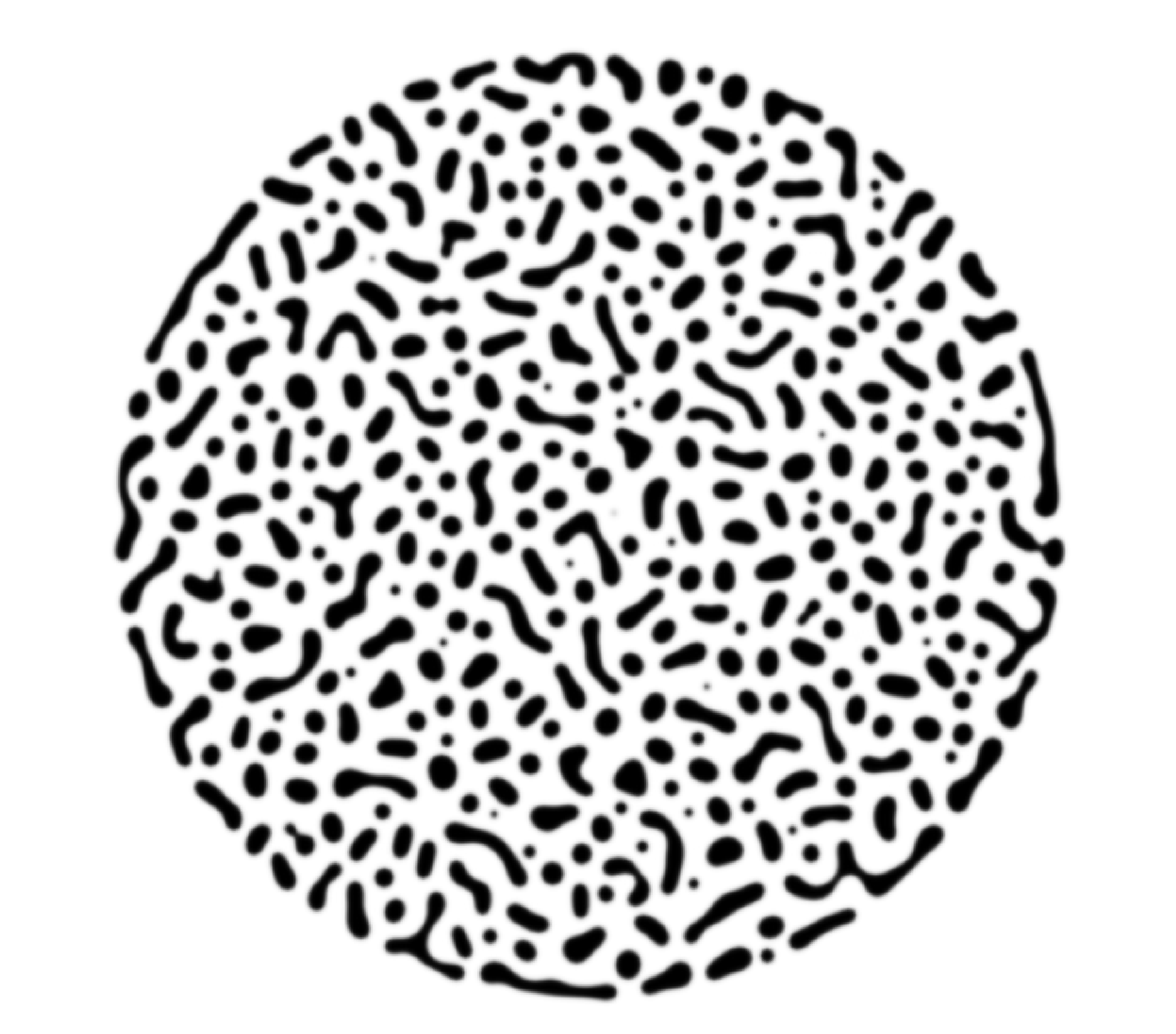} &
        \includegraphics[width=0.23\textwidth]{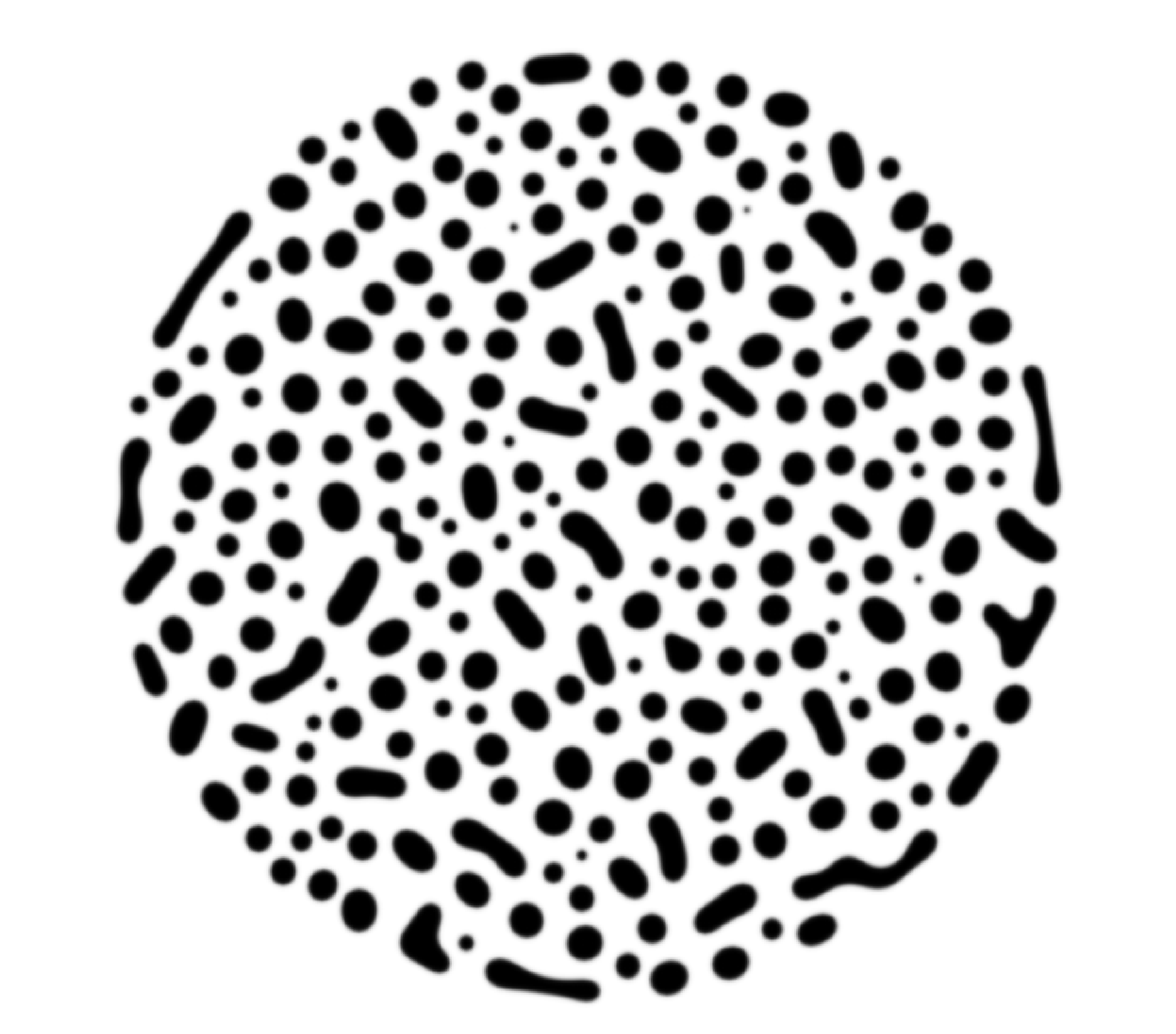} \\
        \Large (A) $\Bar{t}=0.00$ & \Large (B) $\Bar{t}=1.03$ & \Large (C) $\Bar{t}=3.87$ & \Large (D) $\Bar{t}=11.97$ 
    \end{tabular}
    \caption{
        Evolution of the phase field $\varphi$ at the time points $\Bar{t}=0.00$ (A), 1.03 (B), 3.87 (C), 11.97 (D) with $\ell_{\varphi}=1.333\times10^{-4}$, $\ell_{2}=1.5$, $\ell^{\varphi}_{r}=4240$, $\ell^{\sigma}_{r}=1.3$, and $k_{2}=1.0$. The remaining parameters have the values listed in Table~\ref{tab:Parameter_List}. The combined reaction-parameter change produces more fragments in this example than in Figure~\ref{fig:finger_formation}.
    }
    \label{fig:higher_reaction_rate}
\end{figure}

Figure~\ref{fig:higher_reaction_rate} retains $\ell_\varphi$ and $\ell_2$ from Figure~\ref{fig:finger_formation}, while increasing $k_1$ from $0.9$ to $4.0$ and decreasing $k_2$ from $8.0$ to $1.0$. The tabulated reaction coefficients are $\ell_r^\varphi=4240$ and $\ell_r^\sigma=1.27$ (rounded to $1.3$ in the captions). This combined parameter change produces faster cyto-phase loss and more fragments in the illustrated case. Appendix~\ref{sc:local_threshold} separates the amplitude and sensitivity effects analytically and shows that the reaction-strength ordering of these parameter sets depends on exposure.

\subsubsection{Varying cytotoxic-phase coupling}

The coefficient $\ell_2=\beta/\alpha$ multiplies $\dot\varphi$ in the cytotoxic-phase balance. During cyto-phase loss, this contribution is negative, and increasing $\ell_2$ strengthens the instantaneous coupling to cytotoxic-phase depletion.

\begin{figure}[htbp]
    \centering
    \begin{tabular}{ccccc}
        \Large $\varphi(\bs{x},t)$ & \Large $\varphi(\bs{x},t)$ & \Large $\varphi(\bs{x},t)$ & \Large $\varphi(\bs{x},t)$ \\
        \includegraphics[width=0.23\textwidth]{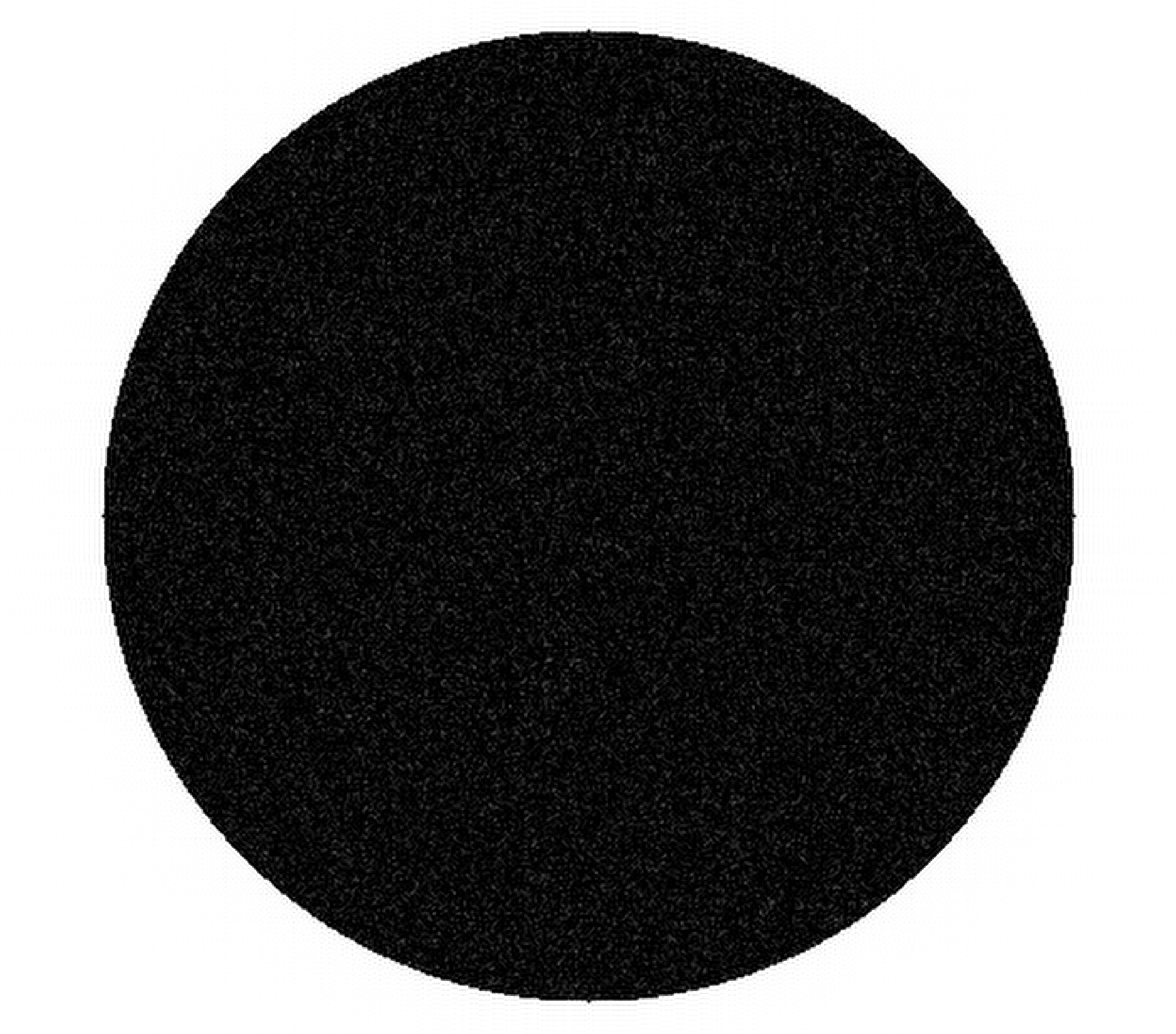} &
        \includegraphics[width=0.23\textwidth]{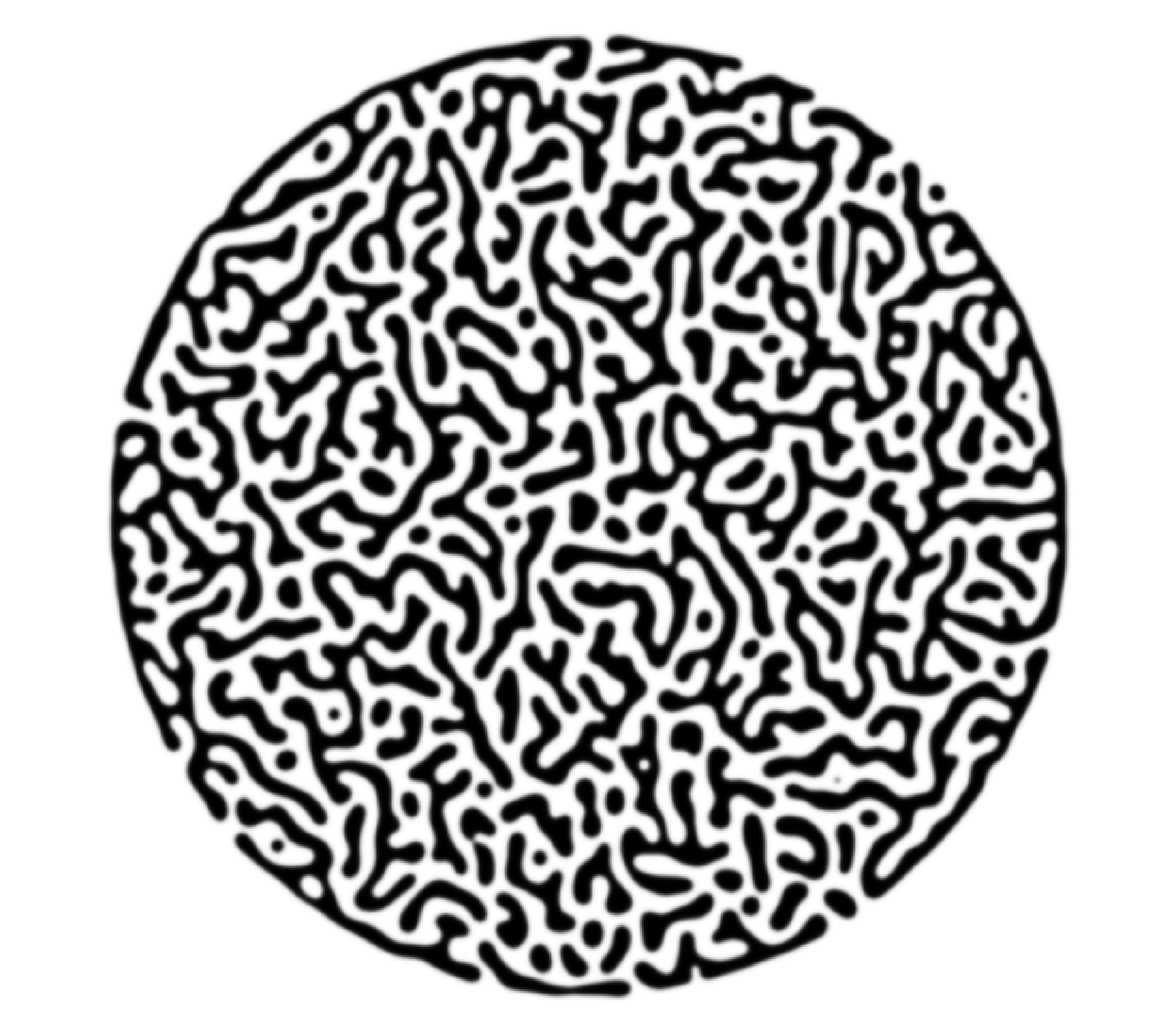} &
        \includegraphics[width=0.23\textwidth]{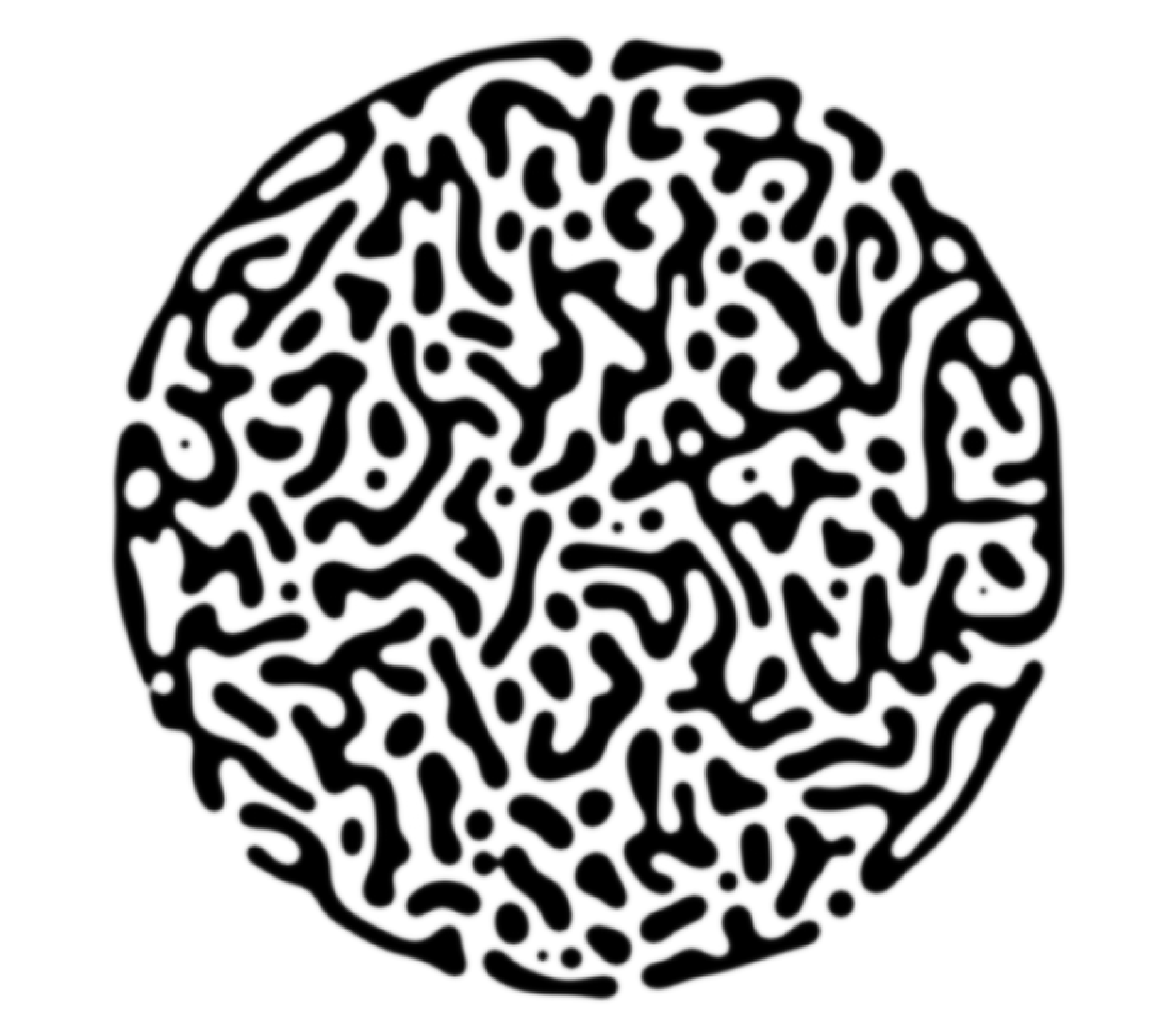} &
        \includegraphics[width=0.23\textwidth]{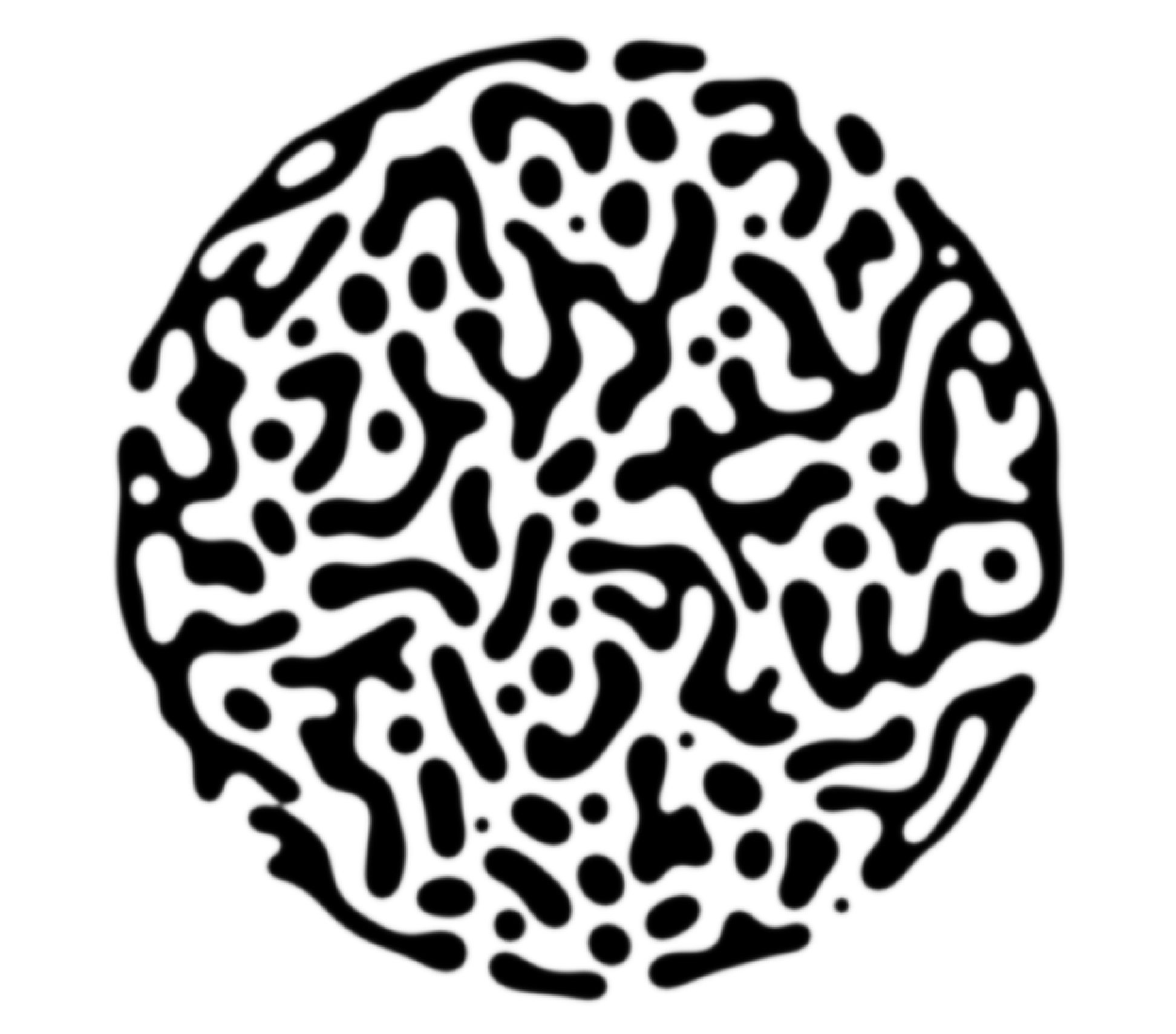} \\
        \Large (A) $\Bar{t}=0.00$ & \Large (B) $\Bar{t}=2.88$ & \Large (C) $\Bar{t}=4.32$ & \Large (D) $\Bar{t}=5.28$
    \end{tabular}
    \caption{
        Evolution of the phase field $\varphi$ at the time points $\Bar{t}=0.00$ (A), 2.88 (B), 4.32 (C), 5.28 (D) with $\ell_{\varphi}=1.333\times10^{-4}$, $\ell_{2}=2.0$, $\ell^{\varphi}_{r}=4240$, $\ell^{\sigma}_{r}=1.3$, and $k_{2}=1.0$. The remaining parameters are listed in Table~\ref{tab:Parameter_List}. Larger fragments are formed with $\ell_2=1.5$ and no nucleation.
    }
    \label{fig:higher_reaction_rate_varied_beta}
\end{figure}

Figure~\ref{fig:higher_reaction_rate_varied_beta} shows an apoptotic simulation where $\ell_{2}=2.0$, $k_1=4.0$ ($\ell_{r}^{\varphi}=4240$ and $\ell_{r}^{\sigma}=1.27$), $k_2=1.0$ and $\ell_{\varphi}=1.33\times10^{-4}$. Compared with Figure~\ref{fig:higher_reaction_rate}, fragmentation is less advanced in this example with stronger cytotoxic-phase coupling. 
\begin{figure}[htbp]
    \centering
    \begin{tabular}{cccc}
        \Large $\varphi(\bs{x},t)$ & \Large $\varphi(\bs{x},t)$ & \Large $\varphi(\bs{x},t)$ & \Large $\varphi(\bs{x},t)$ \\
        \includegraphics[width=0.23\textwidth]{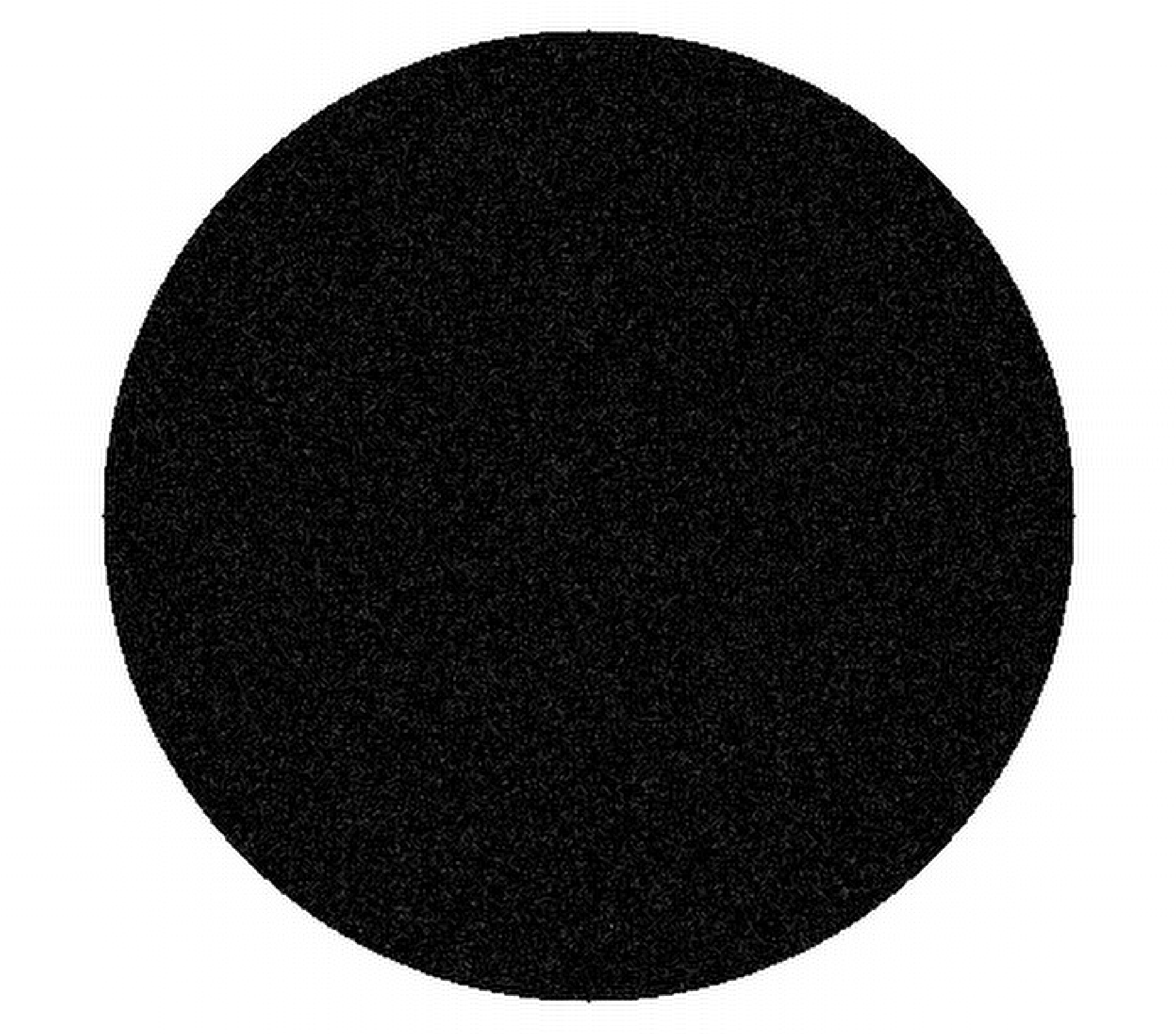} &
        \includegraphics[width=0.23\textwidth]{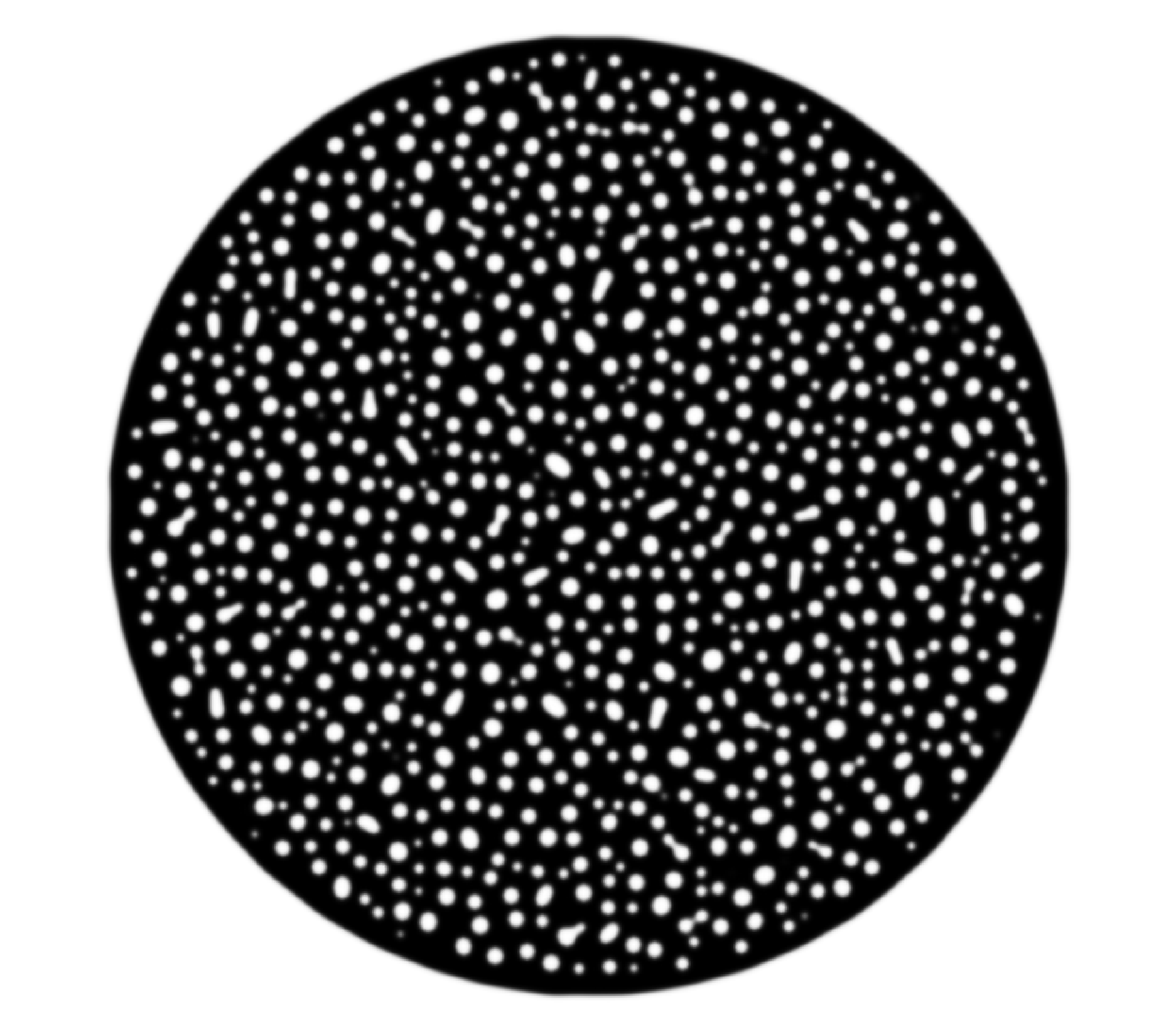} &
        \includegraphics[width=0.23\textwidth]{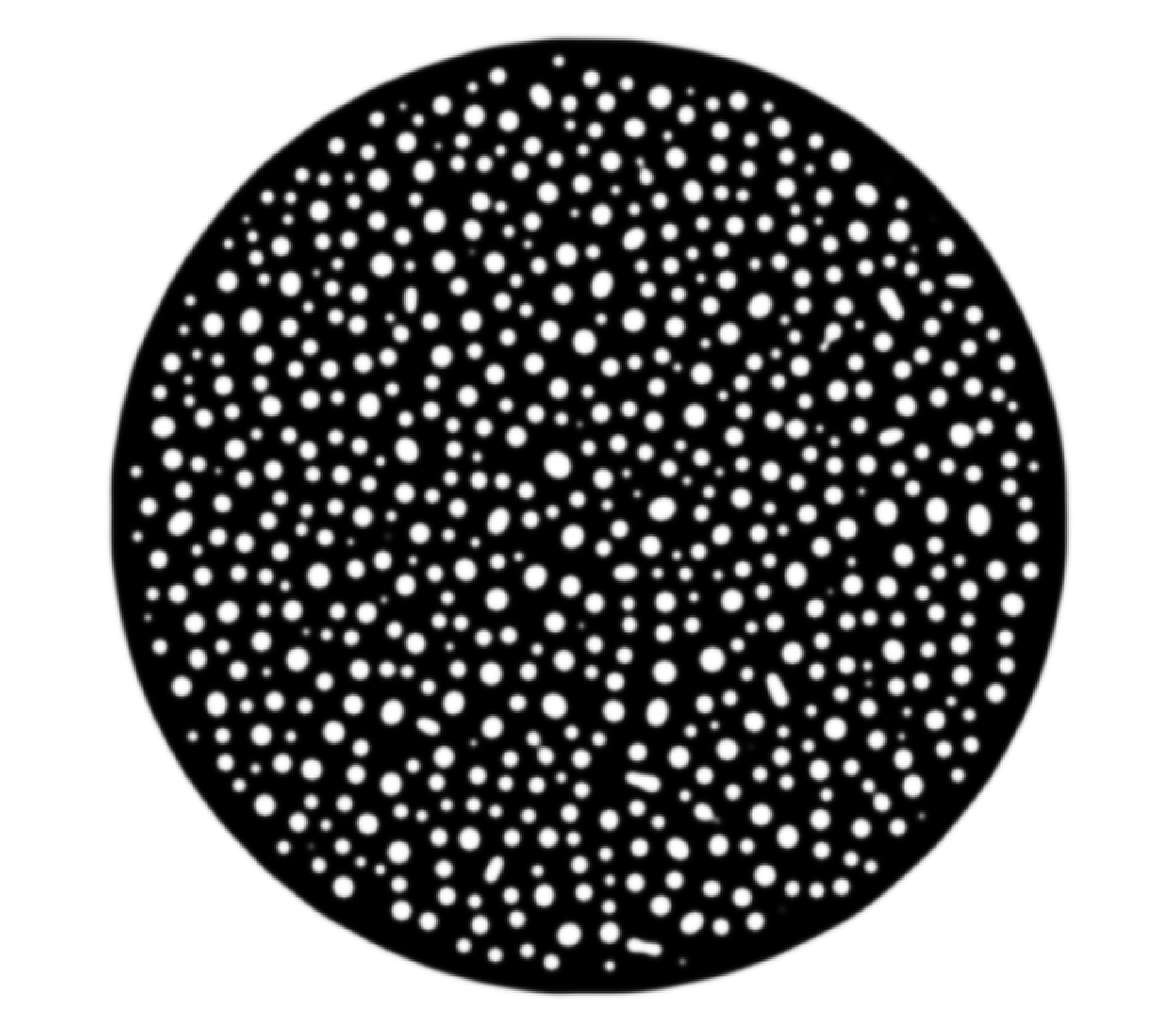} &
        \includegraphics[width=0.23\textwidth]{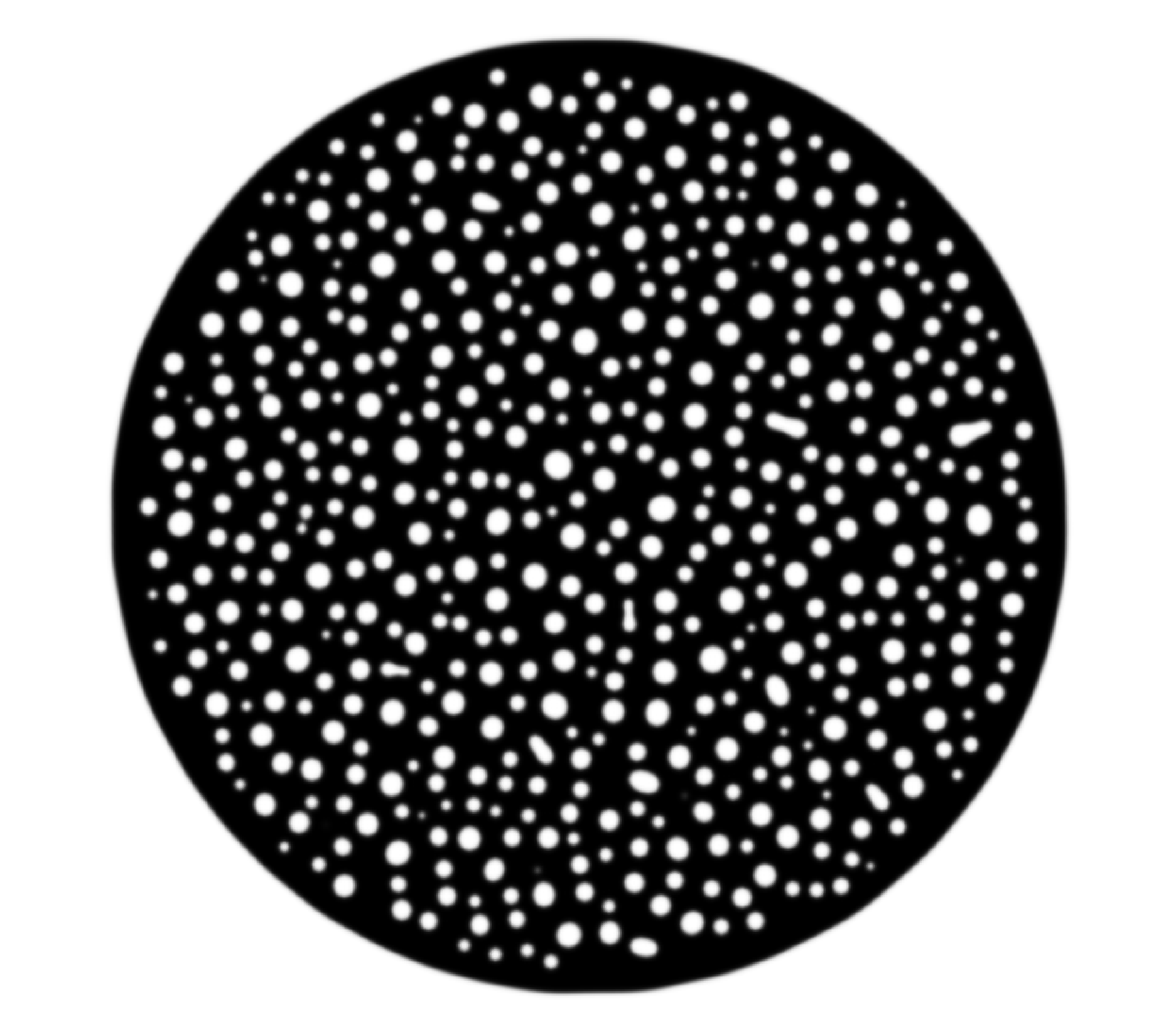} \\
        \Large (A) $\Bar{t}=0.00$ & \Large (B) $\Bar{t}=2.26$ & \Large (C) $\Bar{t}=4.37$ & \Large (D) $\Bar{t}=5.98$ 
    \end{tabular}
    \caption{
        Evolution of the phase field $\varphi$ at the time points $\Bar{t}=0.00$ (A), 2.26 (B), 4.37 (C), 5.98 (D) of the phase field $\varphi$ with $\ell_{\varphi}=1.333\times10^{-4}$, $\ell_{2}=5.0$, $\ell^{\varphi}_{r}=4240$, $\ell^{\sigma}_{r}=1.3$, and $k_{2}=1.0$. The remaining parameters are listed in the Table~\ref{tab:Parameter_List}. Nucleation is increased, and fragmentation is decreased compared to $\ell_2=1.5$ in Figure~\ref{fig:higher_reaction_rate}.
    }
    \label{fig:higher_reaction_rate_varied_beta_max}
\end{figure}

Comparing Figure~\ref{fig:higher_reaction_rate_varied_beta_max} ($\ell_{2}=5.0$) with Figure~\ref{fig:higher_reaction_rate} ($\ell_{2}=1.5$), we see that the use of $\ell_{2}=5.0$ results in an increased number of nucleation sites and no fragmentation taking place. Furthermore, when compared to Figure~\ref{fig:finger_formation} ($\ell_2=1.5$, $\ell_r^\varphi=960$, $\ell_r^\sigma=0.28$ and $k_2=8.0$) nucleation can be seen in both simulations. However, finger formation is only present in Figure~\ref{fig:finger_formation}, and fragmentation is not present in either simulation. For $\ell_{2}=2.0$ (Figure~\ref{fig:higher_reaction_rate_varied_beta}), we observe that both nucleations and fragmentations are present. At the time point of $\bar{t}=2.88$, the number of fragments is greater, and the nucleations are not well defined. However, as phase separation progresses and at $\bar{t}=5.28$, large, well-defined nucleation sites are observed. This indicates that nucleations and fragments can coexist and fragments can transition into nucleations through phase separation.  

\subsection{Internal configurational forces}

Configurational forces describe the energetic driving of interface motion and deformation in this continuum model. Interfacial free energy therefore has a mechanical role even when membrane and cytoskeletal constituents are not resolved separately, consistent with the microforce framework of Gurtin~\cite{gurtin1996generalized}. In the constant-coefficient isotropic case, the interfacial contribution to the variational derivative is $-\epsilon^2\Delta\varphi$. For a local profile $\varphi=q(d)$, with $d$ a smooth signed-distance function and primes denoting derivatives of $q$ with respect to $d$,
\begin{equation}
 \Delta\varphi=q''(d)+\kappa q'(d),\qquad \kappa=\Div\nabla d.
\end{equation}
Thus the interfacial energy produces a curvature-dependent response. This provides an effective description of membrane deformation, without establishing equivalence to a curvature-squared bending energy with an independently specified bending modulus. Du et al.~\cite{du2005spontaneous} demonstrate how a specifically prescribed bending and spontaneous-curvature energy can be represented by phase fields.

Appendix~\ref{sc:interfacial_tension} makes the connection to effective interfacial tension explicit: integrating the cyto-phase configurational stress across a planar equilibrium interface gives the surface stress associated with tension and, in the thin-interface approximation, its curvature force. The derivation treats smooth anisotropic orientations and separates the cytotoxic-phase stress. Quantitative comparison with membrane or cortical tension requires dimensional calibration. We evaluate Equation~\eqref{equ:internal_config_force} around the cyto-phase boundary and cavities to interpret the simulated morphological transitions.
\begin{figure}[htbp]
    \centering
    \begin{tabular}{cccc}
        \Large $\varphi(\bs{x},t), \ \bs{f}(\bs{x},t)$ & \Large $\sigma(\bs{x},t), \ \bs{f}(\bs{x},t)$ & \Large $\varphi(\bs{x},t), \ \bs{f}(\bs{x},t)$ & \Large $\sigma(\bs{x},t), \ \bs{f}(\bs{x},t)$  \\
        \includegraphics[width=0.23\textwidth]{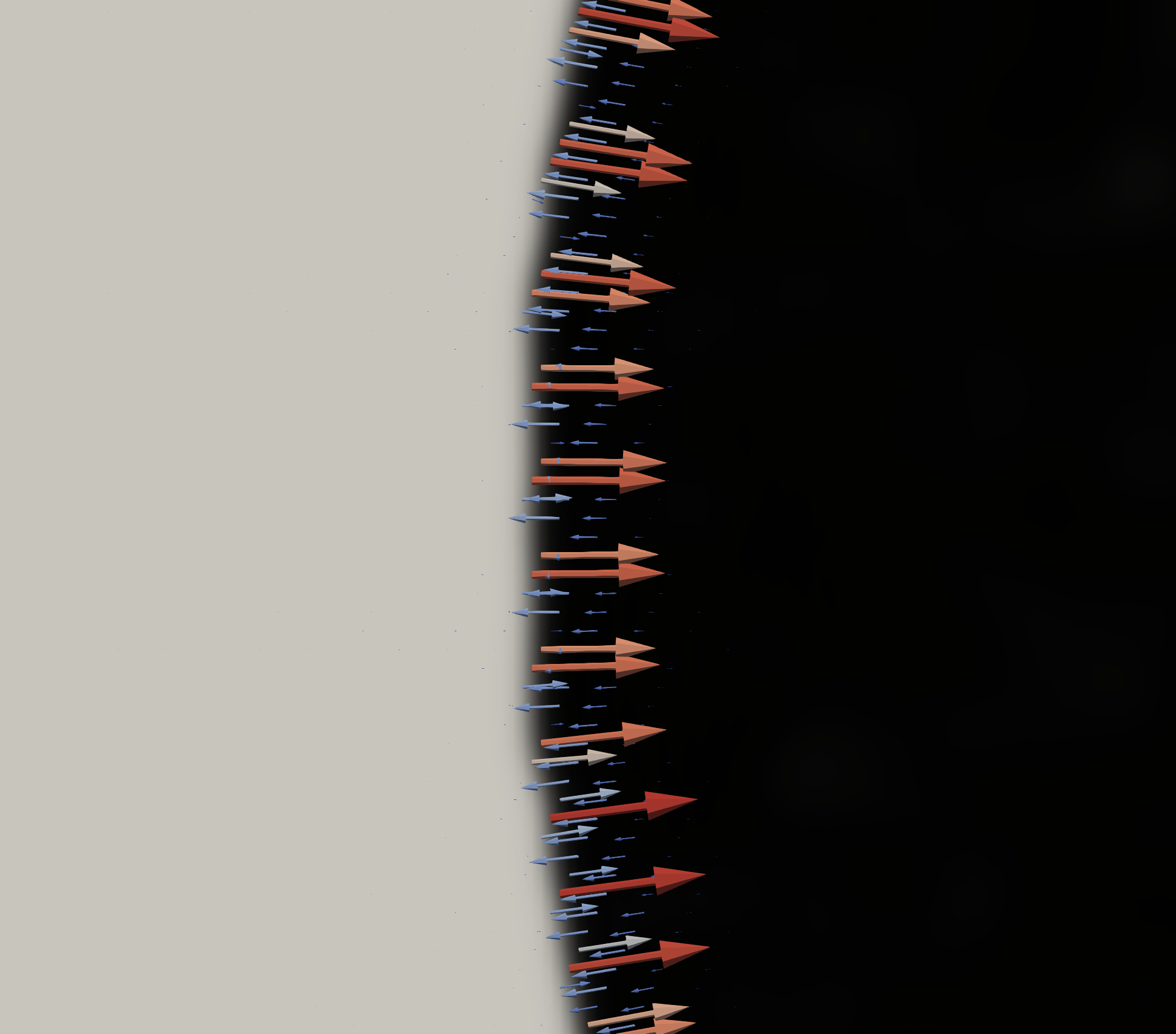} &
        \includegraphics[width=0.23\textwidth]{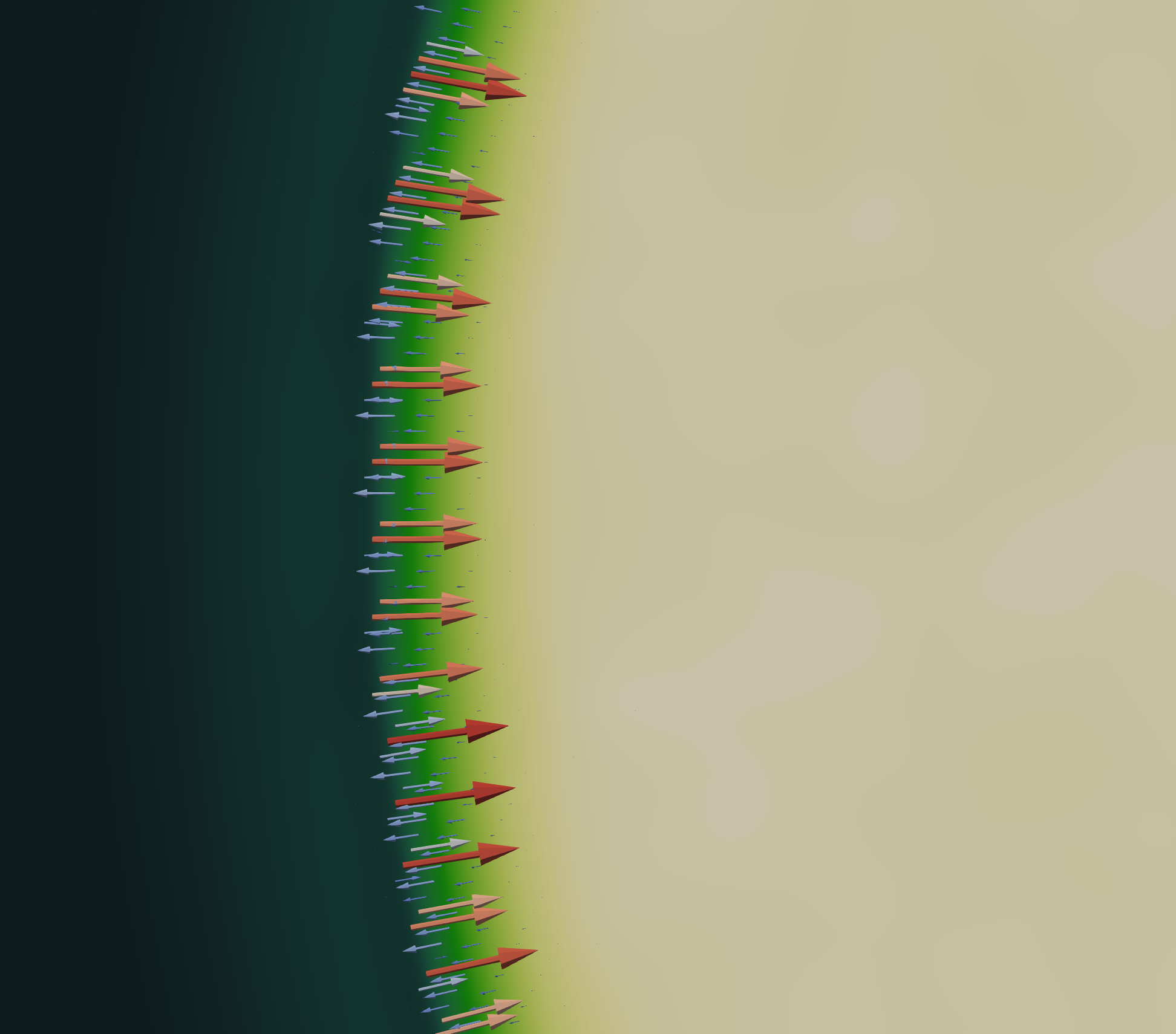} &
        \includegraphics[width=0.23\textwidth]{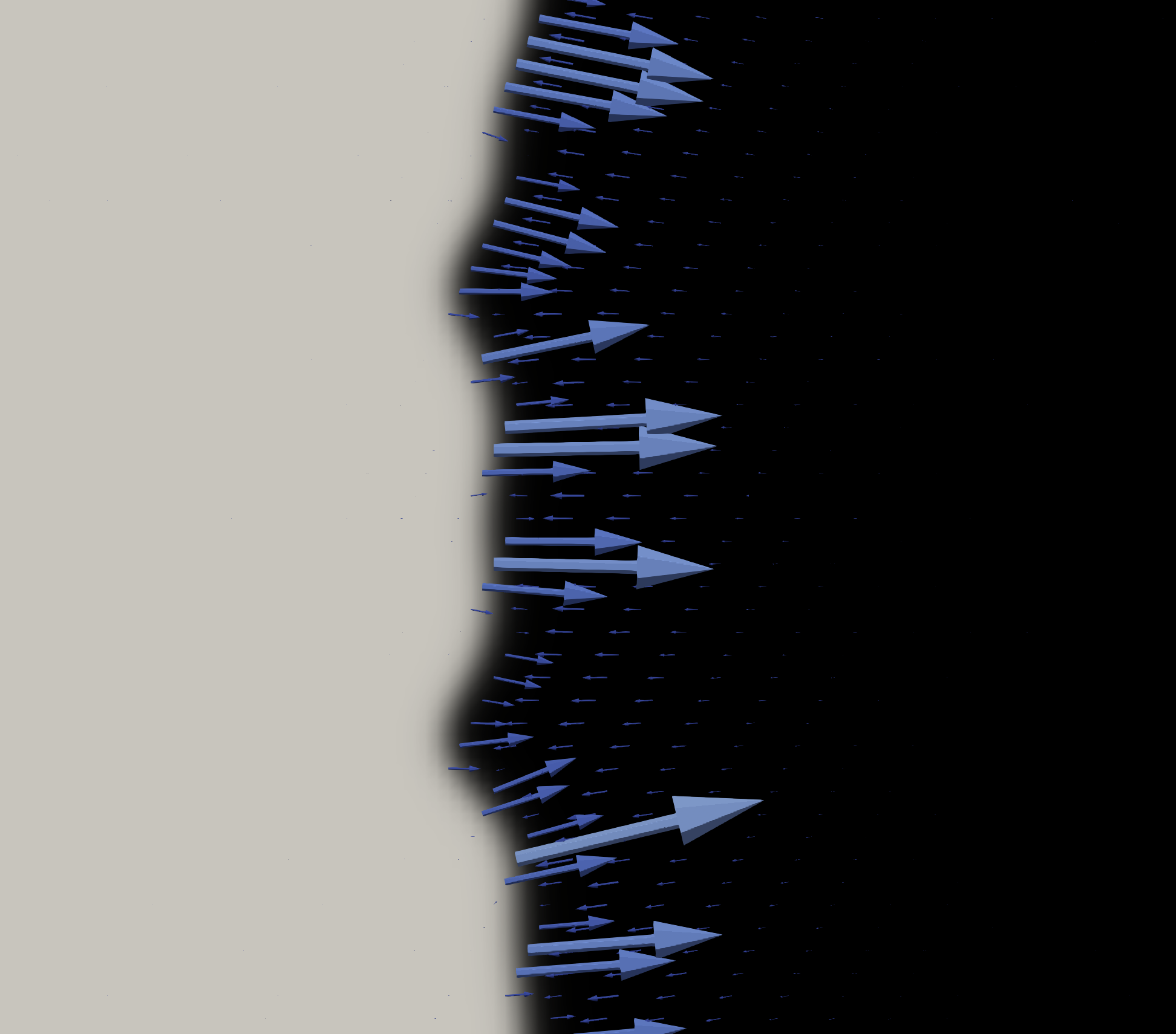} &
        \includegraphics[width=0.23\textwidth]{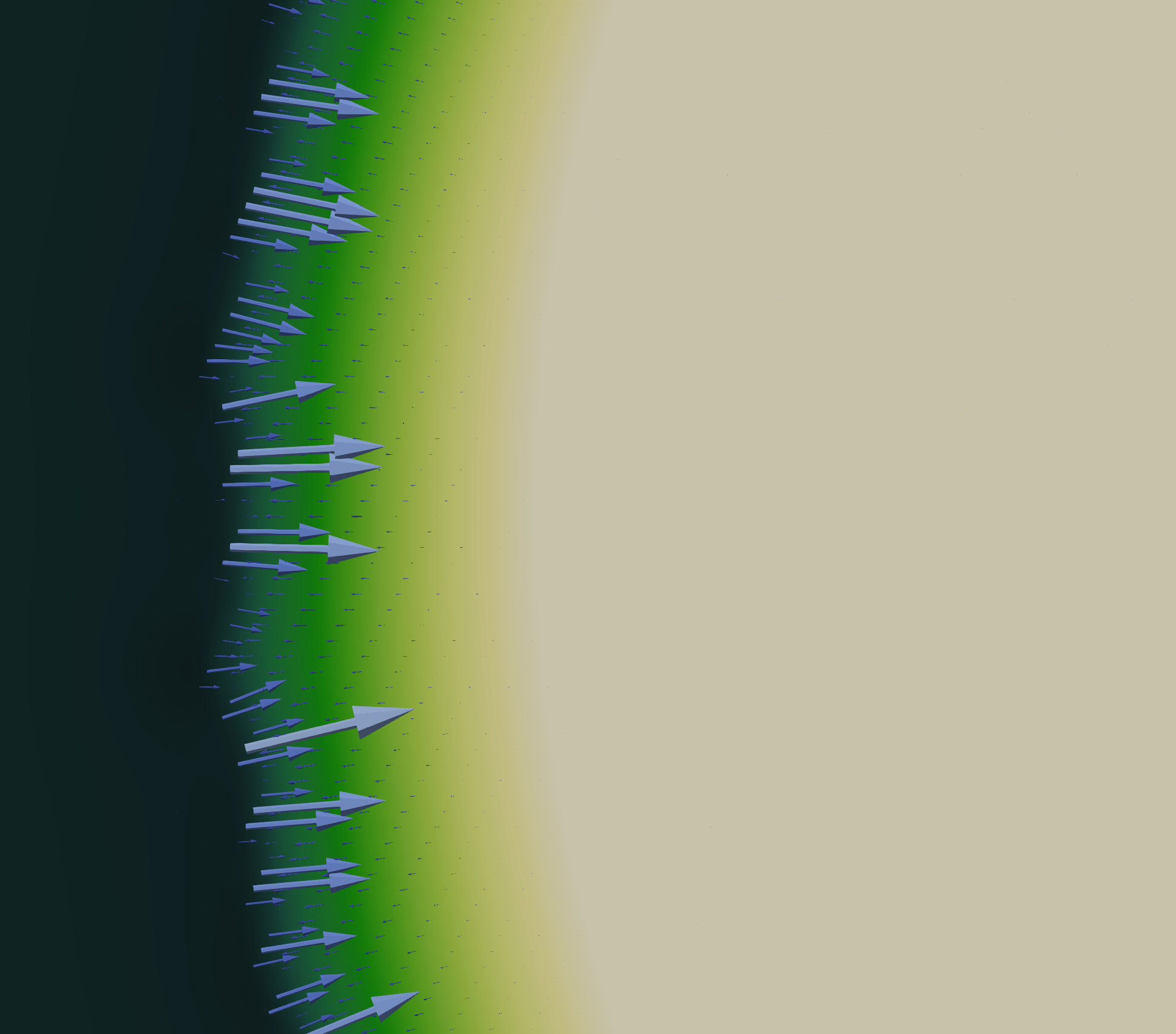} \\
        \multicolumn{2}{c}{\Large (A) $\Bar{t}=0.19$} & \multicolumn{2}{c}{\Large (B) $\Bar{t}=1.20$} \vspace{0.25cm} \\
        \includegraphics[width=0.23\textwidth]{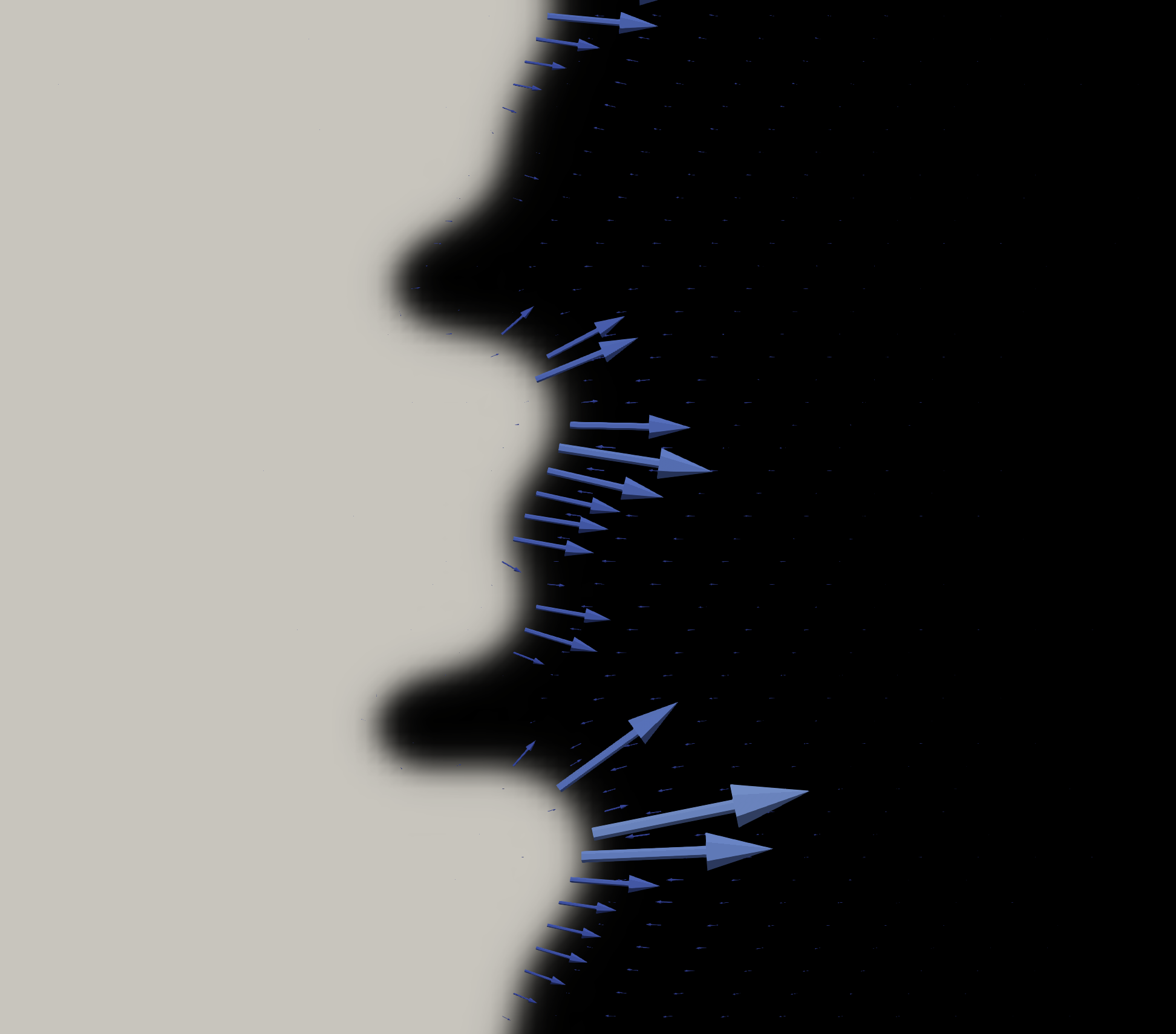} &
        \includegraphics[width=0.23\textwidth]{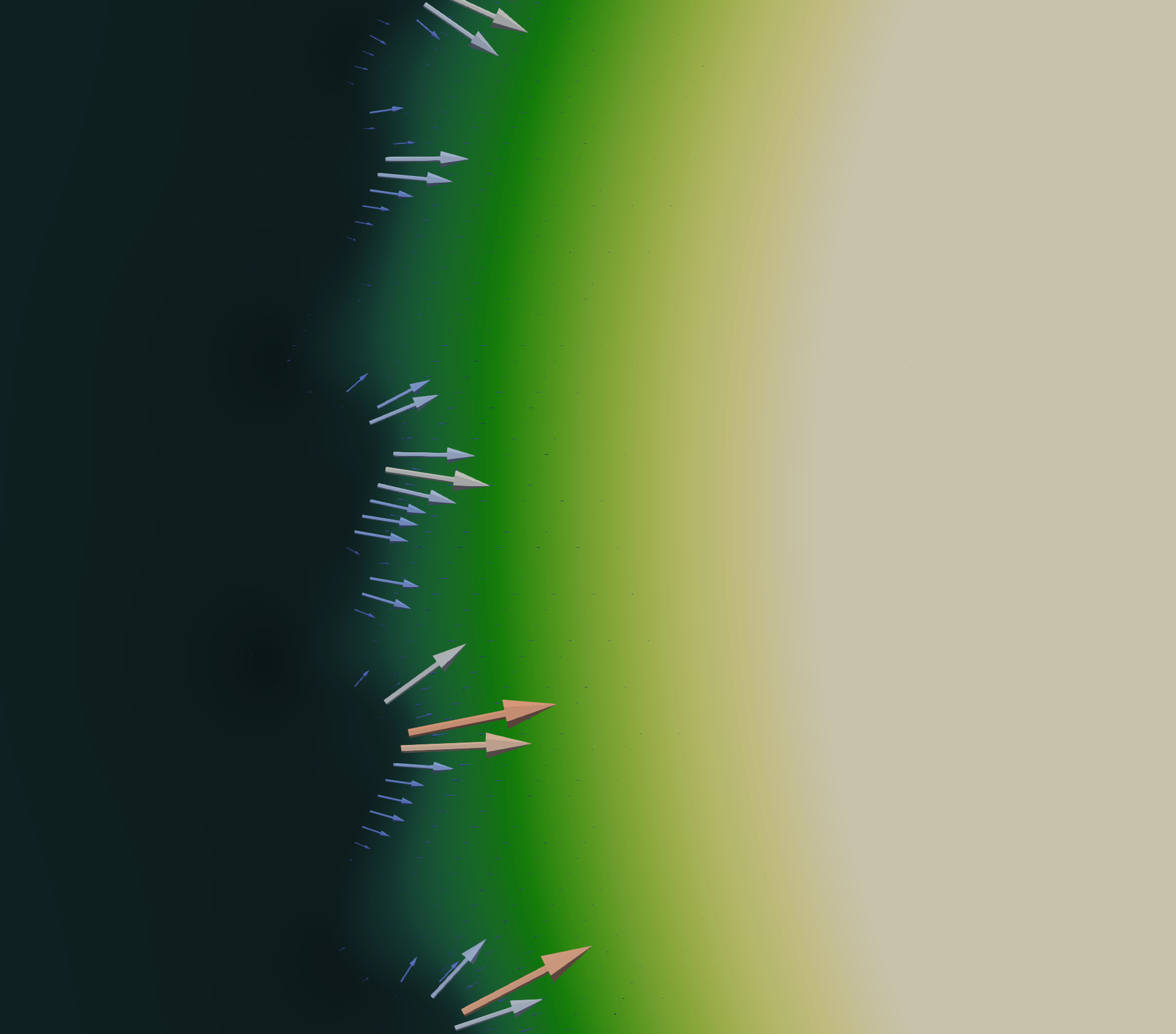} &
        \includegraphics[width=0.23\textwidth]{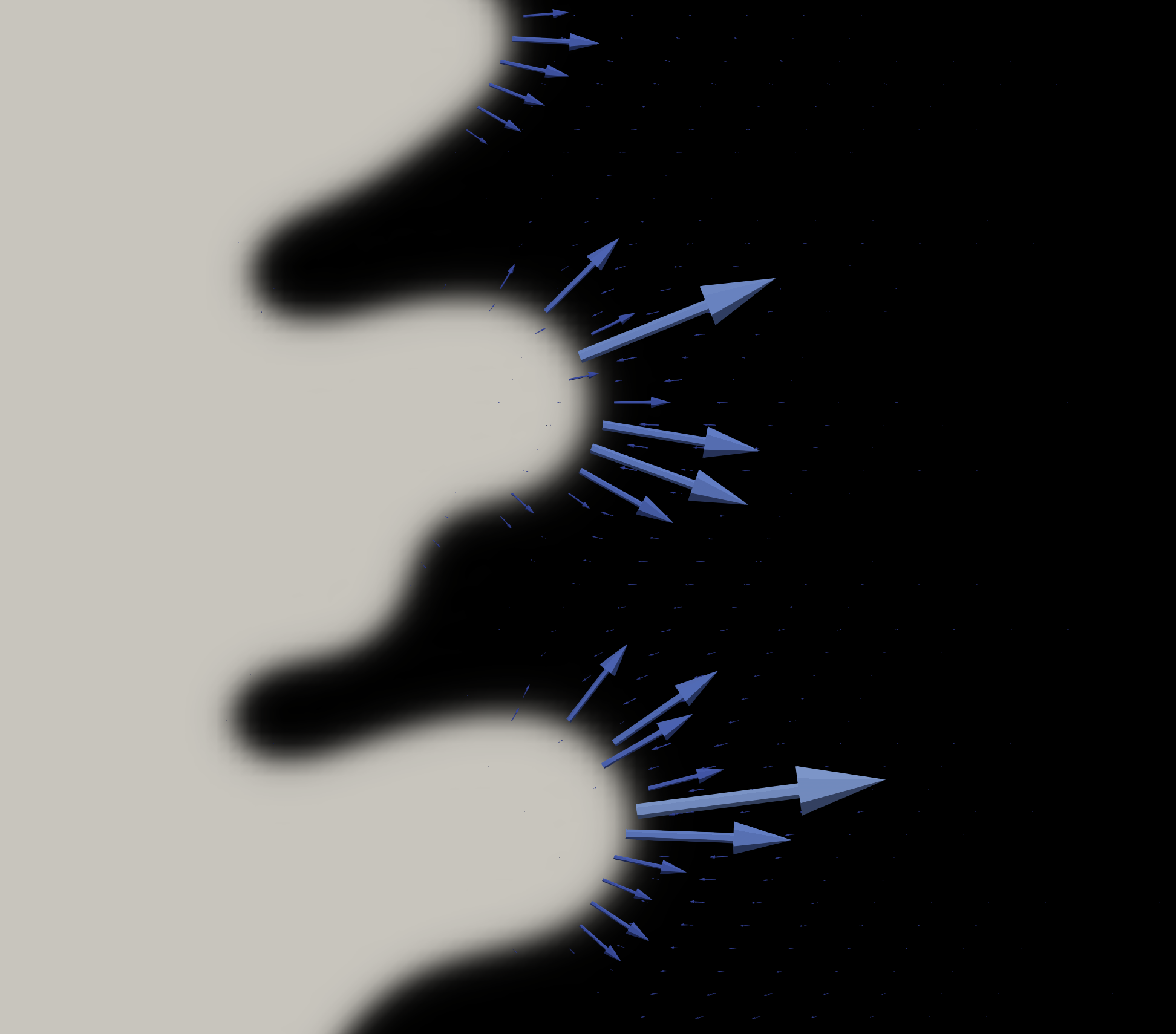} &
        \includegraphics[width=0.23\textwidth]{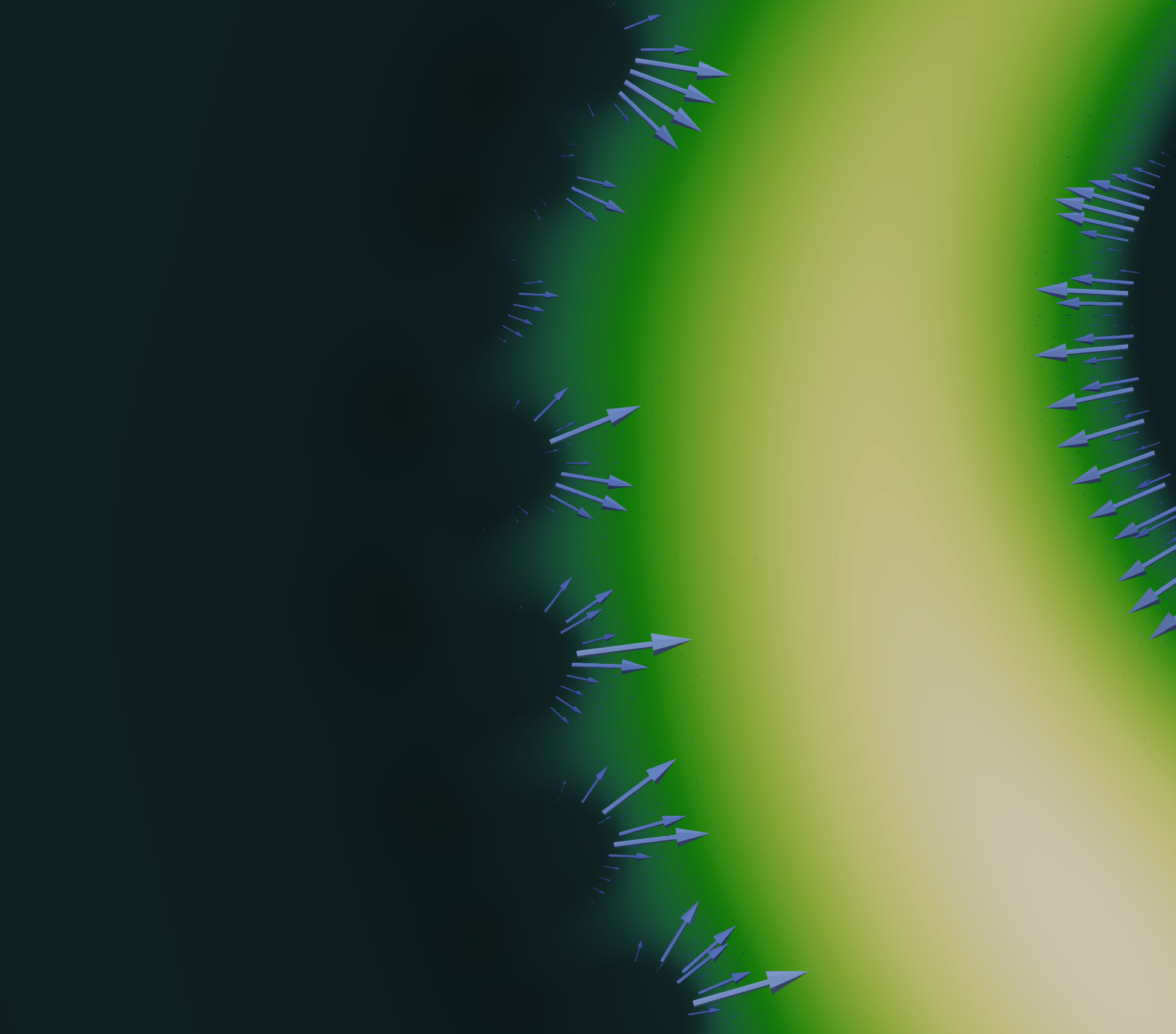} \\
        \multicolumn{2}{c}{\Large (C) $\Bar{t}=2.64$} & \multicolumn{2}{c}{\Large (D) $\Bar{t}=3.84$} \vspace{0.25cm}\\
        \Large $\varphi(\bs{x},t)$ & \Large $\sigma(\bs{x},t)$ & \Large  $\varphi(\bs{x},t)$ & \Large $\sigma(\bs{x},t)$ \\
        \includegraphics[width=0.23\textwidth]{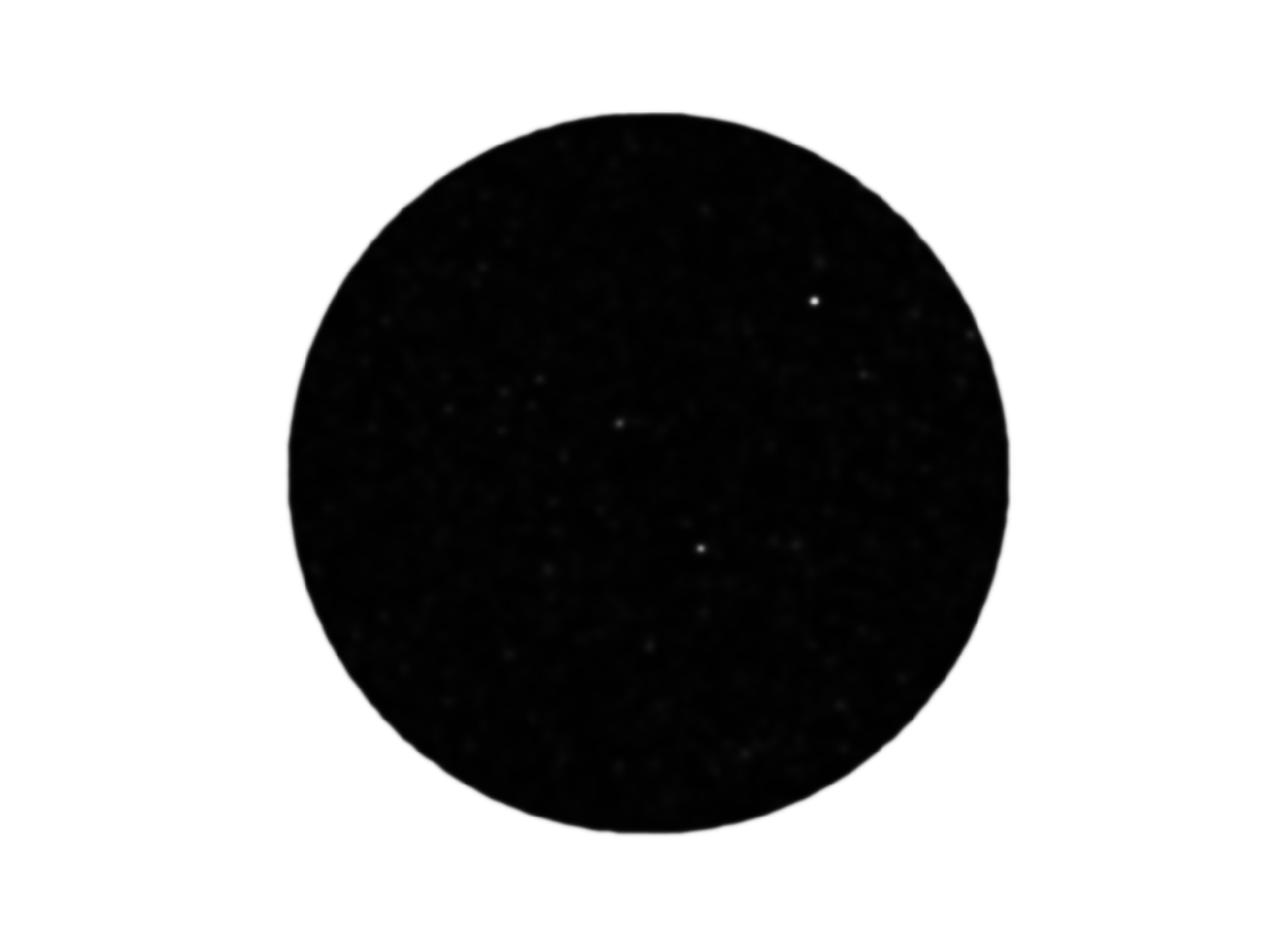} &
        \includegraphics[width=0.23\textwidth]{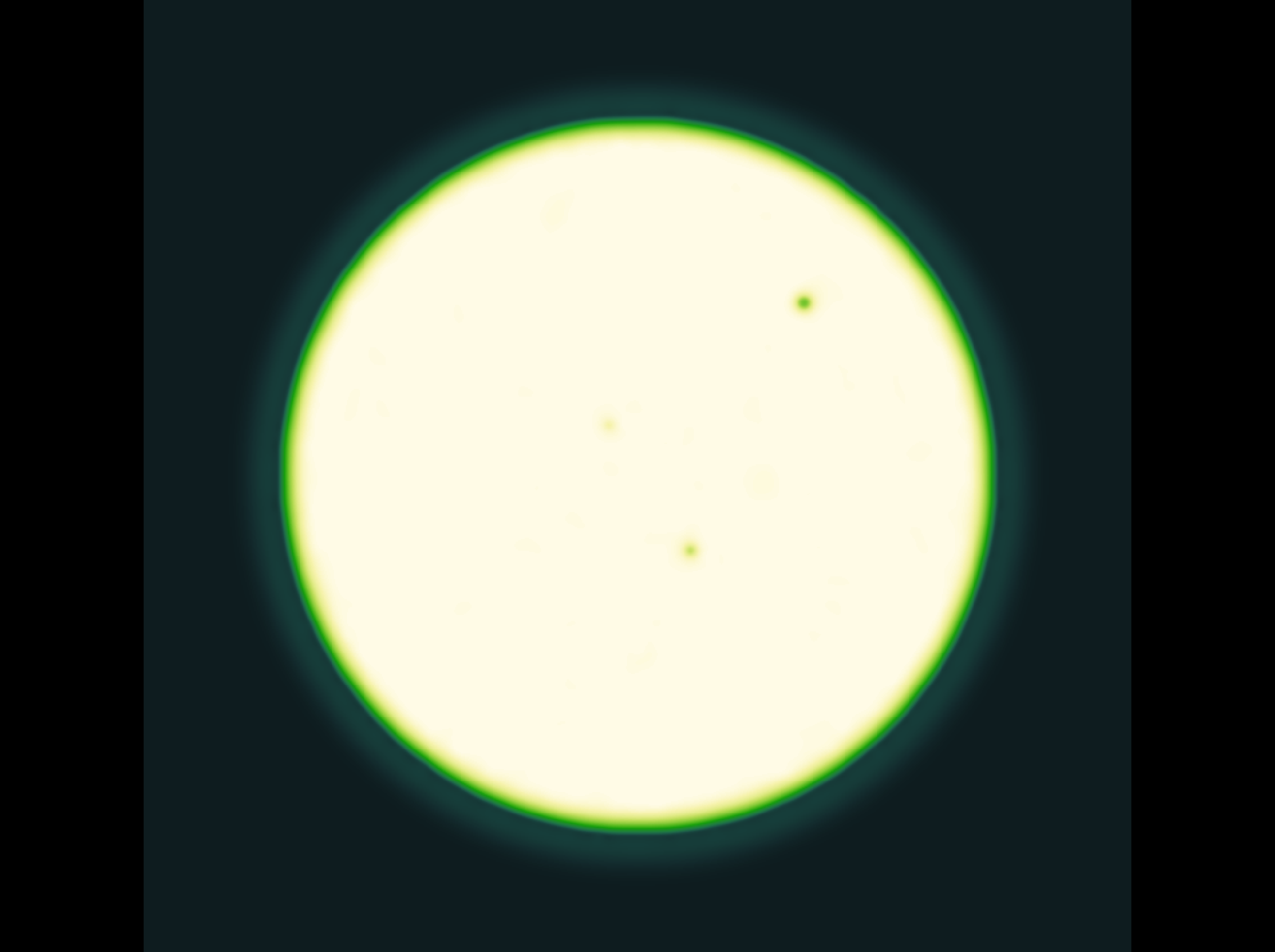} &
        \includegraphics[width=0.23\textwidth]{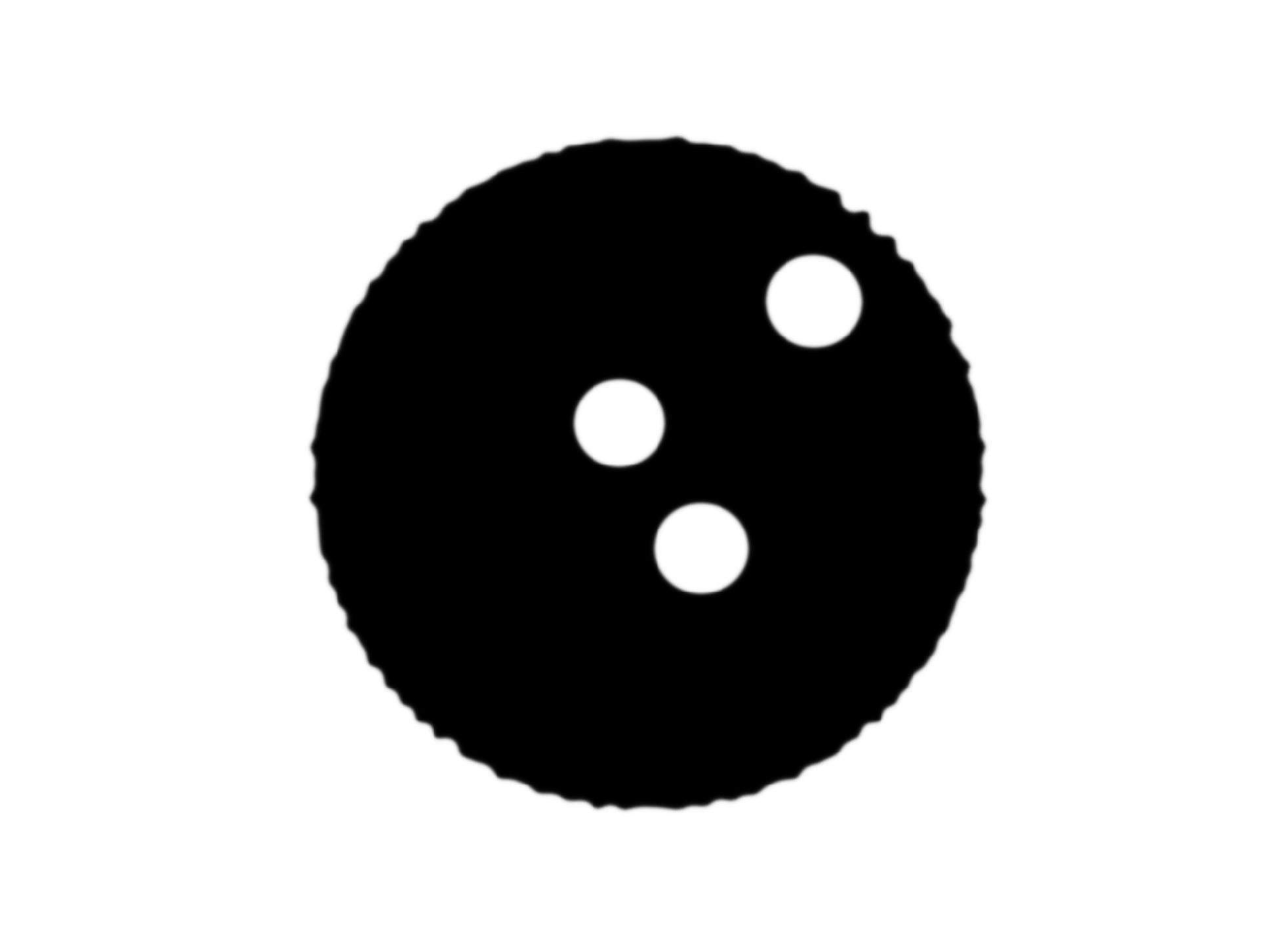} &
        \includegraphics[width=0.23\textwidth]{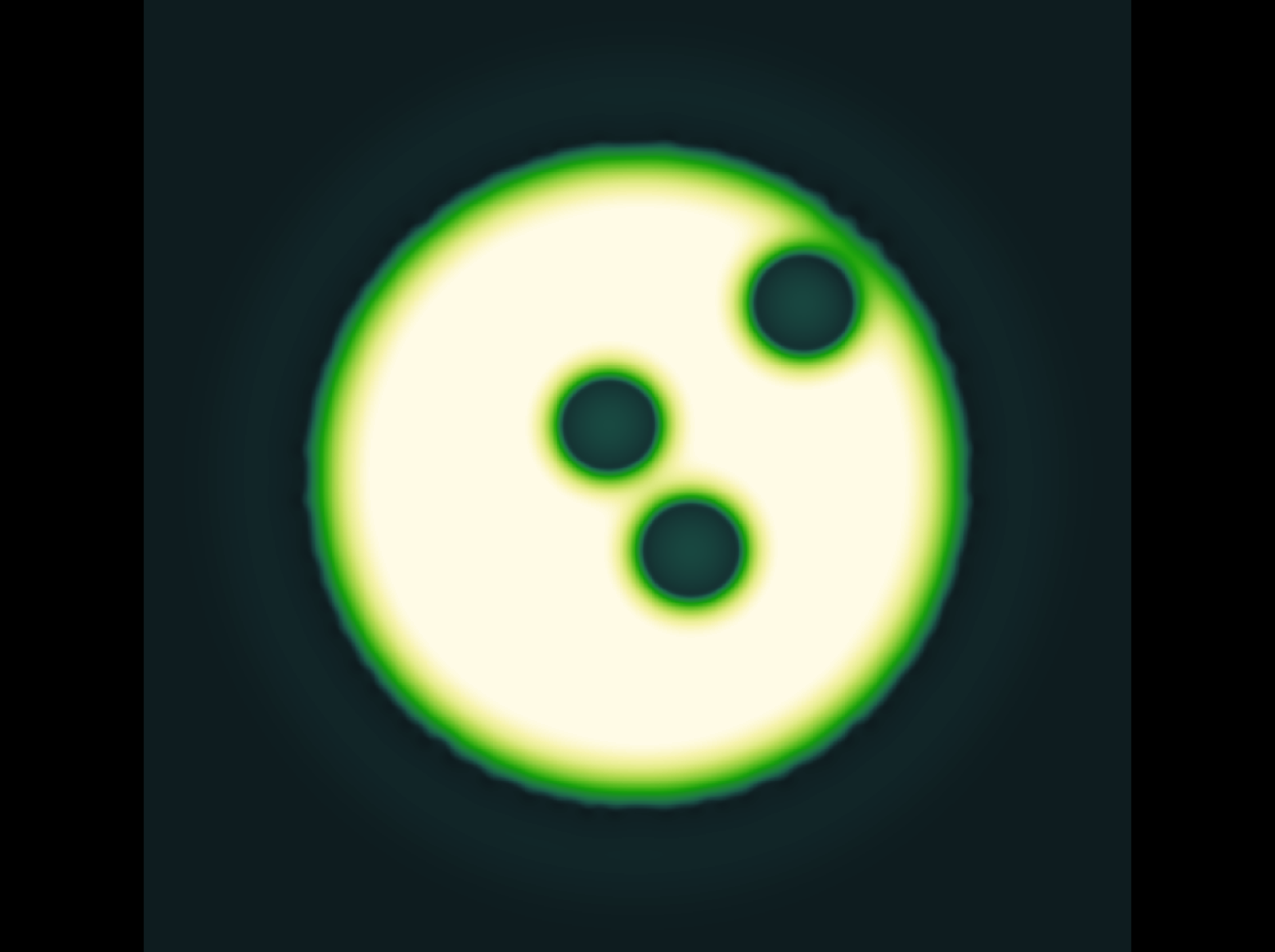} \\
         \multicolumn{2}{c}{\Large (E) (i) $\Bar{t}=0.19$} & \multicolumn{2}{c}{\Large (E) (ii) $\Bar{t}=1.20$} \\
        \includegraphics[width=0.23\textwidth]{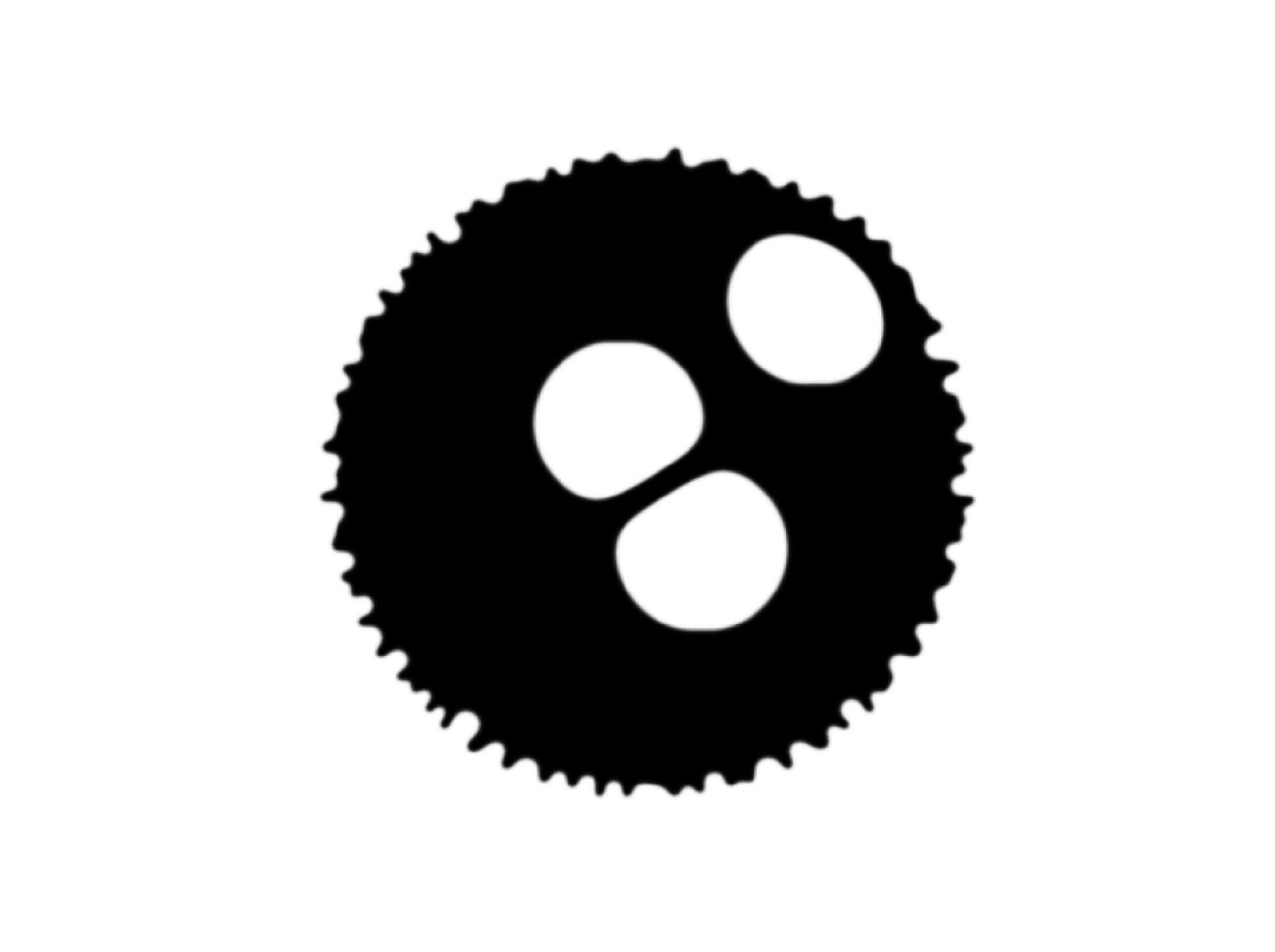} &
        \includegraphics[width=0.23\textwidth]{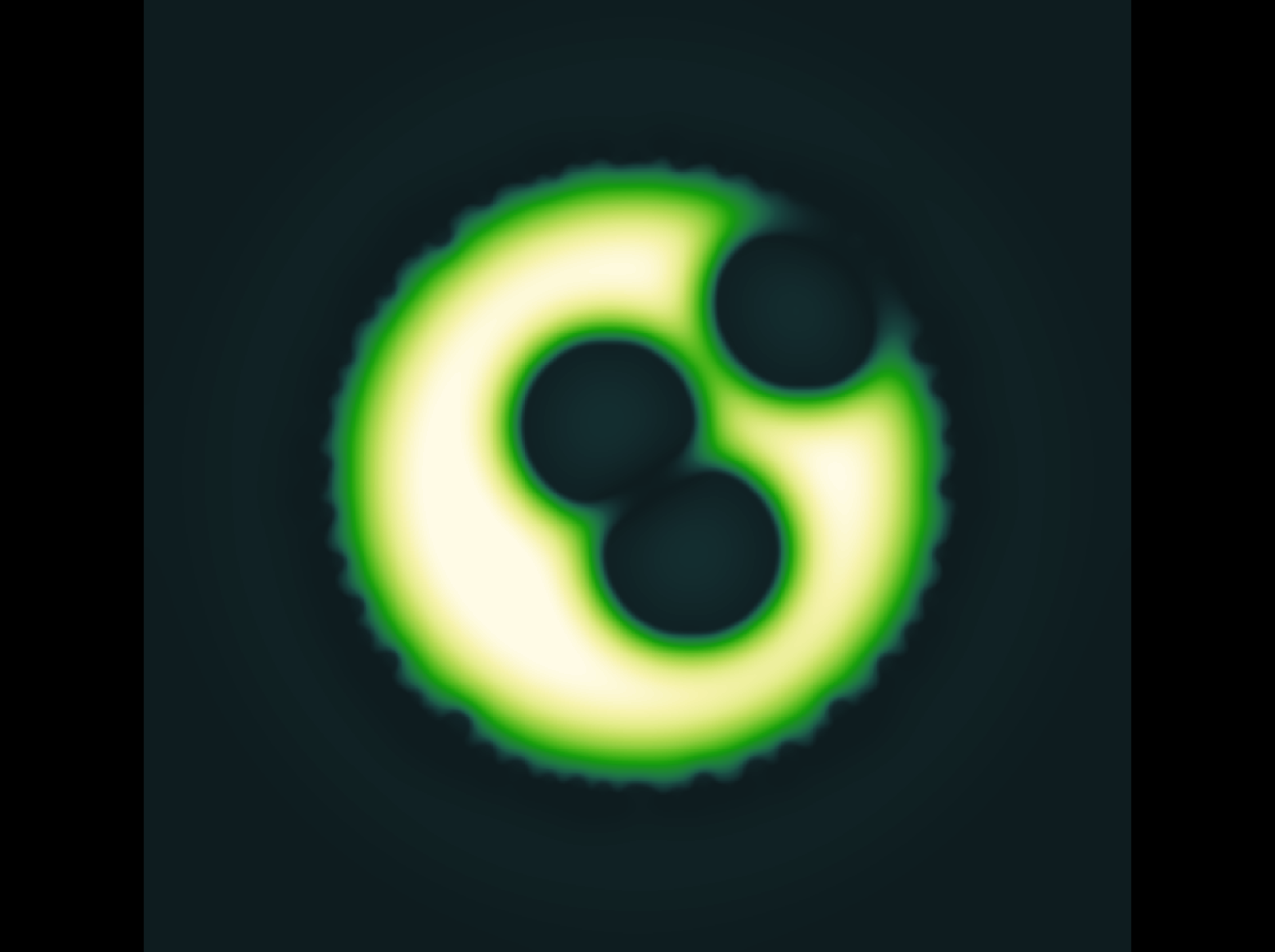} &
        \includegraphics[width=0.23\textwidth]{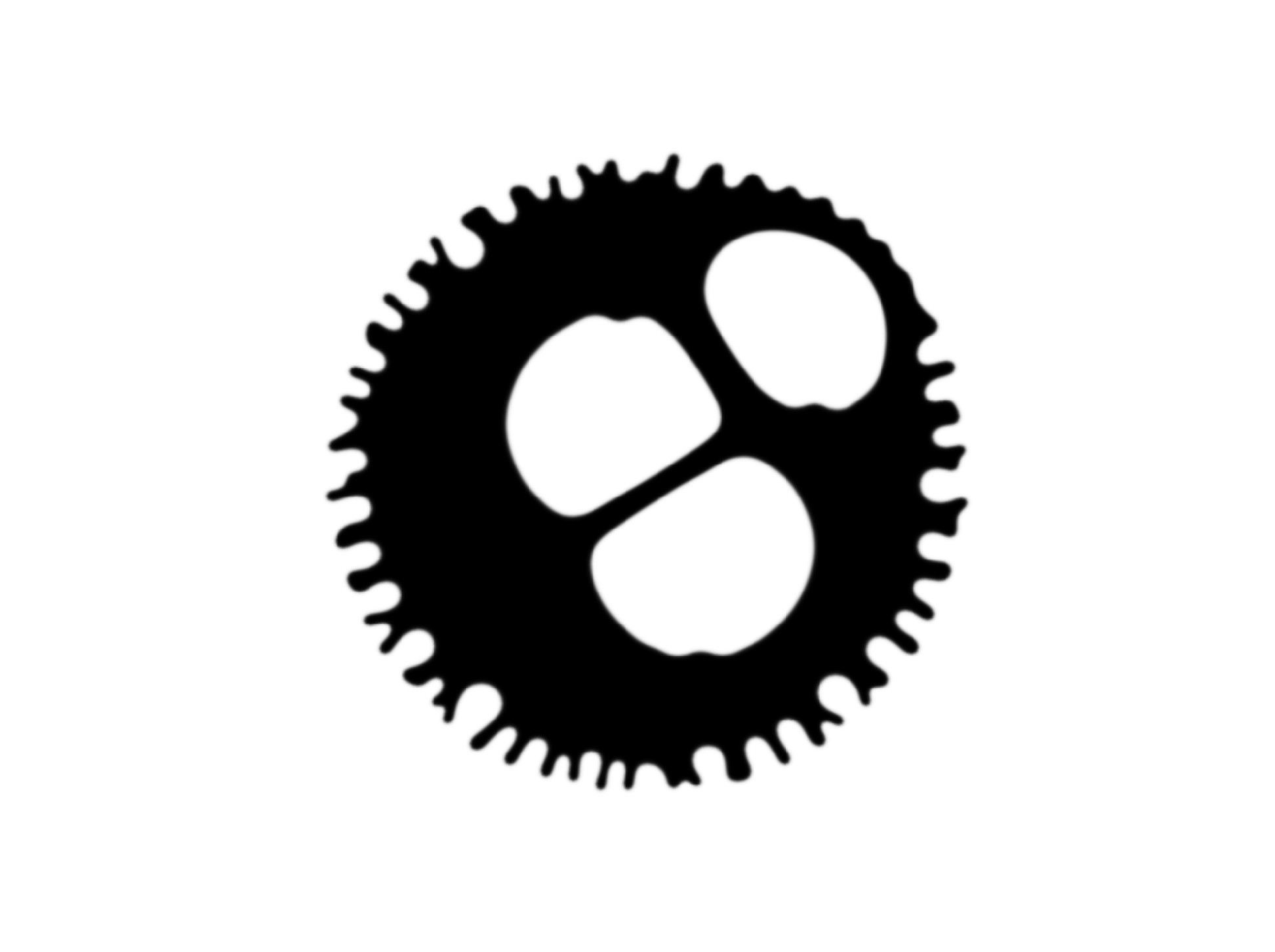} &
        \includegraphics[width=0.23\textwidth]{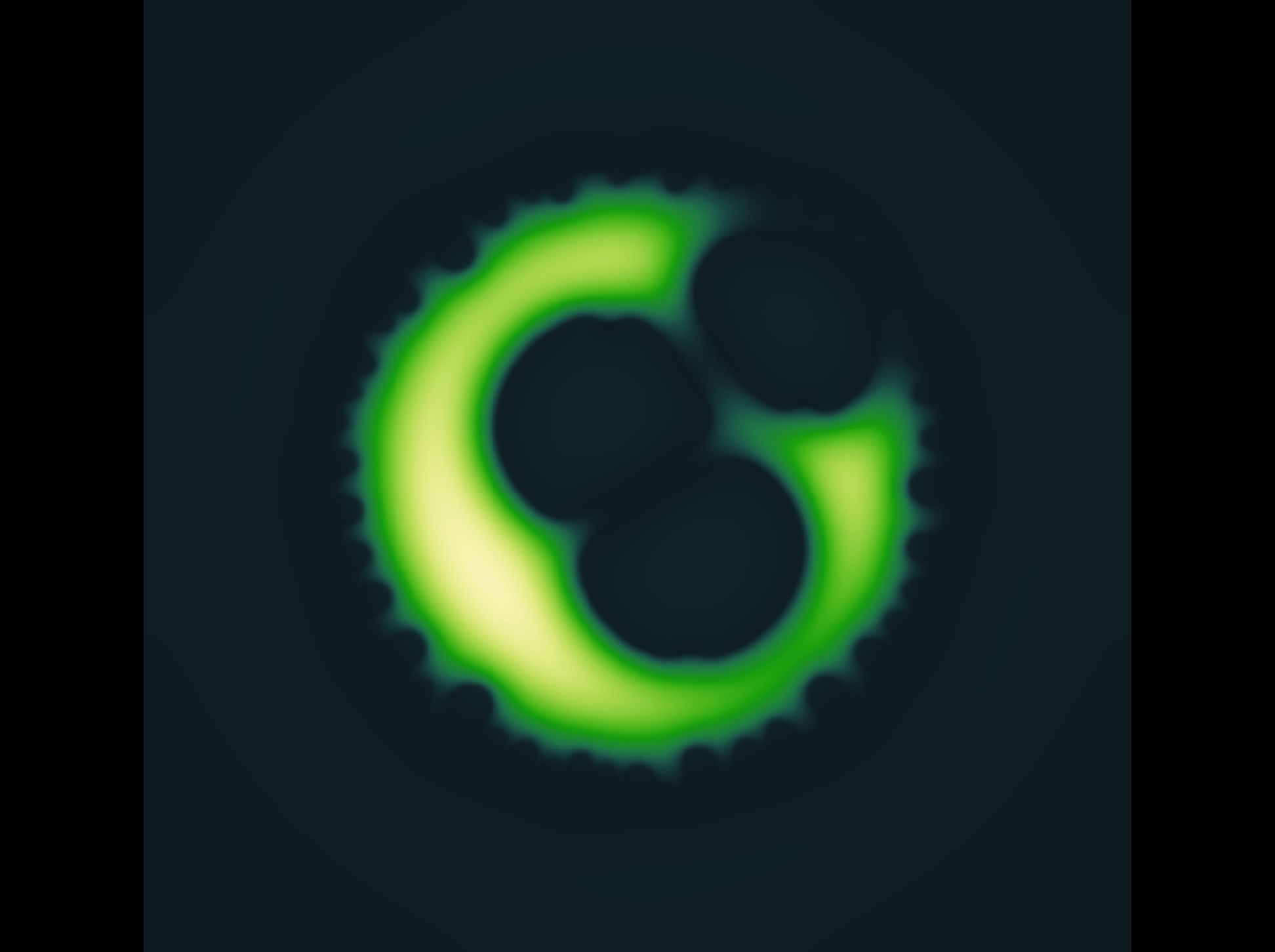} \\
        \multicolumn{2}{c}{\Large (E) (iii) $\Bar{t}=2.64$} & \multicolumn{2}{c}{\Large (E) (iv) $\Bar{t}=3.84$} \\
    \end{tabular}
    \caption{
        Evolution of a section of the interface for the phase fields $\varphi$ and $\sigma$, and the internal configurational forces $\bs{f}$, and $|\bs{f}|$ at the time points $\Bar{t}$=0.19 (A, E(i)), 1.20 (B, E(ii)), 2.64 (C, E(iii)), and 3.84 (D, E(iv)). The parameters selected for this simulation as listed in Table~\ref{tab:Parameter_List}, $\ell_{\varphi}=1.333\times10^{-4}$, $\ell_{2}=1.5$, $\ell^{\varphi}_{r}=960$, $\ell^{\sigma}_{r}=0.28$, and $k_{2}=8.0$. The arrows in images (A), (B), (C), and (D) indicate the direction of $\bs{f}$, while the colour represents the magnitude throughout the simulation—red denotes higher values than blue. In E(i-iv), the evolution of the phase fields $\varphi$ and $\sigma$ have been plotted across the domain, illustrating the global degradation of these phases.
    }
    \label{fig:internal_config_force_interface}
\end{figure}

Figure~\ref{fig:internal_config_force_interface} shows details of the interface encompassing the cyto phase field $\varphi$ and the cytotoxic phase field $\sigma$ with arrows representing the internal configurational forces. At $\Bar{t}=0.19$, in Figure~\ref{fig:internal_config_force_interface} (A), the internal configurational force $\bs{f}$ is nearly uniform in space along the interface. Therefore, the interfaces for the phase fields $\varphi$ and $\sigma$ degrade uniformly. As time progresses to $\Bar{t}=1.20$, Figure~\ref{fig:internal_config_force_interface} (B), the internal configurational force oscillates along the interface. Thus, it produces the early stages of finger formation. At $\Bar{t}=2.64 \ \text{and} \ 3.84$, Figures~\ref{fig:internal_config_force_interface} (C) and (D), distinct fingers form in the phase field $\varphi$. Furthermore, the magnitude of the internal configurational forces increases at the basin of the opening that forms between two fingers.
\begin{figure}[htbp]
    \centering
    \begin{tabular}{cccc}
        \Large $\varphi(\bs{x},t), \ \bs{f}(\bs{x},t)$ & \Large $\sigma(\bs{x},t), \ \bs{f}(\bs{x},t)$ & \Large $\varphi(\bs{x},t), \ \bs{f}(\bs{x},t)$ & \Large $\sigma(\bs{x},t), \ \bs{f}(\bs{x},t)$ \\
        \includegraphics[width=0.23\textwidth]{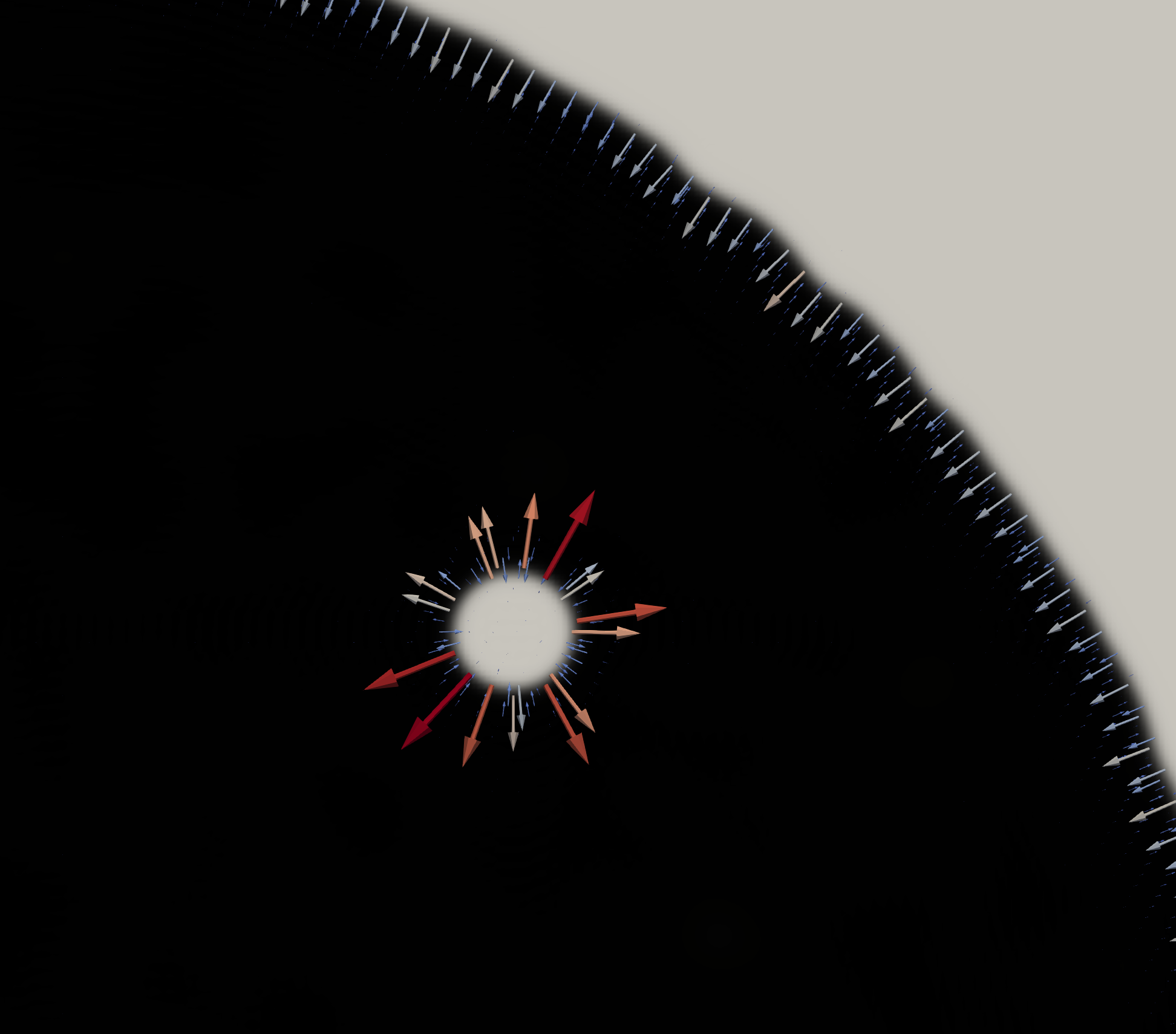} &
        \includegraphics[width=0.23\textwidth]{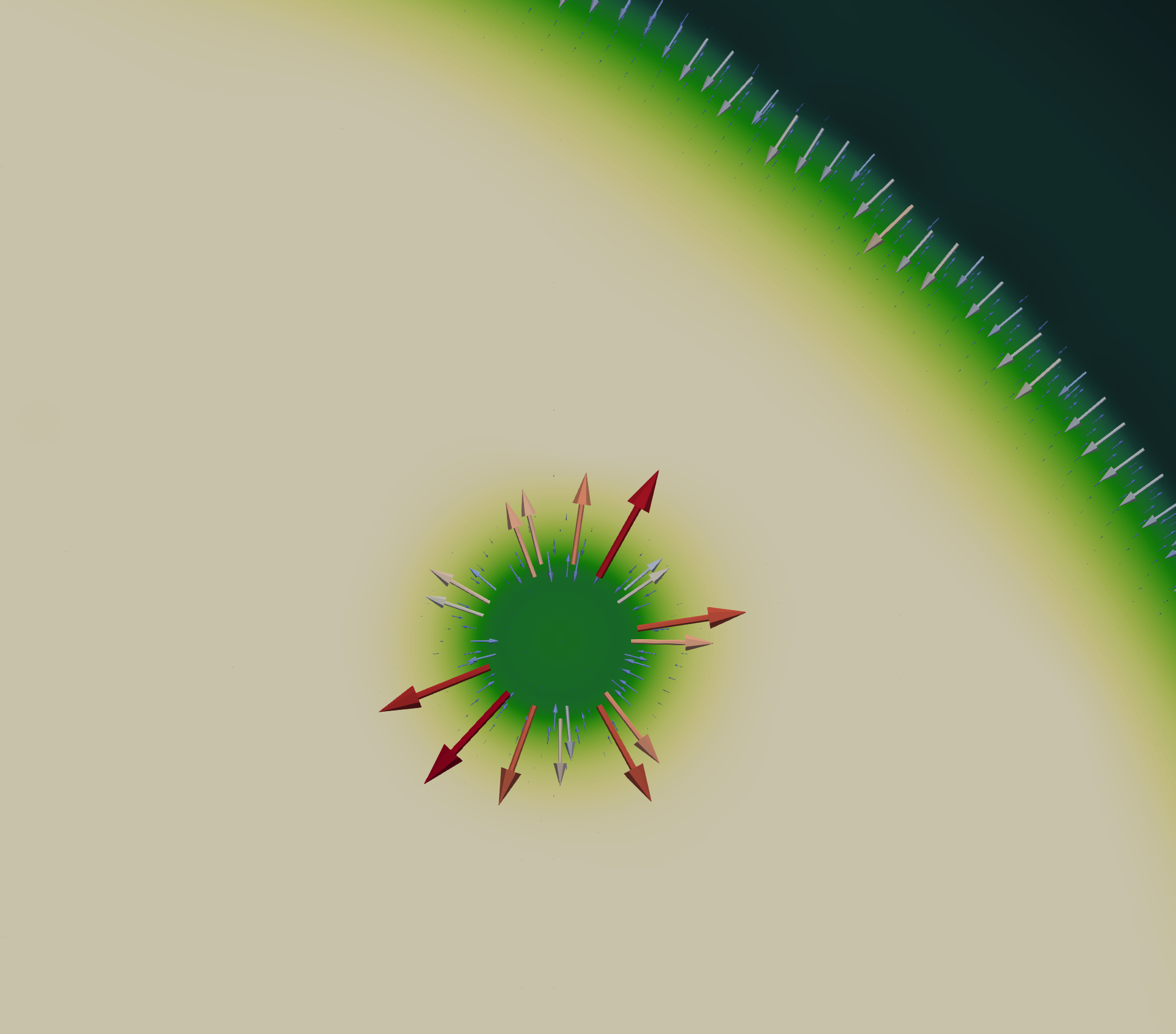} &
        \includegraphics[width=0.23\textwidth]{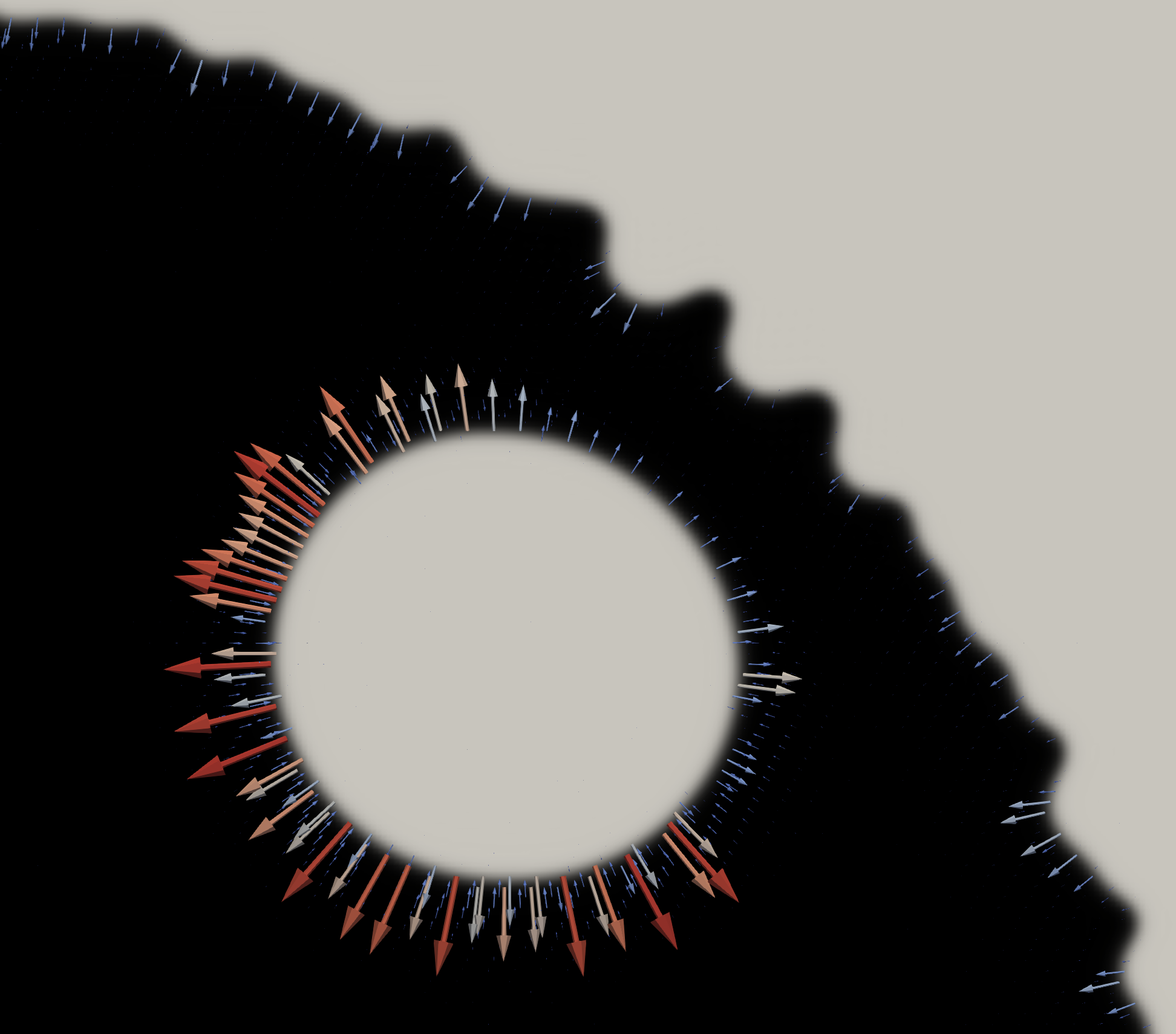} &
        \includegraphics[width=0.23\textwidth]{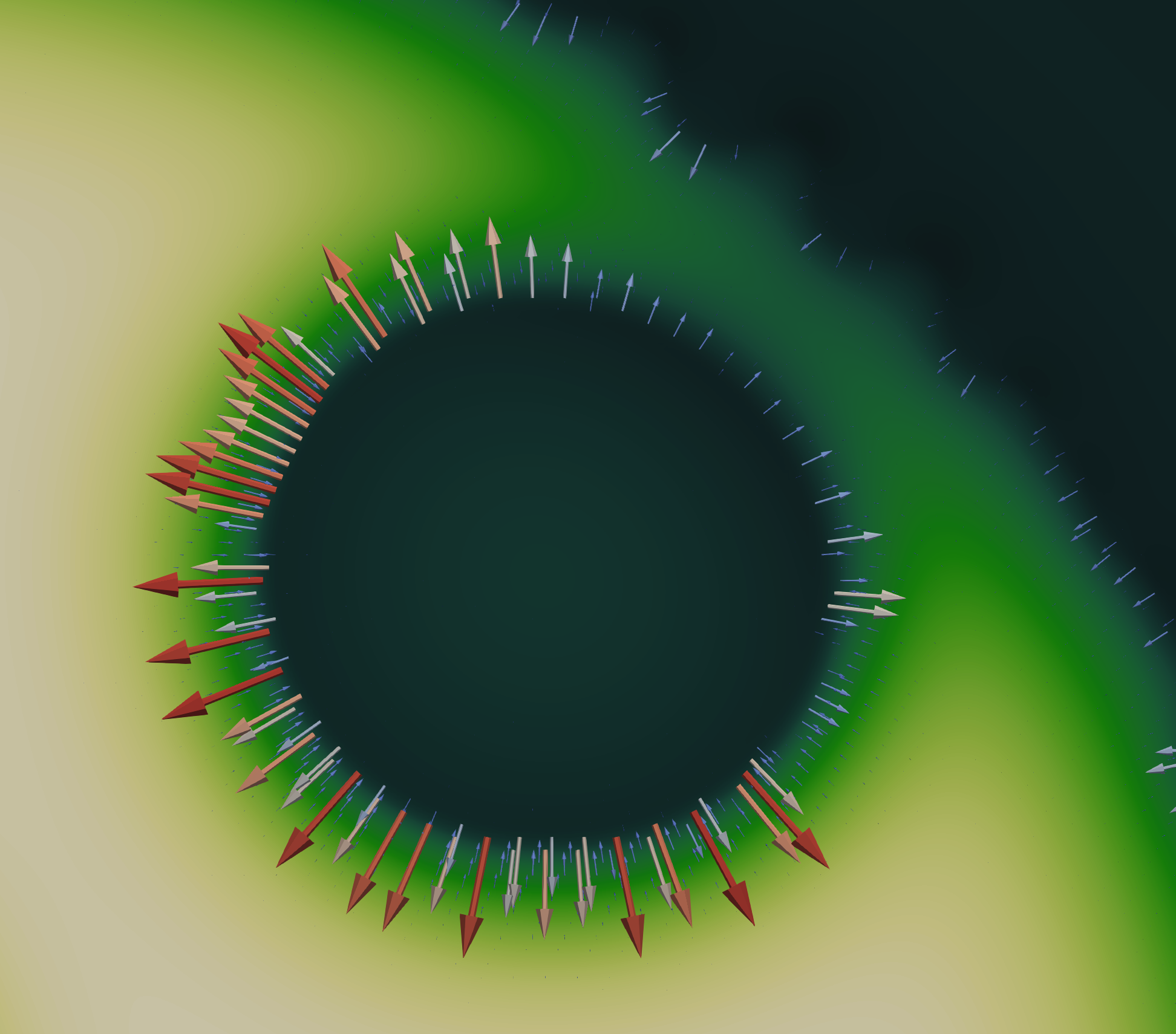} \\
        \multicolumn{2}{c}{\Large (A) $\Bar{t}=0.41$} & \multicolumn{2}{c}{\Large (B) $\Bar{t}=1.44$} \vspace{0.25cm} \\
        \includegraphics[width=0.23\textwidth]{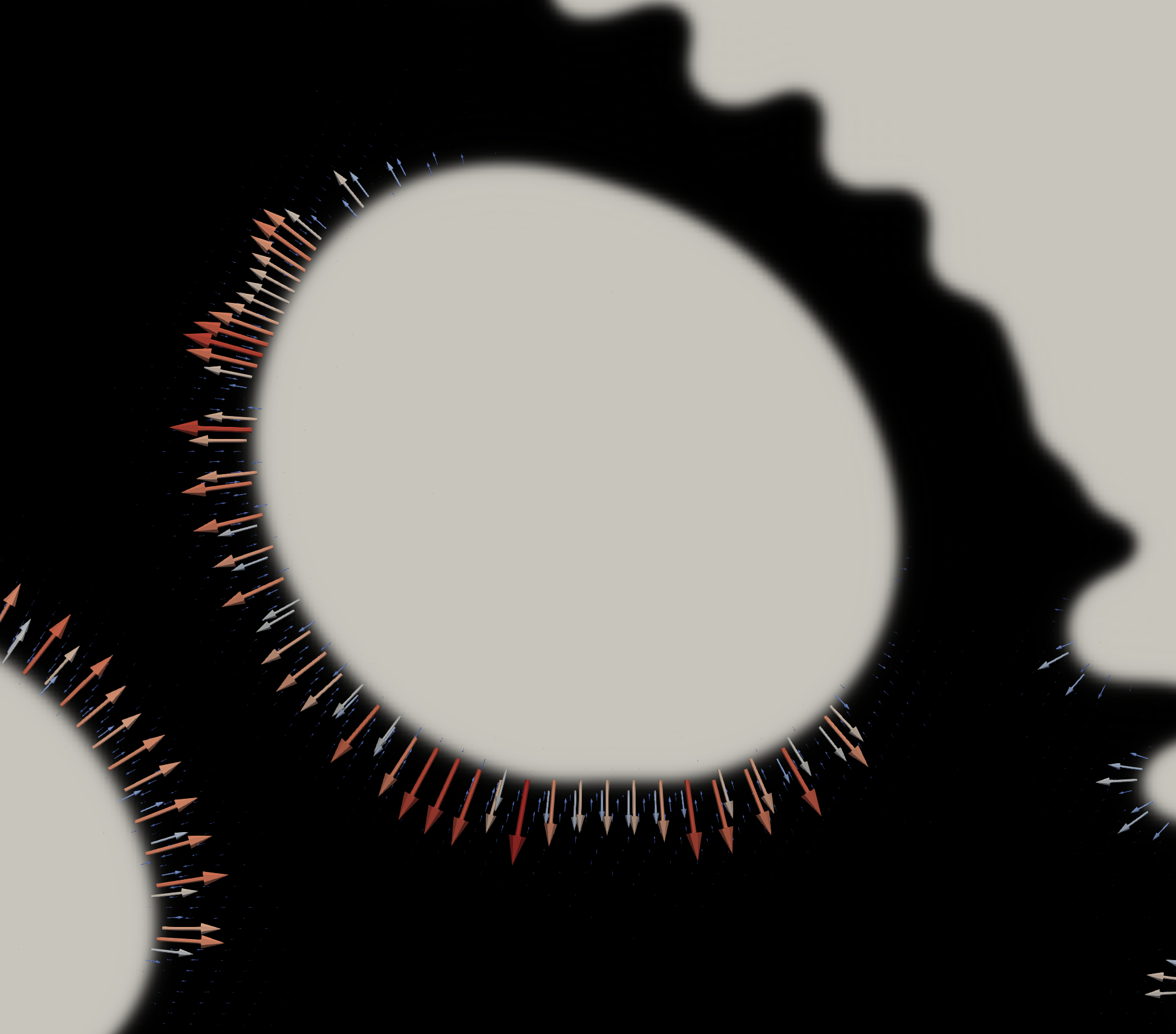} &
        \includegraphics[width=0.23\textwidth]{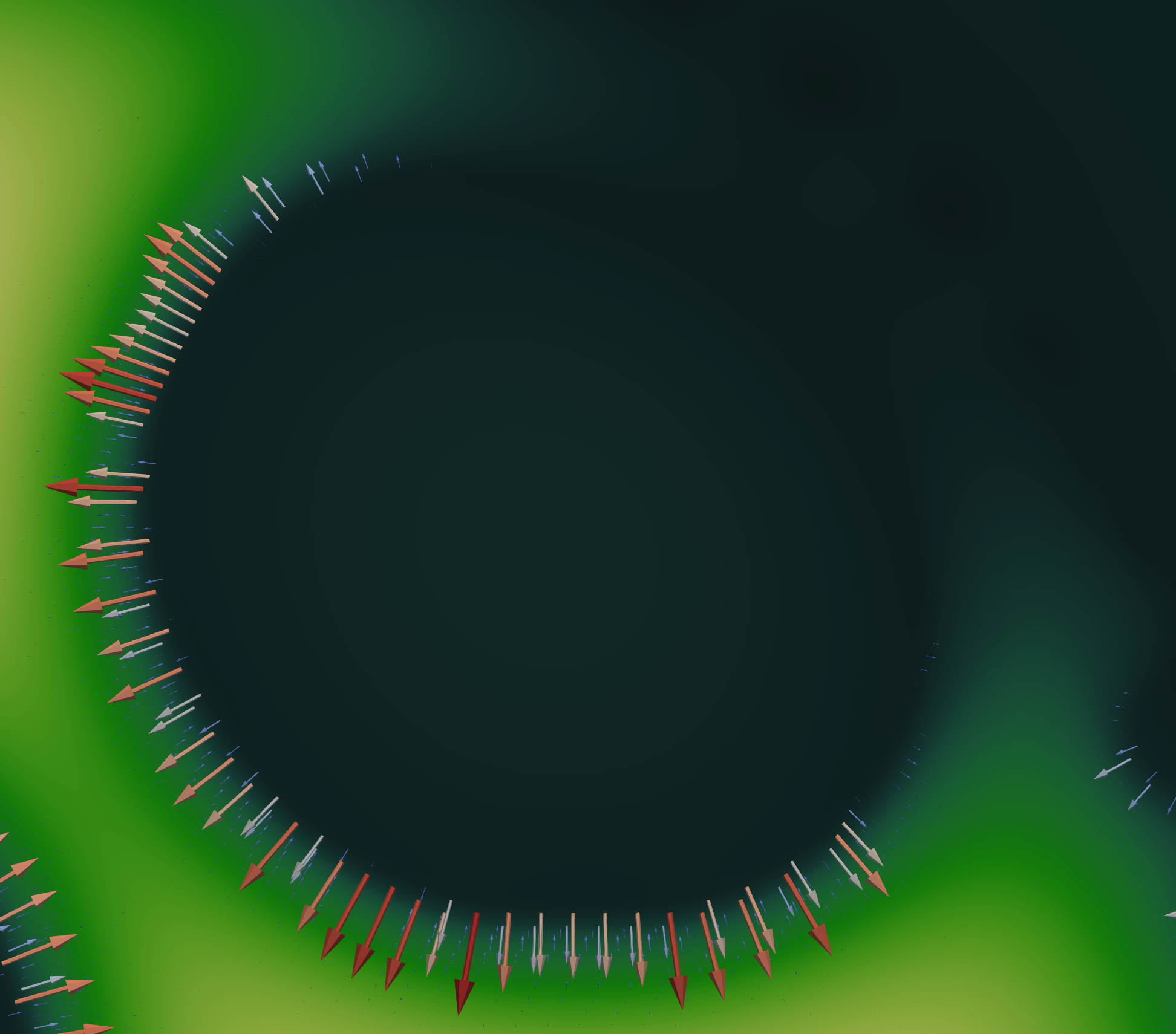} &
        \includegraphics[width=0.23\textwidth]{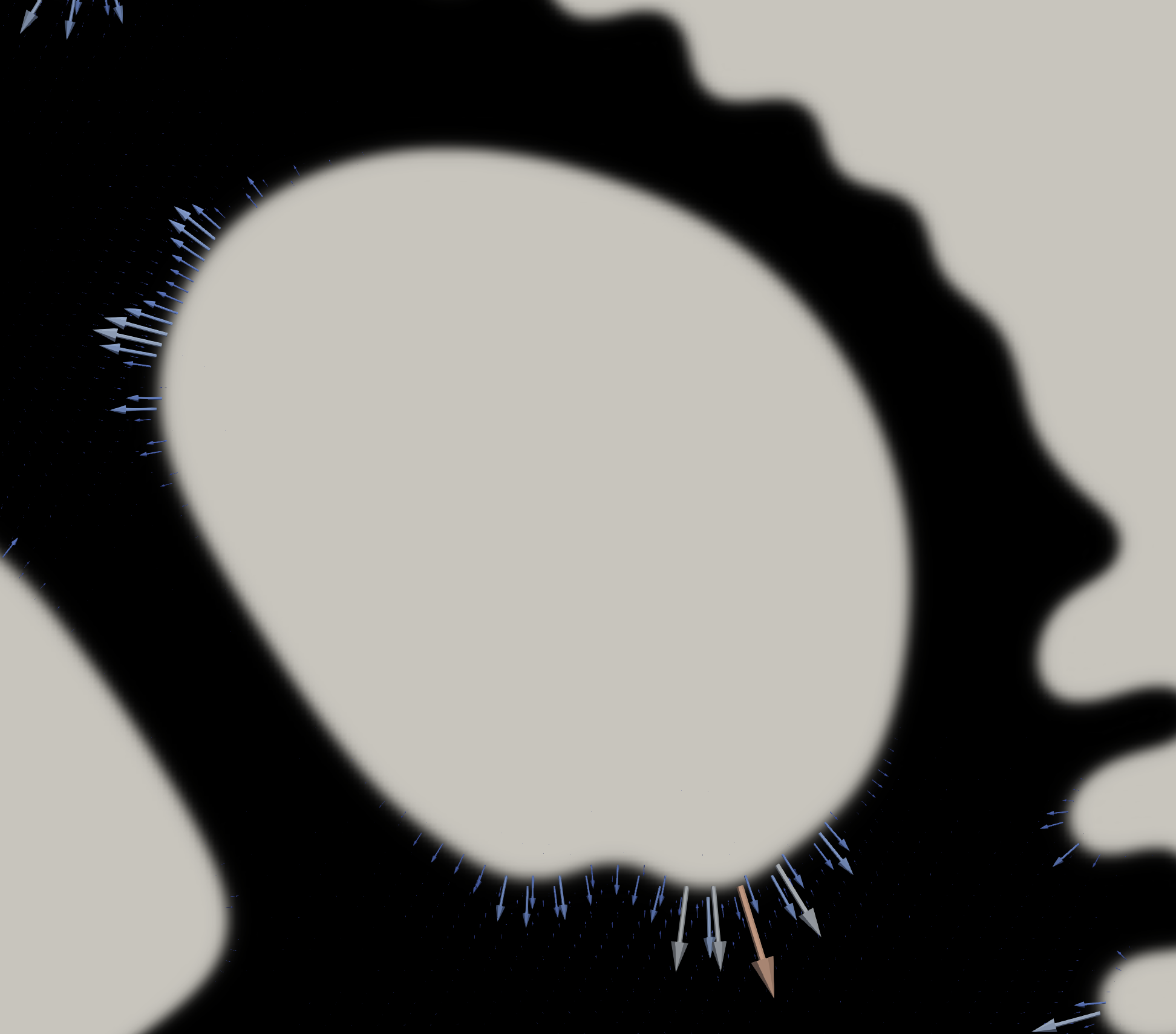} &
        \includegraphics[width=0.23\textwidth]{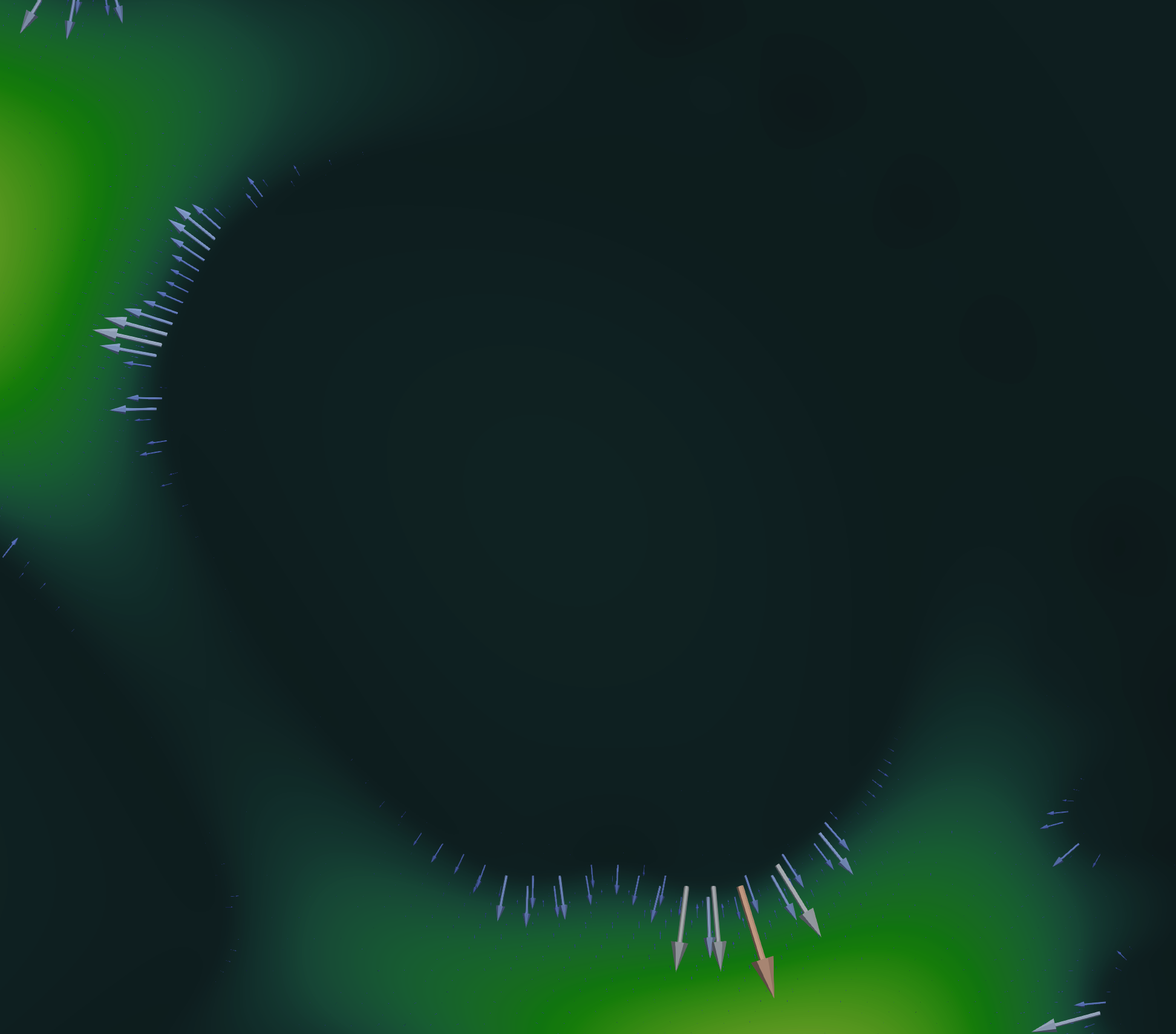} \\
        \multicolumn{2}{c}{\Large (C) $\Bar{t}=2.64$} & \multicolumn{2}{c}{\Large (D) $\Bar{t}=3.84$} \\
        \Large $\varphi(\bs{x},t)$ & \Large  $\sigma(\bs{x},t)$ &  \Large  $\varphi(\bs{x},t)$ & \Large  $\sigma(\bs{x},t)$ \\
        \includegraphics[width=0.23\textwidth]{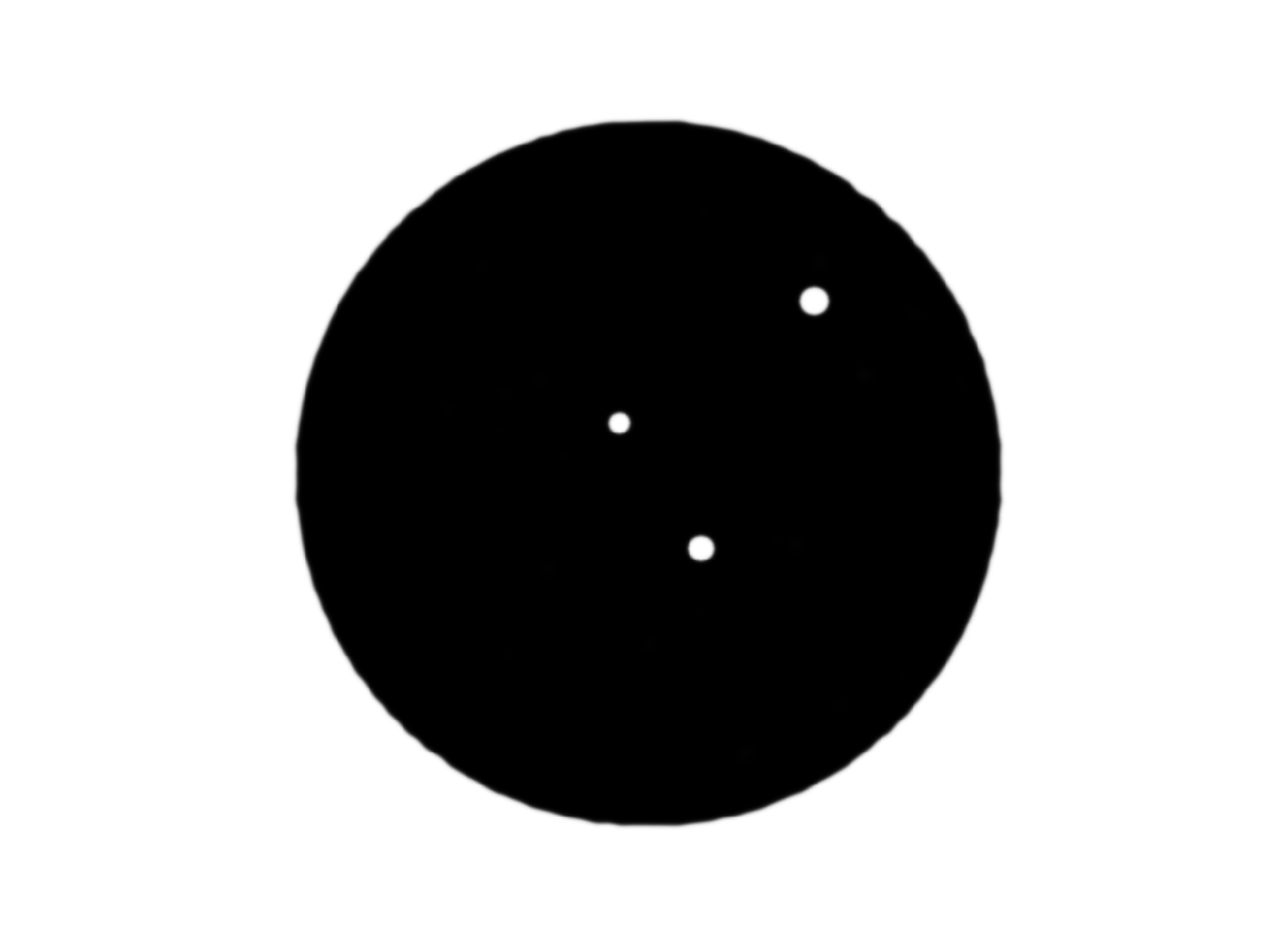} &
        \includegraphics[width=0.23\textwidth]{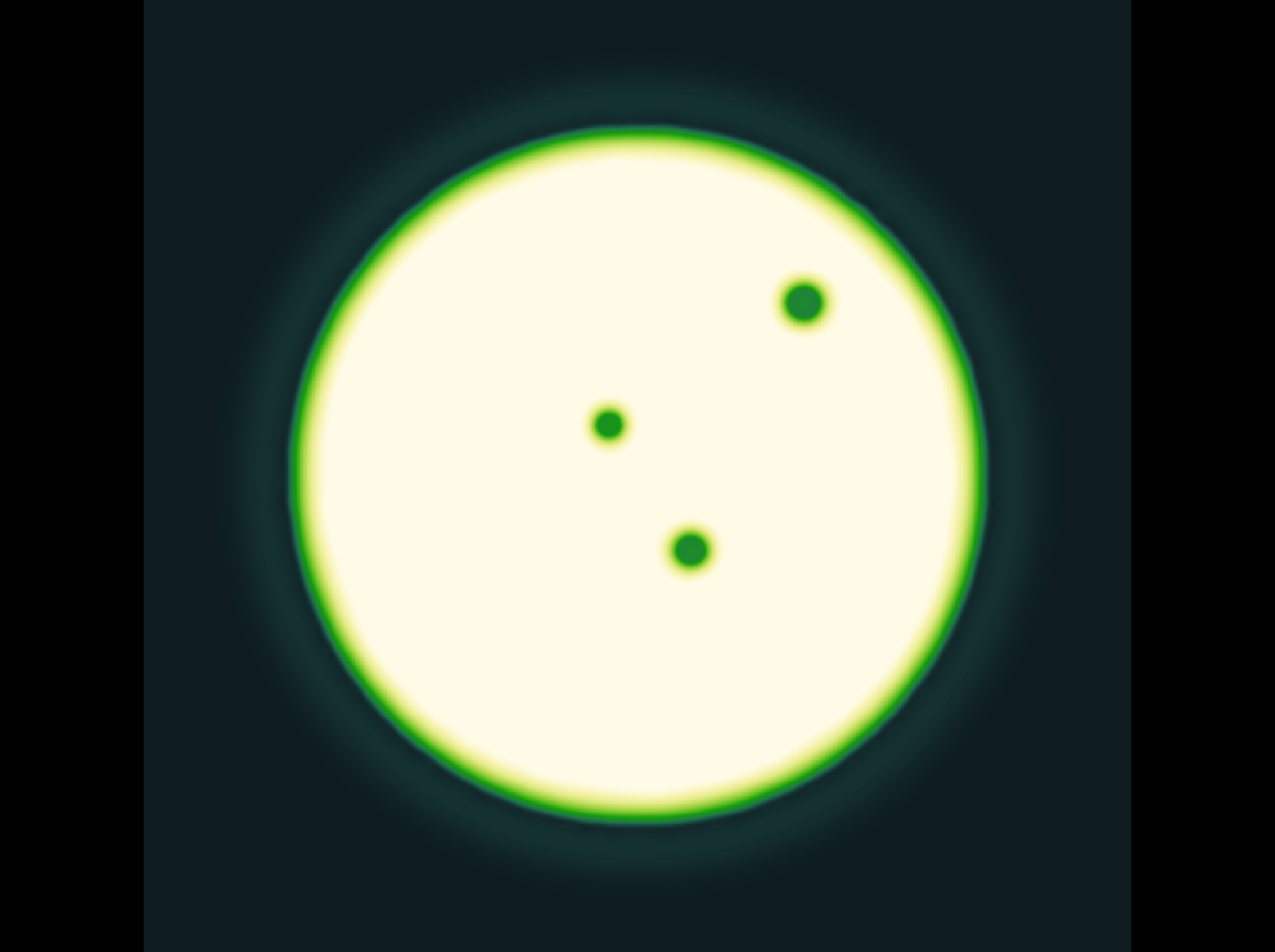} &
        \includegraphics[width=0.23\textwidth]{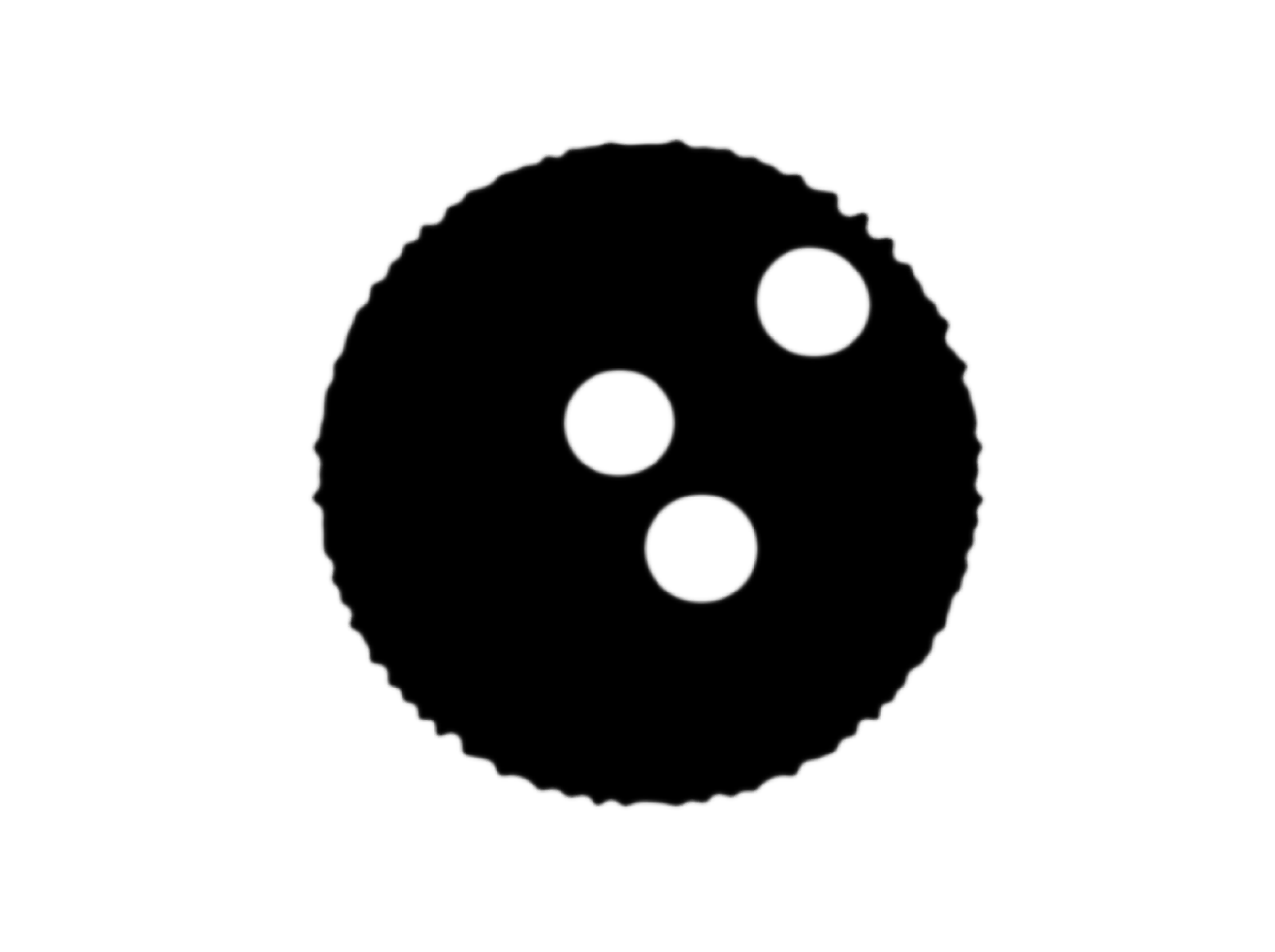} &
        \includegraphics[width=0.23\textwidth]{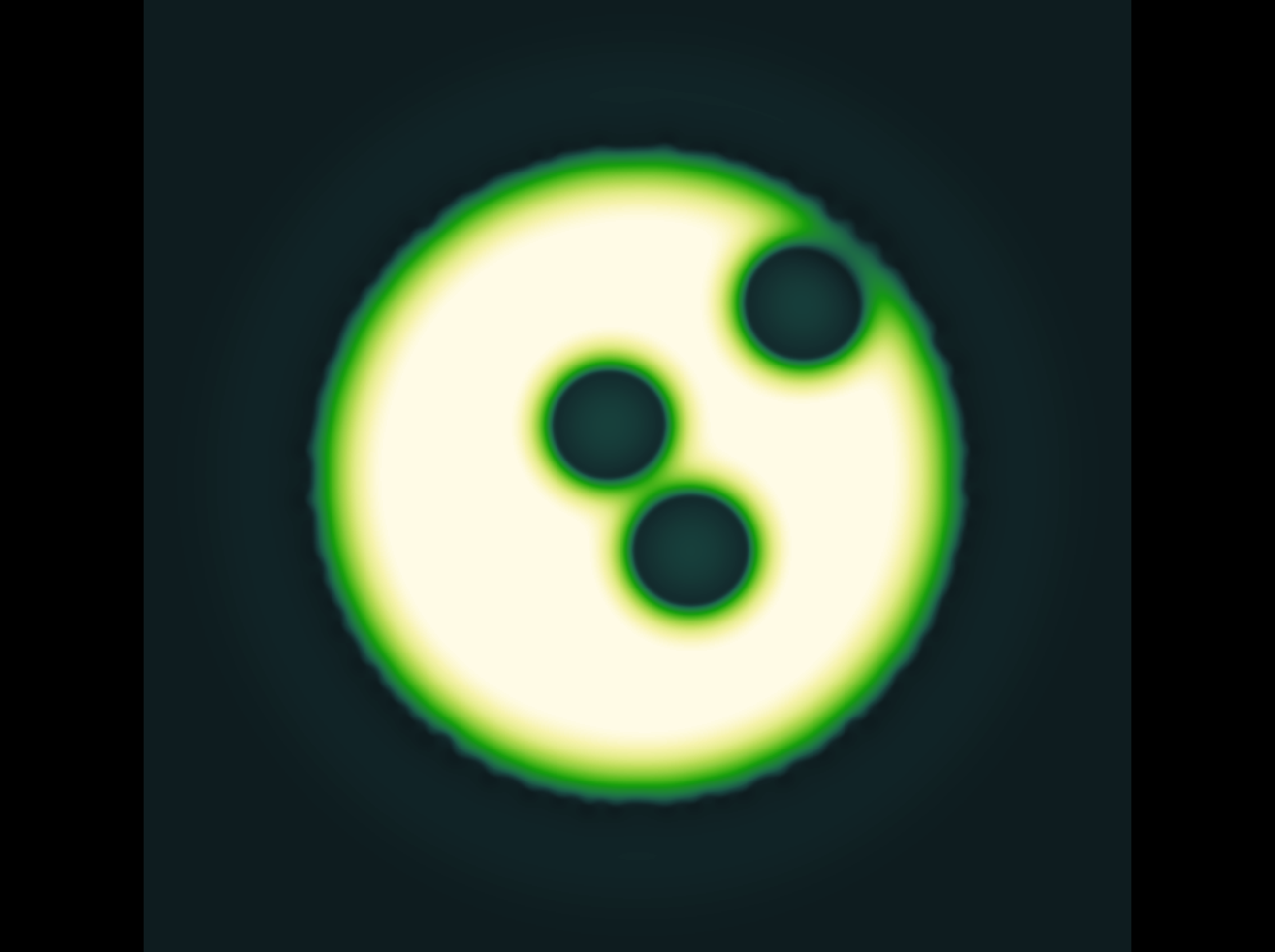} \\
        \multicolumn{2}{c}{\Large (E) (i) $\Bar{t}=0.41$} & \multicolumn{2}{c}{\Large (E) (ii) $\Bar{t}=1.44$} \\
        \includegraphics[width=0.23\textwidth]{Figures/Apoptotic/phi_full_config_force/phi.0100.png} &
        \includegraphics[width=0.23\textwidth]{Figures/Apoptotic/sigma_full_config_force/sigma.0100.png} &
        \includegraphics[width=0.23\textwidth]{Figures/Apoptotic/phi_full_config_force/phi.0150.png} &
        \includegraphics[width=0.23\textwidth]{Figures/Apoptotic/sigma_full_config_force/sigma.0150.png} \\
        \multicolumn{2}{c}{\Large (E) (iii) $\Bar{t}=2.64$} & \multicolumn{2}{c}{\Large (E) (iv) $\Bar{t}=3.84$} \\
    \end{tabular}
    \caption{
        Evolution of a nucleation in $\varphi$ and $\sigma$. The internal configurational force $\bs{f}$ is present at the nucleation interface. The nucleation is at $\Bar{t}$=0.41 (A, E(i)), 1.44 (B, E(ii)), 2.64 (C, E(iii)), and 3.84 (D, E(iv)). The parameters selected for this simulation as listed in Table~\ref{tab:Parameter_List}, $\ell_{\varphi}=1.333\times10^{-4}$, $\ell_{2}=1.5$, $\ell^{\varphi}_{r}=960$, $\ell^{\sigma}_{r}=0.28$, and $k_{2}=8.0$. The arrows in images (A), (B), (C), and (D) indicate the direction of $\bs{f}$, while the colour represents the magnitude throughout the simulation (red denotes higher values than blue). In E(i-iv), the evolution of the phase fields $\varphi$ and $\sigma$ have been plotted across the domain, illustrating the global degradation of these phases.
    }
    \label{fig:internal_config_force_cavity}
\end{figure}

Figure~\ref{fig:internal_config_force_cavity} depicts the development of internal configurational forces during nucleation as the phase fields $\varphi$ and $\sigma$ degrade. At $\Bar{t}=0.41$, Figure~\ref{fig:internal_config_force_cavity} (A), the internal configurational force has a greater magnitude at the nucleation interface than at the main interface surrounding the phase field $\varphi$. The same behaviour can be observed in the phase field $\sigma$. Even though the nucleation appears to be circular, in Figure~\ref{fig:internal_config_force_cavity} (A), the magnitude of the internal configurational force along the interface surrounding the nucleation is not uniform. Despite the disuniformity in the internal configurational force magnitude, the cavity expands rapidly in all directions. As the simulation progresses to $\Bar{t}$=1.44 (Figure~\ref{fig:internal_config_force_cavity} (B)), 2.64 (Figure~\ref{fig:internal_config_force_cavity} (C)), and 3.84 (Figure~\ref{fig:internal_config_force_cavity} (D)),  the internal configuration force disappears at the nucleation interface closest to the interface encompassing the bulk of phase fields $\varphi$ and $\sigma$. Therefore, the reduction in the degradation of the phase field $\varphi$ results from the phase field $\sigma$ no longer being present in the space between the nucleation and the main interface. At the later time points $\Bar{t}$=1.44 (Figure~\ref{fig:internal_config_force_cavity} (B)), 2.64 (Figure~\ref{fig:internal_config_force_cavity} (C)), and 3.84 (Figure~\ref{fig:internal_config_force_cavity} (D)), the cytotoxic phase $\sigma$ is depleted in the displayed regions. In these snapshots, a reduced cytotoxic-phase value accompanies weaker configurational forces. At $\sigma=0$, the reaction vanishes, while the potential and interfacial terms remain in the cyto-phase equation. In the lower area of the nucleation at $\Bar{t}=3.84$ (Figure~\ref{fig:internal_config_force_cavity} (D)), it is observed that the magnitude of the internal configurational forces develops in two distinct regions on either side of a finger forming.

\section{Mechanistic interpretation of morphological transitions}

The simulations exhibit boundary protrusions, internal cavities and detached fragments under different reaction and interfacial parameters. Initial heterogeneity permits interior reactions inducing changes in topology such as cavity formation and detachment. Appendix~\ref{sc:local_threshold} separates the effects of reaction amplitude and sensitivity to the cytotoxic phase through analytical recovery boundaries.

Appendix~\ref{sc:interface_motion} derives a local interface velocity that separates curvature-driven motion from reaction-driven recession and shows how orientation affects the speed. An isotropic circular reference damps absolute boundary perturbations, although their amplitude relative to the shrinking radius can increase. Appendix~\ref{sc:controlled_simulations} complements this reference with matched numerical cases: absolute early modes grow under both interfacial laws, and extensive fragmentation can coexist with a nearly circular cloud envelope.

\section{Comparison of electron microscopy images and simulations}
\label{sc:experiments}
 
We compare the simulated cyto phase $\varphi$ with binarised electron microscopy images of KML001-treated PC3, DU145 and LNCaP prostate cancer cells from You~et~al.~\cite{you2015kml001}, whose microscopy figure was subsequently corrected~\cite{you2019correction} (see \href{https://github.com/Nonlinear-Phenomena-Lab/Phase-field-Apoptosis}{repository} for original images and code used to convert the experimental images). The comparison concerns protrusions, detached regions and internal cavities. The experimental panels sample different cells at specified exposure times. Experimental exposure times and dimensionless simulation times are therefore displayed separately, without calibrating correspondence. The cytotoxic phase $\sigma$ represents the effective cytotoxic action and is not measured in these images.

You et al. reports both apoptosis and autophagic vacuoles. Here, a cavity denotes a region of reduced cyto phase enclosed by cyto material but does not identify a biological death pathway. The comparison assesses whether the model reproduces morphological features observed in these treated cells, without assigning all observed vacuoles specifically to apoptosis.

\begin{figure}
    \begin{tabular}{cc}
        \Large Electron Microscopy Image & \Large Phase field $\varphi(\bs{x},t)$ \\
        \includegraphics[width=0.38\linewidth]{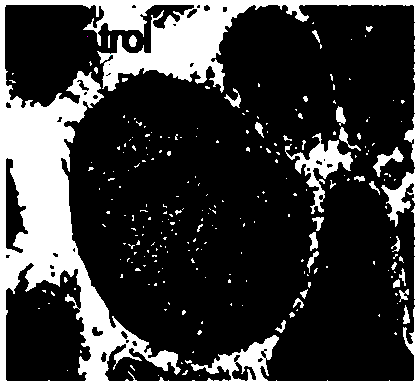} & \includegraphics[width=0.38\textwidth]{Figures/Apoptotic/finger_formation/phi.0000.png} \\
        \Large (A)~(i) 0~hr  & \Large (B)~(i) $\bar{t}=0.00$ \\
        \includegraphics[width=0.38\linewidth]{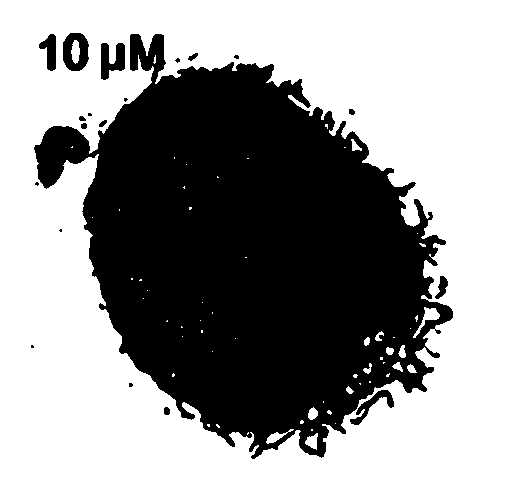} & \includegraphics[width=0.38\textwidth]{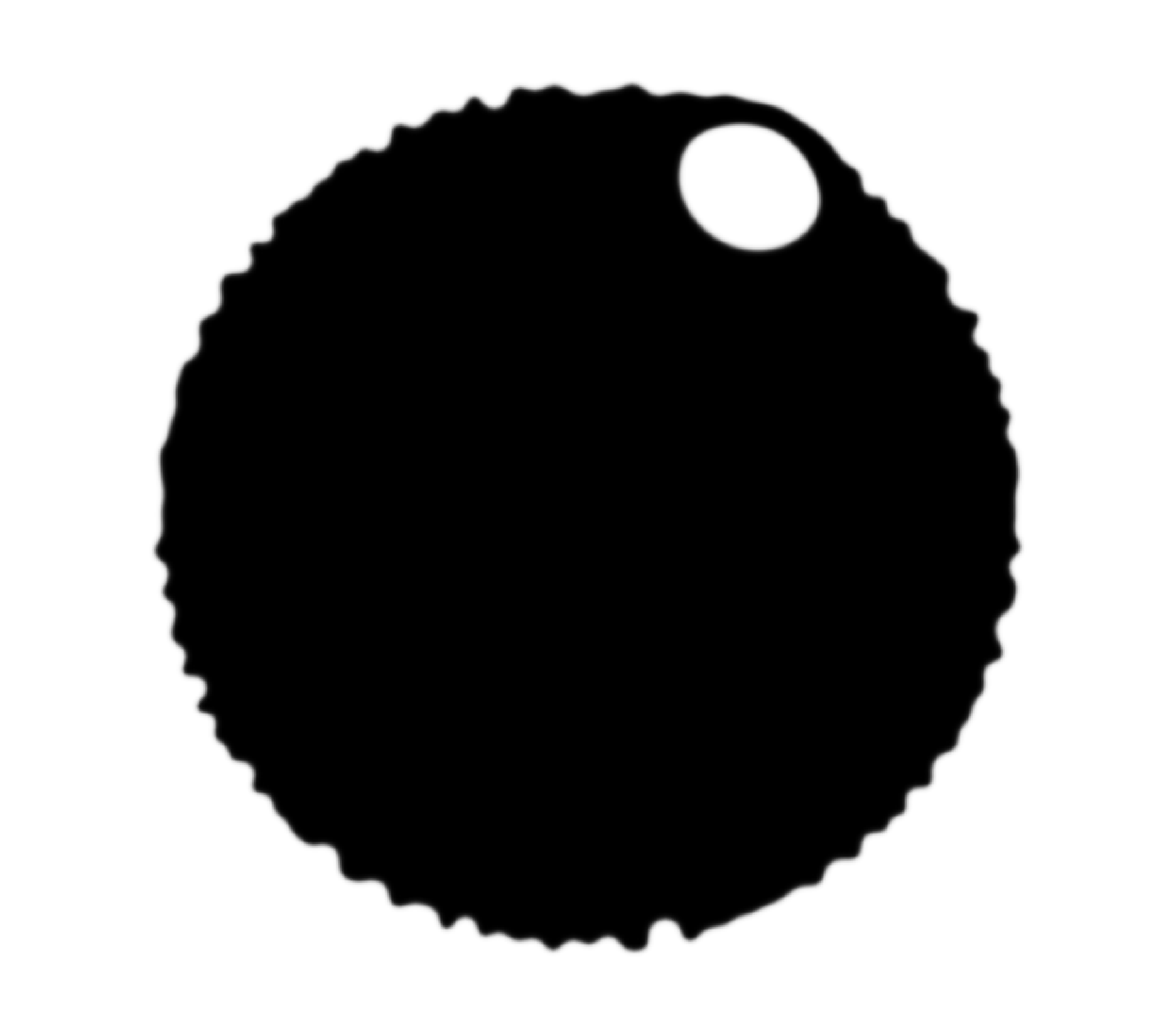} \\
        \Large (A)~(ii) 24~hr  & \Large (B)~(ii) $\bar{t}=1.90$ \\
        \includegraphics[width=0.38\linewidth]{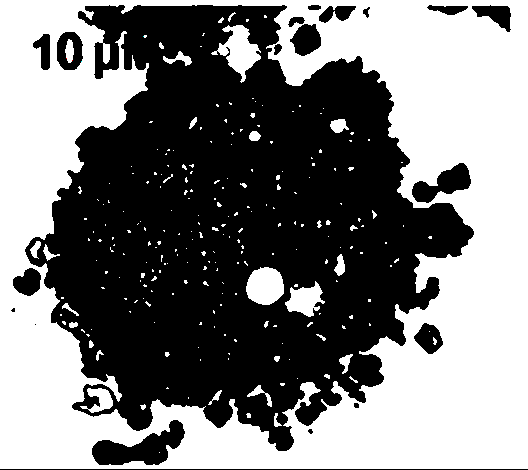} & \includegraphics[width=0.38\textwidth]{Figures/Apoptotic/finger_formation/phi.0150.png} \\
        \Large (A)~(iii) 48~hr  & \Large (B)~(iii) $\bar{t}=4.14$ 
    \end{tabular}
    \caption{
        (A)~(i-iii) Electron microscopy images by You~et~al.~\cite{you2015kml001} of PC3 prostate cancer cells at 0, 24 and 48~h, with a treatment concentration of 10~$\mu$M KML001; (B)~(i-iii) Time evolution of the phase field $\varphi$ undergoing apoptosis induced by the phase field $\sigma$ illustrating finger formation (see Figure~\ref{fig:finger_formation} and Table~\ref{tab:Parameter_List} for parameter values). Experimental panels show separate cells.
    }
    \label{fig:mem_bleb_finger_form}
\end{figure}

Figure~\ref{fig:mem_bleb_finger_form} shows boundary protrusions and internal cavities in both the microscopy panels and the simulation. The experimental panels exhibit stronger branching at 24~h and more detached regions at 48~h than the paired simulation. The model also produces fragmentation in Figures~\ref{fig:finger_formation_diffuse}, \ref{fig:higher_reaction_rate} and \ref{fig:higher_reaction_rate_varied_beta}. These examples demonstrate the range of morphologies accessible to the model.

\begin{figure}
    \begin{tabular}{cc}
        \Large Electron Microscopy Image & \Large Phase field $\varphi(\bs{x},t)$ \\
        \includegraphics[width=0.38\linewidth]{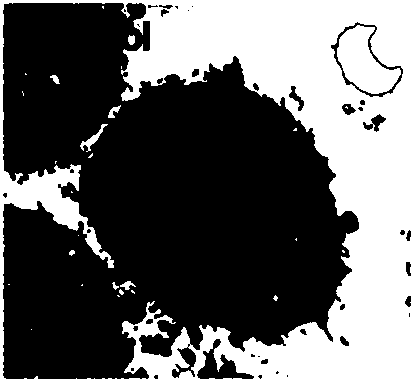} & \includegraphics[width=0.38\textwidth]{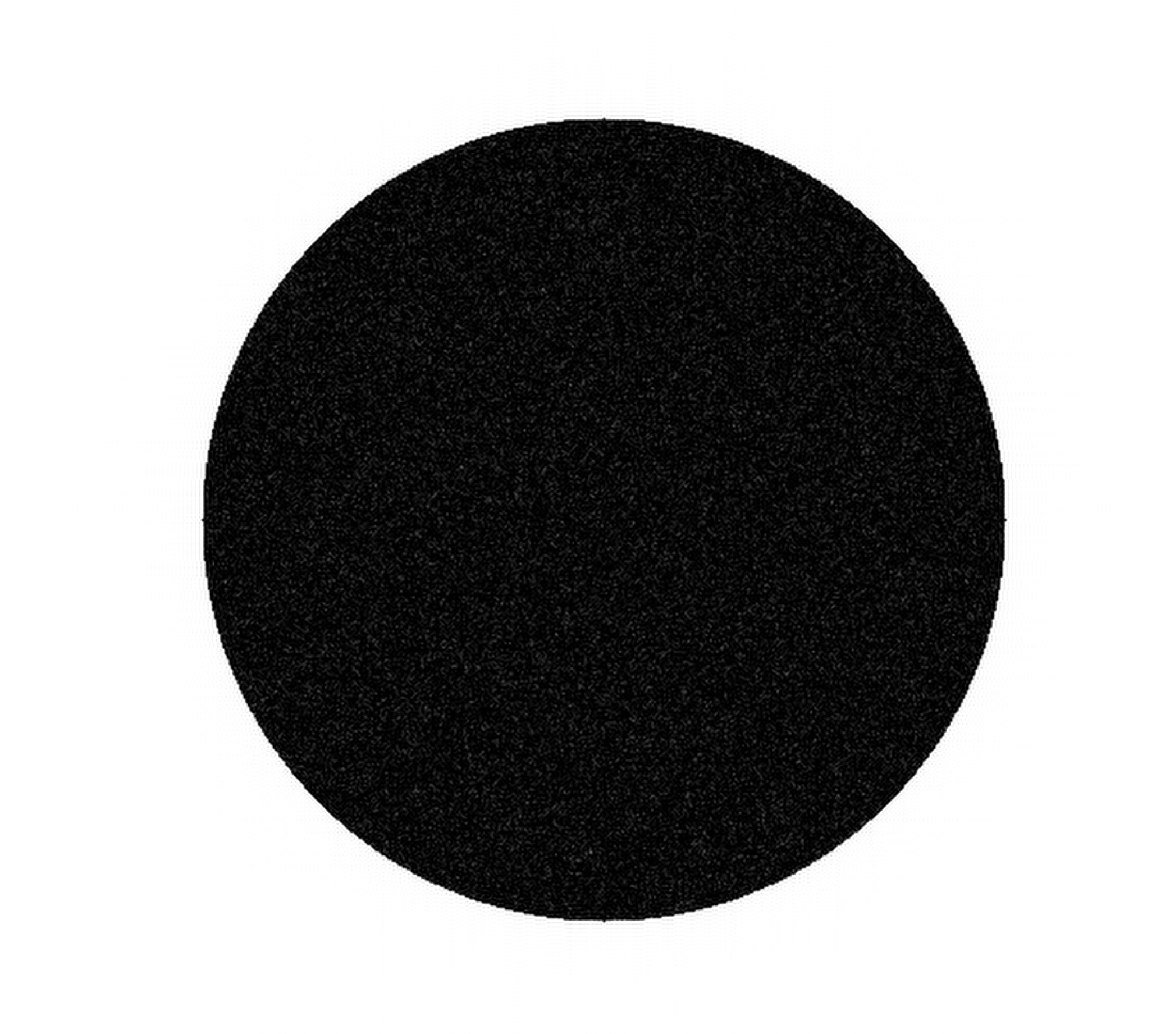} \\
        \Large (A)~(i) 0~hr  & \Large (B)~(i) $\bar{t}=0.00$ \\ 
        \includegraphics[width=0.38\linewidth]{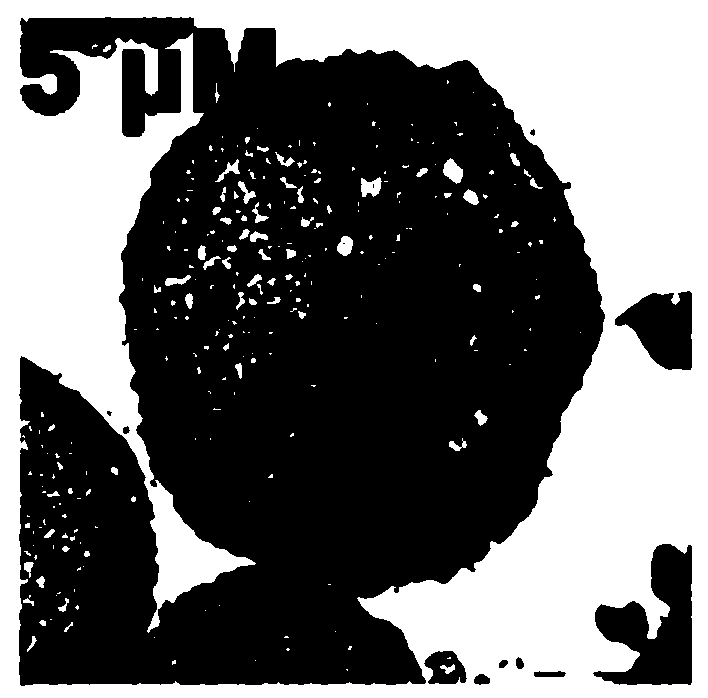} & \includegraphics[width=0.38\textwidth]{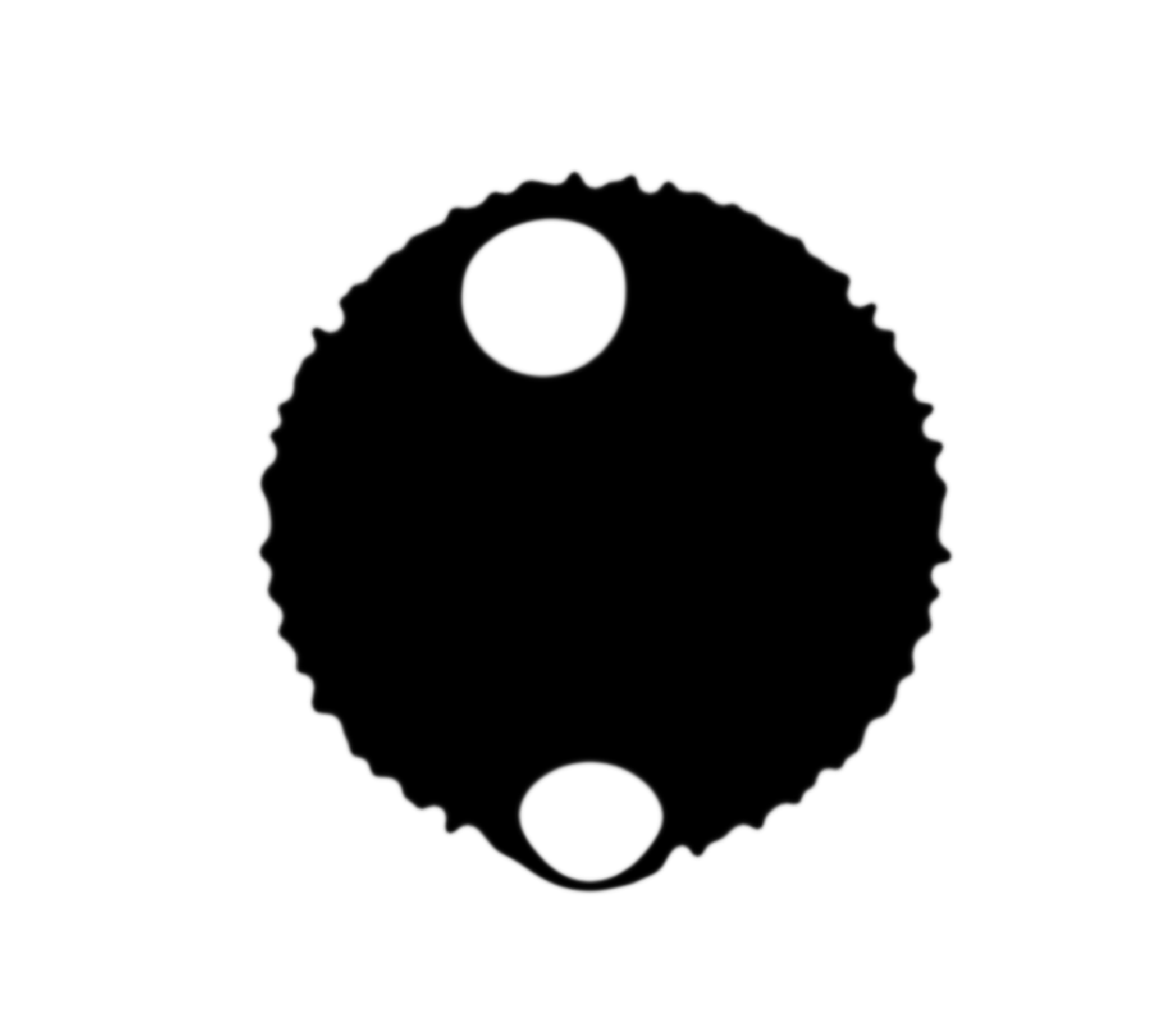} \\
        \Large (A)~(ii) 24~hr  & \Large (B)~(ii) $\bar{t}=2.21$ \\ 
        \includegraphics[width=0.38\linewidth]{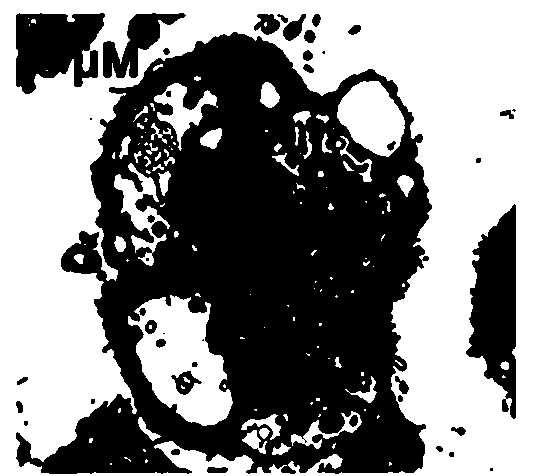} & \includegraphics[width=0.38\textwidth]{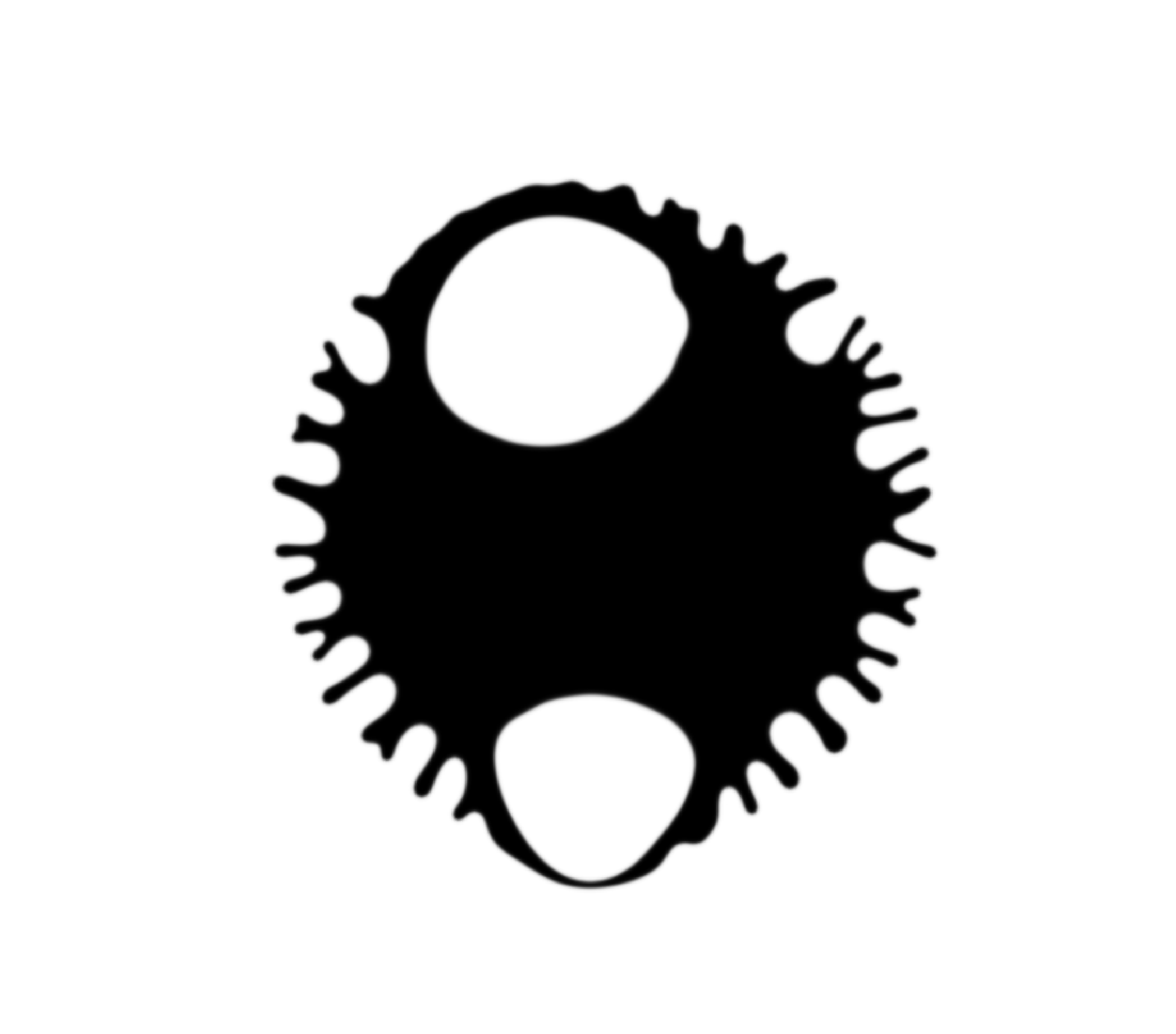} \\
        \Large (A)~(iii) 48~hr  & \Large (B)~(iii) $\bar{t}=4.87$ \\
    \end{tabular}
    \caption{
        (A)~(i-iii) Electron microscopy images by You~et~al.~\cite{you2015kml001} of DU145 prostate cancer cells at 0, 24 and 48~h, with a treatment concentration of 5~$\mu$M KML001; (B) Time evolution of the phase field $\varphi$ undergoing apoptosis induced by the phase field $\sigma$ illustrating nucleation and finger formation. This simulation utilised different noise in the initial condition than the images present in Figure~\ref{fig:finger_formation}. Experimental panels show separate cells.
    }
    \label{fig:cavity-formation}
\end{figure}

Figure~\ref{fig:cavity-formation} compares cavity-like regions in DU145 cells with simulated cavities. Both display cyto material surrounding internal voids, although the experimental features are more numerous and irregular than the prominent smooth simulated cavities. In the simulation, depletion of $\sigma$ suppresses the reaction contribution around a cavity, while potential-driven and interfacial evolution remain active.

\begin{figure}
    \begin{tabular}{cc}
        \Large Electron Microscopy Image & \Large Phase field $\varphi(\bs{x},t)$ \\
        \includegraphics[width=0.38\linewidth]{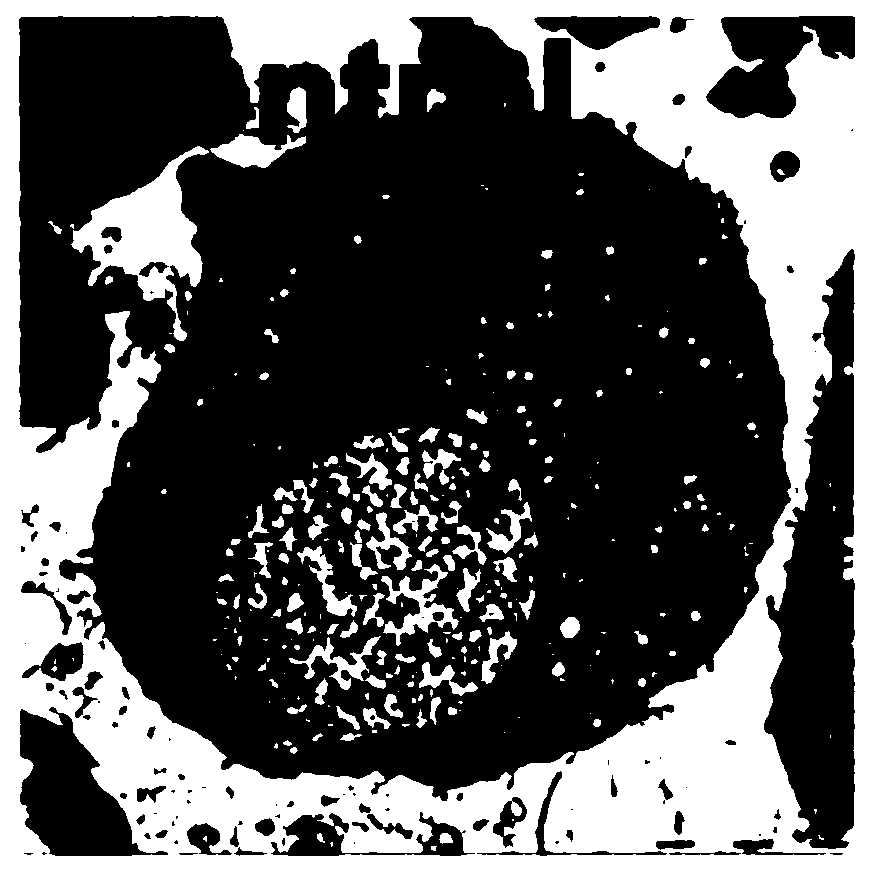} &  \includegraphics[width=0.38\textwidth]{Figures/Apoptotic/higher_reaction_rate_varied_beta_max/phi.0000.png} \\
        \Large (A)~(i) 0~hr  & \Large (B)~(i) $\bar{t}=0.00$ \\
        \includegraphics[width=0.38\linewidth]{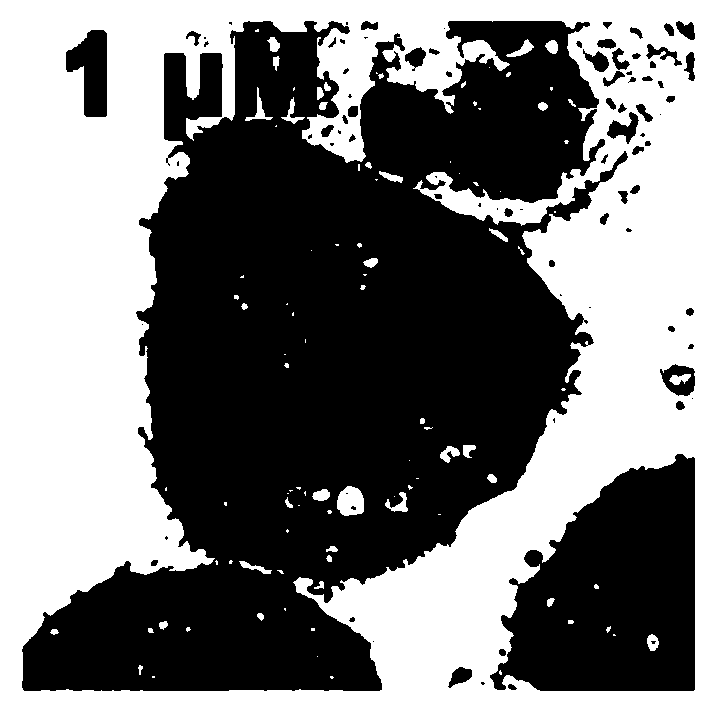} &  \includegraphics[width=0.38\textwidth]{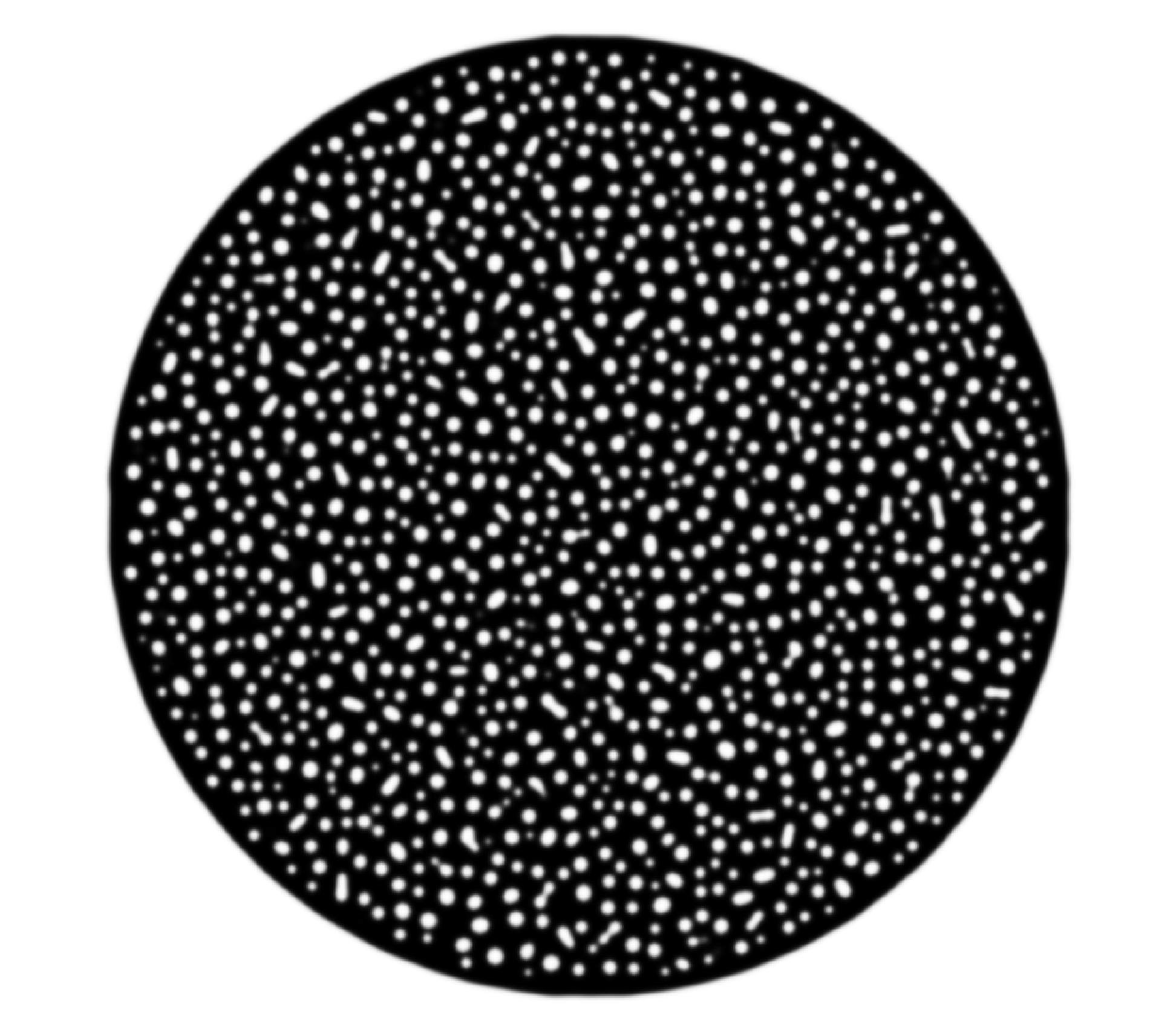} \\
        \Large (A)~(ii) 24~hr  & \Large (B)~(ii) $\bar{t}=1.22$ \\
        \includegraphics[width=0.38\linewidth]{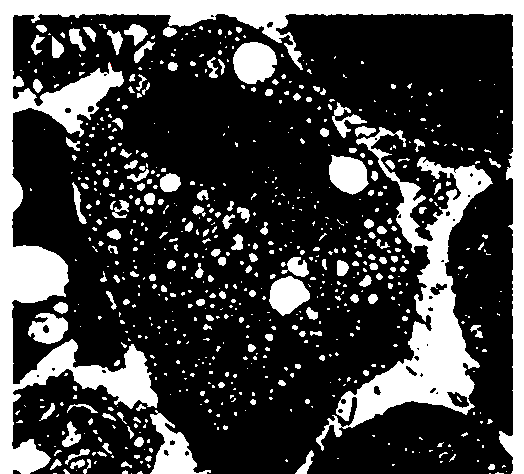} &  \includegraphics[width=0.38\textwidth]{Figures/Apoptotic/higher_reaction_rate_varied_beta_max/phi.0275.png} \\
        \Large (A)~(iii) 48~hr  & \Large (B)~(iii) $\bar{t}=5.98$ 
    \end{tabular}
    \caption{
        (A)~(i-iii) Electron microscopy images by You~et~al.~\cite{you2015kml001} showing internal voids in LNCaP cells at 0, 24 and 48~h, with a treatment concentration of 1~$\mu$M KML001. (B)~(i-iii) Simulated cavities (parameters are given in Figure~\ref{fig:higher_reaction_rate_varied_beta_max}). Experimental panels show separate cells.
    }
    \label{fig:greater-number-of-cavity-formation}
\end{figure}

Figure~\ref{fig:greater-number-of-cavity-formation} compares internal voids in LNCaP cells with multiple cavities in the simulation from Figure~\ref{fig:higher_reaction_rate_varied_beta_max}. Cavity number and spatial distribution differ, particularly between the 24~h panel and $\bar t=1.22$. The shared feature is the occurrence of multiple internal cavities.

Taken together, the comparisons show that the model reproduces several morphological features associated with cytotoxic cell death, including features observed during apoptosis. This qualitative correspondence supports the model's capacity to describe such shape changes, without identifying a unique biochemical mechanism from morphology alone.

\newpage
\subsection{\texorpdfstring{Simulated area evolution}{Simulated area evolution}}
\label{sc:simulated_area}
\begin{figure}
    \centering
    \begin{tabular}{c c}
        \rotatebox{90}{\Large $A/A_{0}$} &
        \begin{tabular}{c}
            \Large (A) \\[0.3em]
            \includegraphics[width=0.5\linewidth]{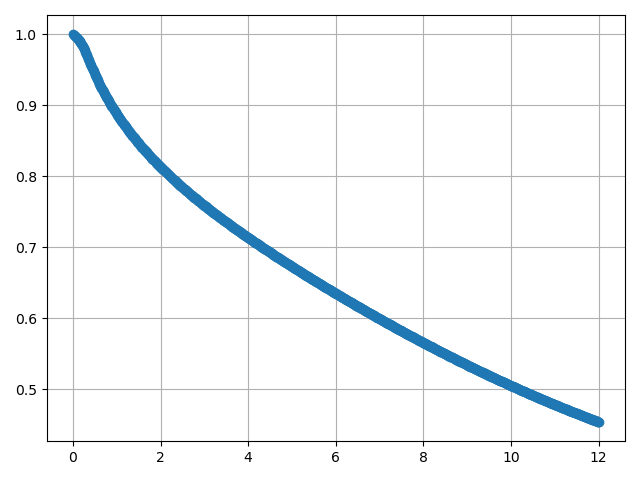} \\[0.4em]
            \Large Time $(\bar{t})$
        \end{tabular}
        \vspace{0.5cm}\\ 

        \rotatebox{90}{\Large $A/A_{0}$} &
        \begin{tabular}{c}
            \Large (B) \\[0.3em]
            \includegraphics[width=0.5\linewidth]{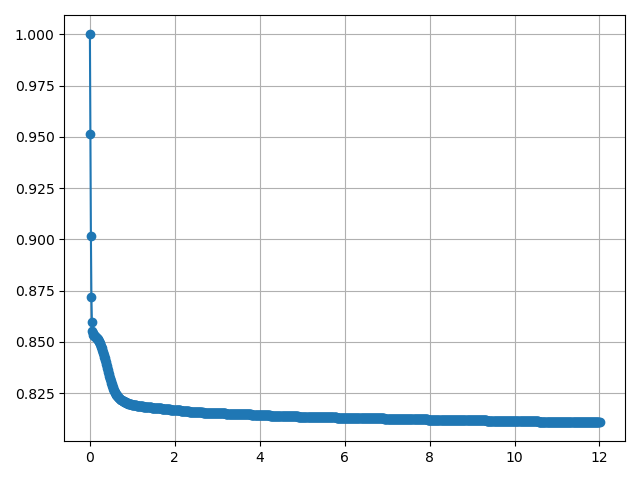} \\[0.4em]
            \Large Time $(\bar{t})$
        \end{tabular}
    \end{tabular}
    \caption{Normalised simulated area fraction $A(t)/A_0$ as a function of dimensionless time. Both panels contain simulation data only. (A) Simulation with parameter values $\ell_{\varphi}=1.333\times10^{-4}$, $\ell_{2}=1.5$, $\ell^{\varphi}_{r}=960$, $\ell^{\sigma}_{r}=0.28$, and $k_{2}=8.0$ (see Table~\ref{tab:Parameter_List} for complete list of parameters). (B) same as (A) with the following parameter values $\ell_{\varphi}=1.333\times10^{-4}$, $\ell_{2}=5.0$, $\ell^{\varphi}_{r}=4240$, $\ell^{\sigma}_{r}=1.3$, and $k_{2}=1.0$ (see Table~\ref{tab:Parameter_List} for full list of parameter values).}
    \label{fig:Area_over_time}
\end{figure}

We quantify shrinkage in the simulations through the domain-averaged cyto phase field. This is an effective area fraction in the diffuse-interface description, defined by
\begin{equation}\label{equ:area_sim}
    A(t)=\frac{1}{|\Omega|}\int_{\Omega}\varphi(\bs{x},t)\dv.
\end{equation}
The corresponding effective cell area is $|\Omega|A(t)$. We report the normalised quantity $A(t)/A_0$, where $A_0=A(0)$. On a fixed domain, the factor $|\Omega|^{-1}$ cancels in this ratio, so the normalised area fraction and normalised effective area are identical. Figure~\ref{fig:Area_over_time} presents simulation results only as no experimental area measurements are available for comparison.

The two panels correspond to the simulations shown in Figures~\ref{fig:finger_formation} and \ref{fig:higher_reaction_rate_varied_beta_max}, respectively. The area measure quantifies the loss of cyto phase but does not distinguish blebbing, cavity formation and fragmentation. These features depend on the spatial distribution and connectivity of the phase field, as well as on its initial condition.

\section{Conclusions}

The model represents apoptotic activation by a spatially distributed cytotoxic phase coupled to the cyto phase, focusing on morphological evolution without resolving individual biochemical pathways. The coupled phase-field model, implemented in Dedalus~\cite{burns2020dedalus}, produces shrinkage, finger formation, cavities and fragmentation. Changes in interfacial scale, reaction coefficients and rate coupling produce distinct morphological trajectories. In this framework the free energy in Equation~\eqref{equ:Free_energy} does not include all biological processes associated with apoptosis, such as adhesion, area and volume constraints.

The configurational balance describes the energetic driving of interface evolution. Appendix~\ref{sc:interfacial_tension} relates the integrated cyto-phase stress to effective interfacial tension and its curvature force. Appendix~\ref{sc:local_threshold} establishes homogeneous recovery regimes for fixed and evolving cytotoxic phase. Reaction saturation allows sufficiently intact homogeneous material to recover for every nonnegative cytotoxic-phase history in the first reported parameter set.

The analytical recovery diagram separates reaction amplitude from sensitivity to the cytotoxic phase. The matched numerical comparisons in Appendix~\ref{sc:controlled_simulations} show that cavities, protrusions and fragmentation persist under the energy-matched isotropic law. Early boundary modes grow in absolute amplitude, while extensive fragmentation can leave a nearly circular gross envelope.

Qualitative comparison of the simulations in Figures~\ref{fig:finger_formation} and \ref{fig:higher_reaction_rate_varied_beta_max} with electron microscopy from You et al.~\cite{you2015kml001,you2019correction} identifies shared morphological features, including boundary protrusions, cavities and detached regions. The normalised cyto-phase area quantifies simulated shrinkage. Together, the simulations and analytical results describe how reaction kinetics and interfacial energetics govern morphological change and local recovery.

\section{Acknowledgments}
The Engineering and Physical Sciences Research Council (EPSRC) supported the project's funding via the Centre of Doctoral Training (CDT) in Transformative Pharmaceutical Technologies [EP/S023054/1]. We are grateful for access to the University of Nottingham's Ada HPC service. [LE] This work was partially supported by the Flexible Interdisciplinary Research Collaboration Fund at the University of Nottingham, Project ID 7466664.

\section{Data Availability}
The original numerical solver is available in the \href{https://github.com/Nonlinear-Phenomena-Lab/Phase-field-Apoptosis}{https://github.com/Nonlinear-Phenomena-Lab/Phase-field-Apoptosis} repository. The controlled-run parameters, initialisation and measurement definitions are specified in Appendix~\ref{sc:controlled_simulations}.


\appendix

\section{Configurational mechanics}
\label{sc:configurational.mechanics}

Following Gurtin~\cite{gurtin1999configurational}, Fried~\cite{fried2006relationship} and Espath \& Calo~\cite{espath2021phase}, we write the local configurational balance as
\begin{equation}\label{equ:local_config_balance}
    \Div \bs{C} + \bs{f} = \bs{0},
\end{equation}
Here $\bs{1}$ denotes the identity tensor, $\bs{C}$ is the configurational stress tensor and $\bs{f}$ is the internal configurational force density.

For $\psi=\psi(\varphi,\sigma,\nabla\varphi,\nabla\sigma)$ with no explicit spatial dependence, define $\bs{\xi}_u=\partial\psi/\partial\nabla u$, where $u\in\{\varphi,\sigma\}$. The chain and product rules give
\begin{equation}\label{equ:config_cancellation}
{
 \Div\left(\psi\bs{1}-\sum_{u\in\{\varphi,\sigma\}}\nabla u\otimes\bs{\xi}_u\right)
 =\sum_{u\in\{\varphi,\sigma\}}
 \left(\frac{\partial\psi}{\partial u}-\Div\bs{\xi}_u\right)\nabla u.
}
\end{equation}
The Hessian products $(\nabla^2u)\bs{\xi}_u$ cancel. Using the field balances $\tau\dot\varphi-(\varsigma_\varphi^p-\varsigma_\varphi^r)r=-\delta\Psi/\delta\varphi$ and $\alpha\dot\sigma-\beta\dot\varphi-\varsigma_\sigma r=-\delta\Psi/\delta\sigma$ gives the configurational stress and internal force:
\begin{equation}\label{equ:configurational_stress_tensor}
    \bs{C} =  \psi\bs{1} - \nabla\varphi\otimes\frac{\partial\psi}{\partial\nabla\varphi} - \nabla\sigma\otimes\frac{\partial\psi}{\partial\nabla\sigma},
\end{equation}
and
\begin{equation}\label{equ:internal_config_force}
    \bs{f} = (\tau \dot{\varphi} - (\varsigma_\varphi^p - \varsigma_\varphi^r) r) \nabla \varphi + (\alpha \dot{\sigma} - \beta \dot{\varphi} - \varsigma_\sigma r) \nabla \sigma.
\end{equation}

\subsection{\texorpdfstring{Correspondence with interfacial tension}{Correspondence with interfacial tension}}
\label{sc:interfacial_tension}
\begingroup
For $\psi_\varphi=f(\varphi)+\epsilon(\theta)^2|\nabla\varphi|^2/2$ and $\bs{C}_\varphi=\psi_\varphi\bs{1}-\nabla\varphi\otimes\bs{\xi}_\varphi$, take a planar equilibrium profile $\varphi=q(z)$ at fixed orientation, with $q(-\infty)=0$, $q(+\infty)=1$ and $q'>0$. Here, $z$ is signed normal distance and primes on $q$ denote $z$ derivatives. Stationarity of $\int_{\bb{R}}[f(q)+\epsilon(\theta)^2(q')^2/2]\di{z}$ gives $f'(q)-\epsilon(\theta)^2q''=0$. This is the reference equilibrium of the cyto-phase energy used to define tension. Multiplication by $q'$ and integration, using $q'\to0$ and $f(q)\to0$ in the bulk, give equality of the gradient and potential energy densities (equipartition) and the interfacial tension:
\begin{equation}\label{equ:tension_profile}
\frac{\epsilon(\theta)^2}{2}(q')^2=f(q)\qquad\text{and}\qquad\Upsilon(\theta):=\int_{\bb{R}}\psi_\varphi\di{z}=\epsilon(\theta)\int_0^1\sqrt{2f(q)}\di{q}.
\end{equation}
Here, $\epsilon>0$ and $\Upsilon$ is energy per unit interfacial area, or per unit length in two dimensions. For $f(q)=q^2(1-q)^2/4$, $c_f:=\int_0^1\sqrt{2f(q)}\di{q}=1/(6\sqrt{2})$, so $\Upsilon=c_f\epsilon$.

For constant isotropic $\epsilon$, write $\bs{n}=\nabla\varphi/|\nabla\varphi|$ and $\bs{P}=\bs{1}-\bs{n}\otimes\bs{n}$. The identity $\epsilon^2(q')^2=2f(q)$ gives
\begin{equation}\label{equ:tension_stress_iso}
\bs{C}_\varphi=\epsilon^2(q')^2\bs{P}\qquad\text{and}\qquad\bs{S}_\Gamma:=\int_{\bb{R}}\bs{C}_\varphi\di{z}=\Upsilon\bs{P}.
\end{equation}
The subscript $\Gamma$ denotes the limiting interface, and $\bs{S}_\Gamma$ is the integrated stress. This is the surface stress of an interface under tension. For constant $\Upsilon$ and interface thickness small compared with the radii of curvature, the leading local-equilibrium capillary force is
\begin{equation}\label{equ:tension_force_iso}
\bs{t}_{\rm cap}=\Divs \bs{S}_\Gamma=-\Upsilon\kappa\bs{n},\qquad\text{where } \kappa=\Divs \bs{n}.
\end{equation}
Here $\Divs\bs{P}=-(\Divs\bs{n})\bs{n}$. The force is per unit area (per unit length in two dimensions) and points inward on a convex interface, independently of the choice of normal orientation.

For smooth anisotropy in two dimensions, let $\bs{t}$ be $\bs{n}$ rotated anticlockwise by $\pi/2$, and let primes on $\Upsilon$ denote orientation derivatives. Then,
\begin{equation}\label{equ:tension_stress_aniso}
\bs{\xi}_\varphi=q'\bigl(\epsilon^2\bs{n}+\epsilon\epsilon_\theta\bs{t}\bigr)\qquad\text{and}\qquad\bs{S}_\Gamma=\Upsilon\bs{t}\otimes\bs{t}-\Upsilon'\bs{n}\otimes\bs{t}.
\end{equation}
Indeed, $\int\epsilon^2(q')^2\di{z}=\Upsilon$ and $\int\epsilon\epsilon_\theta(q')^2\di{z}=\Upsilon'$. For arclength $s$ increasing along $\bs{t}$, $\partial_s\bs{n}=\kappa\bs{t}$ and $\partial_s\bs{t}=-\kappa\bs{n}$, so
\begin{equation}\label{equ:tension_force_aniso}
\bs{t}_{\rm cap}=\partial_s(\bs{S}_\Gamma\bs{t})=-\bigl(\Upsilon+\Upsilon''\bigr)\kappa\bs{n}.
\end{equation}
For $\epsilon=\widetilde\epsilon(\vartheta+\cos\theta)$, $\Upsilon+\Upsilon''=c_f\widetilde\epsilon\vartheta$: the capillary stiffness is constant on each smooth angular branch.

For the additive energy $\psi=\psi_\varphi+\gamma^2|\nabla\sigma|^2/2$, the cytotoxic phase contributes separately through
\begin{equation}\label{equ:activation_config_stress}
\bs{C}_\sigma=\frac{\gamma^2}{2}|\nabla\sigma|^2\bs{1}-\gamma^2\nabla\sigma\otimes\nabla\sigma,\qquad\bs{C}=\bs{C}_\varphi+\bs{C}_\sigma.
\end{equation}
Comparison with measured tension requires a dimensional energy scale and, for a planar reduction, a thickness convention (line tension $h\Upsilon$ for uniform thickness $h$). This effective tension does not separate membrane, cortical and cytoskeletal contributions or specify an independent bending modulus. The latter requires an explicit curvature energy, such as a Helfrich-type bending energy~\cite{du2005spontaneous}.
\endgroup

\clearpage

\section{\texorpdfstring{Exposure thresholds and recovery}{Exposure thresholds and recovery}}
\label{sc:local_threshold}
\begingroup
We analyse spatially homogeneous states, using $\ell_r^\varphi>0$ for the degradation magnitudes listed in Table~\ref{tab:Parameter_List}. We assume $\ell_f,k_2>0$ and $\sigma\ge0$. Time $\bar t$ is dimensionless, and recovery means $\varphi(\bar t)\to1$ after a downward perturbation.

\subsection{Reduction and fixed-exposure threshold}
\label{sc:fixed_exposure_threshold}

Differentiating $f(\varphi)=\varphi^2(1-\varphi)^2/4$ gives $f'(\varphi)=\varphi(\varphi-1)(\varphi-1/2)$. Omitting spatial terms therefore gives
\begin{equation}\label{equ:local_threshold}
\dot\varphi=-\ell_f f'(\varphi)-\ell_r^\varphi\varphi(1-\varphi)\arctan(k_2\sigma)=\ell_f\varphi(1-\varphi)\left[\varphi-\frac12-\frac{\ell_r^\varphi}{\ell_f}\arctan(k_2\sigma)\right].
\end{equation}
The cyto-phase threshold is
\begin{equation}\label{equ:critical_cyto_phase}
 \varphi_c(\sigma):=\frac12+\frac{\ell_r^\varphi}{\ell_f}\arctan(k_2\sigma).
\end{equation}
For fixed $\sigma$, the sign of $\dot\varphi$ in $0<\varphi<1$ is the sign of $\varphi-\varphi_c$. Thus, when $\varphi_c<1$, initial states below $\varphi_c$ decay to zero and those above it recover to one, while $\varphi_c$ itself is an unstable equilibrium. If $\varphi_c\ge1$, every strictly interior initial state decays.

To examine small perturbations of the intact state, set $\eta=1-\varphi$. Substituting $\varphi=1-\eta$ gives exactly
\begin{equation}\label{equ:recovery_perturbation}
\dot\eta=\ell_f\eta(1-\eta)(\varphi_c(\sigma)-1+\eta)=\left[\ell_r^\varphi\arctan(k_2\sigma)-\frac{\ell_f}{2}\right]\eta+O(\eta^2).
\end{equation}
The perturbation decays if
\begin{equation}\label{equ:local_recovery_condition}
 \arctan(k_2\sigma)<\frac{\ell_f}{2\ell_r^\varphi}.
\end{equation}
Since $\arctan(k_2\sigma)<\pi/2$ for finite $\sigma$, this holds at every finite exposure when $\ell_r^\varphi\le\ell_f/\pi$. Otherwise, solving the equality gives
\begin{equation}\label{equ:critical_exposure}
 \sigma_c=\frac{1}{k_2}\tan\left(\frac{\ell_f}{2\ell_r^\varphi}\right),
 \qquad \ell_r^\varphi>\frac{\ell_f}{\pi},
\end{equation}
with recovery of small perturbations for $\sigma<\sigma_c$. At equality, $\varphi_c=1$ and $\dot\eta=\ell_f\eta^2(1-\eta)>0$: the vanishing linear rate does not imply stability. Subcritical exposure permits recovery but does not rescue fixed-exposure initial states already below $\varphi_c$.

\subsection{Recovery with evolving cytotoxic phase}

With $\varphi_c$ defined by expression \eqref{equ:critical_cyto_phase}, the homogeneous coupled equations are
\begin{equation}
\dot\varphi=\ell_f\varphi(1-\varphi)(\varphi-\varphi_c(\sigma))\qquad\text{and}\qquad\dot\sigma=\ell_2\dot\varphi+\ell_r^\sigma\varphi(1-\varphi)\arctan(k_2\sigma).
\end{equation}
Both rates vanish at $\varphi=1$, so every constant pair $(1,\sigma_*)$ is an equilibrium. Set $\varphi=1-\eta$ and $\sigma=\sigma_*+\delta\sigma$. Considering terms of order $\mathcal{O}(\eta)$,
\begin{equation}\label{equ:coupled_recovery_linearisation}
\begin{aligned}
\dot\eta=\ell_f(\varphi_c(\sigma_*)-1)\eta\qquad\text{and}\qquad\dot{\delta\sigma}=(-\ell_2\ell_f(\varphi_c(\sigma_*)-1)+\ell_r^\sigma\arctan(k_2\sigma_*))\eta.
\end{aligned}
\end{equation}
Terms involving $\delta\sigma$ enter multiplied by $\eta$ and are therefore second order. The cyto-phase perturbation evolves independently of $\delta\sigma$ at this order, so the linear system has eigenvalues $\ell_f(\varphi_c(\sigma_*)-1)$ and zero. The zero eigenvalue reflects the family of equilibria. If $\varphi_c(\sigma_*)<1$, sufficiently small downward cyto-phase perturbations decay, and the coupled state approaches $(1,\sigma_\infty)$, where $\sigma_\infty$ may differ from $\sigma_*$. Indeed, a strict negative cyto-phase eigenvalue persists in a neighbourhood of the equilibrium: $\eta$ decays exponentially there and $\dot\sigma=O(\eta)$ is integrable. The positive cytotoxic-phase reaction term changes the cytotoxic-phase perturbation but not this local recovery threshold.

For finite perturbations, suppose $0\le\sigma(\bar t)\le\bar\sigma$ for all subsequent times and $\varphi_c(\bar\sigma)<\varphi(0)<1$. Since $\varphi_c$ increases with $\sigma$,
\begin{equation}\label{equ:uniform_recovery_bound}
\dot\varphi\ge\ell_f\varphi(1-\varphi)(\varphi-\varphi_c(\bar\sigma))>0.
\end{equation}
The initial inequality persists: the rate is positive above $\varphi_c(\bar\sigma)$, so $\varphi$ cannot decrease to that threshold. Thus $\varphi(\bar t)\ge\varphi(0)$, while $\varphi=1$ remains invariant. For $0<\varphi<1$,
\begin{equation}
\frac{\dot\varphi}{\varphi(1-\varphi)}=\frac{\di{}}{\di{\bar t}}\log\frac{\varphi}{1-\varphi}\ge\ell_f(\varphi(0)-\varphi_c(\bar\sigma))>0.
\end{equation}
Integrating yields
\begin{equation}\label{equ:recovery_integrated_bound}
\log\frac{\varphi(\bar t)}{1-\varphi(\bar t)}\ge\log\frac{\varphi(0)}{1-\varphi(0)}+\ell_f(\varphi(0)-\varphi_c(\bar\sigma))\bar t.
\end{equation}
The right-hand side tends to $+\infty$, which requires $\varphi(\bar t)\to1$. This is a sufficient recovery condition for a bounded cytotoxic-phase history.

Reaction saturation removes the need to prescribe $\bar\sigma$ when $\ell_r^\varphi<\ell_f/\pi$, because
\begin{equation}\label{equ:saturation_recovery}
\varphi_c(\sigma(\bar t))\le\frac12+\frac{\pi\ell_r^\varphi}{2\ell_f}<1.
\end{equation}
Using this upper bound in the same argument proves recovery whenever $\varphi(0)>1/2+\pi\ell_r^\varphi/(2\ell_f)$, for any nonnegative cytotoxic-phase history, including the coupled evolution.

Finally, at fixed $0<\varphi<1$,
\begin{equation}
\pards{\dot\varphi}{\sigma}=-\frac{\ell_r^\varphi k_2\varphi(1-\varphi)}{1+(k_2\sigma)^2}<0.
\end{equation}
Scalar comparison therefore implies that, from identical initial cyto phase, an everywhere larger prescribed cytotoxic-phase history yields no larger cyto-phase trajectory. Sustained exposure can suppress recovery relative to a lower, depleting history. Recovery follows once the cyto phase lies above the threshold associated with an upper bound on all subsequent exposure. The saturation bound supplies such a threshold independently of the history.

\subsection{Dependence on reaction amplitude and sensitivity to the cytotoxic phase}
For fixed $\sigma>0$ and $1/2<\varphi(0)<1$, the boundary between homogeneous recovery and degradation is
\begin{equation}\label{equ:finite_recovery_boundary}
\ell_r^\varphi=\frac{\ell_f(\varphi(0)-1/2)}{\arctan(k_2\sigma)}.
\end{equation}
Recovery occurs below this boundary. The intact-state stability boundary is its limit as $\varphi(0)\to1$. The separate parameter effects follow from
\begin{equation}
\pards{\varphi_c}{\ell_r^\varphi}=\frac{\arctan(k_2\sigma)}{\ell_f}>0\qquad\text{and}\qquad\pards{\varphi_c}{k_2}=\frac{\ell_r^\varphi\sigma}{\ell_f(1+(k_2\sigma)^2)}>0.
\end{equation}
With scaling and stoichiometric coefficients fixed, $\ell_r^\varphi=T(\varsigma_\varphi^r-\varsigma_\varphi^p)k_1/(\pi\tau)$ and $\ell_r^\sigma=T\varsigma_\sigma k_1/(\pi\alpha)$: changing $k_1$ changes both reaction coefficients. The reaction coefficient multiplying $\varphi(1-\varphi)$ in the cyto-phase equation is approximately $\ell_r^\varphi k_2\sigma$ for $k_2\sigma\ll1$ and approaches $\pi\ell_r^\varphi/2$ for large positive exposure. Thus, $k_2$ changes sensitivity, while $k_1$ also changes the saturation level.

\begin{figure}[htbp]

\centering
\setlength{\unitlength}{1mm}
\begin{picture}(145,64)(-12,-9)
\put(0,0){\vector(1,0){121}}
\put(0,0){\vector(0,1){52}}
\put(105,-8){$k_2\sigma$}
\put(-12,54){$\ell_r^\varphi/\ell_f$}
\put(0,-1){\line(0,1){2}}
\put(-1,-5){\makebox(0,0){0}}
\put(24,-1){\line(0,1){2}}
\put(23,-5){\makebox(0,0){2}}
\put(48,-1){\line(0,1){2}}
\put(47,-5){\makebox(0,0){4}}
\put(72,-1){\line(0,1){2}}
\put(71,-5){\makebox(0,0){6}}
\put(96,-1){\line(0,1){2}}
\put(95,-5){\makebox(0,0){8}}
\put(120,-1){\line(0,1){2}}
\put(119,-5){\makebox(0,0){10}}
\put(-1,16.0){\line(1,0){2}}
\put(-3,16.0){\makebox(0,0)[r]{0.5}}
\put(-1,32){\line(1,0){2}}
\put(-3,32){\makebox(0,0)[r]{1}}
\put(-1,48.0){\line(1,0){2}}
\put(-3,48.0){\makebox(0,0)[r]{1.5}}
\qbezier(4.1550,48.0000)(4.5171,44.7160)(4.8791,41.4321)
\qbezier(4.8791,41.4321)(5.2411,39.0294)(5.6031,36.6267)
\qbezier(5.6031,36.6267)(5.9651,34.8008)(6.3271,32.9749)
\qbezier(6.3271,32.9749)(6.6892,31.5463)(7.0512,30.1177)
\qbezier(7.0512,30.1177)(7.4132,28.9738)(7.7752,27.8299)
\qbezier(7.7752,27.8299)(8.1372,26.8966)(8.4992,25.9633)
\qbezier(8.4992,25.9633)(8.8612,25.1898)(9.2233,24.4162)
\qbezier(9.2233,24.4162)(9.5853,23.7665)(9.9473,23.1167)
\qbezier(9.9473,23.1167)(10.3093,22.5646)(10.6713,22.0125)
\qbezier(10.6713,22.0125)(11.0333,21.5387)(11.3954,21.0649)
\qbezier(11.3954,21.0649)(11.7574,20.6546)(12.1194,20.2443)
\qbezier(12.1194,20.2443)(12.4814,19.8861)(12.8434,19.5280)
\qbezier(12.8434,19.5280)(13.2054,19.2133)(13.5674,18.8985)
\qbezier(13.5674,18.8985)(13.9295,18.6200)(14.2915,18.3416)
\qbezier(14.2915,18.3416)(14.6535,18.0938)(15.0155,17.8460)
\qbezier(15.0155,17.8460)(15.3775,17.6243)(15.7395,17.4027)
\qbezier(15.7395,17.4027)(16.1016,17.2034)(16.4636,17.0041)
\qbezier(16.4636,17.0041)(16.8256,16.8242)(17.1876,16.6442)
\qbezier(17.1876,16.6442)(17.5496,16.4810)(17.9116,16.3179)
\qbezier(17.9116,16.3179)(18.2736,16.1693)(18.6357,16.0208)
\qbezier(18.6357,16.0208)(18.9977,15.8851)(19.3597,15.7494)
\qbezier(19.3597,15.7494)(19.7217,15.6250)(20.0837,15.5006)
\qbezier(20.0837,15.5006)(20.4457,15.3862)(20.8078,15.2718)
\qbezier(20.8078,15.2718)(21.1698,15.1664)(21.5318,15.0609)
\qbezier(21.5318,15.0609)(21.8938,14.9634)(22.2558,14.8658)
\qbezier(22.2558,14.8658)(22.6178,14.7754)(22.9798,14.6850)
\qbezier(22.9798,14.6850)(23.3419,14.6009)(23.7039,14.5169)
\qbezier(23.7039,14.5169)(24.0659,14.4386)(24.4279,14.3603)
\qbezier(24.4279,14.3603)(24.7899,14.2873)(25.1519,14.2142)
\qbezier(25.1519,14.2142)(25.5140,14.1458)(25.8760,14.0775)
\qbezier(25.8760,14.0775)(26.2380,14.0134)(26.6000,13.9494)
\qbezier(26.6000,13.9494)(26.9620,13.8892)(27.3240,13.8291)
\qbezier(27.3240,13.8291)(27.6860,13.7725)(28.0481,13.7159)
\qbezier(28.0481,13.7159)(28.4101,13.6627)(28.7721,13.6094)
\qbezier(28.7721,13.6094)(29.1341,13.5591)(29.4961,13.5088)
\qbezier(29.4961,13.5088)(29.8581,13.4613)(30.2202,13.4138)
\qbezier(30.2202,13.4138)(30.5822,13.3688)(30.9442,13.3238)
\qbezier(30.9442,13.3238)(31.3062,13.2812)(31.6682,13.2386)
\qbezier(31.6682,13.2386)(32.0302,13.1982)(32.3923,13.1578)
\qbezier(32.3923,13.1578)(32.7543,13.1193)(33.1163,13.0809)
\qbezier(33.1163,13.0809)(33.4783,13.0444)(33.8403,13.0079)
\qbezier(33.8403,13.0079)(34.2023,12.9731)(34.5643,12.9383)
\qbezier(34.5643,12.9383)(34.9264,12.9052)(35.2884,12.8720)
\qbezier(35.2884,12.8720)(35.6504,12.8404)(36.0124,12.8087)
\qbezier(36.0124,12.8087)(36.3744,12.7785)(36.7364,12.7483)
\qbezier(36.7364,12.7483)(37.0985,12.7194)(37.4605,12.6905)
\qbezier(37.4605,12.6905)(37.8225,12.6628)(38.1845,12.6352)
\qbezier(38.1845,12.6352)(38.5465,12.6087)(38.9085,12.5822)
\qbezier(38.9085,12.5822)(39.2705,12.5568)(39.6326,12.5314)
\qbezier(39.6326,12.5314)(39.9946,12.5071)(40.3566,12.4827)
\qbezier(40.3566,12.4827)(40.7186,12.4593)(41.0806,12.4359)
\qbezier(41.0806,12.4359)(41.4426,12.4134)(41.8047,12.3910)
\qbezier(41.8047,12.3910)(42.1667,12.3694)(42.5287,12.3477)
\qbezier(42.5287,12.3477)(42.8907,12.3269)(43.2527,12.3061)
\qbezier(43.2527,12.3061)(43.6147,12.2861)(43.9767,12.2661)
\qbezier(43.9767,12.2661)(44.3388,12.2468)(44.7008,12.2275)
\qbezier(44.7008,12.2275)(45.0628,12.2089)(45.4248,12.1903)
\qbezier(45.4248,12.1903)(45.7868,12.1723)(46.1488,12.1543)
\qbezier(46.1488,12.1543)(46.5109,12.1370)(46.8729,12.1197)
\qbezier(46.8729,12.1197)(47.2349,12.1029)(47.5969,12.0862)
\qbezier(47.5969,12.0862)(47.9589,12.0700)(48.3209,12.0538)
\qbezier(48.3209,12.0538)(48.6829,12.0382)(49.0450,12.0225)
\qbezier(49.0450,12.0225)(49.4070,12.0074)(49.7690,11.9922)
\qbezier(49.7690,11.9922)(50.1310,11.9776)(50.4930,11.9629)
\qbezier(50.4930,11.9629)(50.8550,11.9487)(51.2171,11.9345)
\qbezier(51.2171,11.9345)(51.5791,11.9207)(51.9411,11.9070)
\qbezier(51.9411,11.9070)(52.3031,11.8936)(52.6651,11.8803)
\qbezier(52.6651,11.8803)(53.0271,11.8674)(53.3891,11.8544)
\qbezier(53.3891,11.8544)(53.7512,11.8419)(54.1132,11.8293)
\qbezier(54.1132,11.8293)(54.4752,11.8171)(54.8372,11.8050)
\qbezier(54.8372,11.8050)(55.1992,11.7931)(55.5612,11.7813)
\qbezier(55.5612,11.7813)(55.9233,11.7698)(56.2853,11.7583)
\qbezier(56.2853,11.7583)(56.6473,11.7471)(57.0093,11.7359)
\qbezier(57.0093,11.7359)(57.3713,11.7251)(57.7333,11.7142)
\qbezier(57.7333,11.7142)(58.0954,11.7036)(58.4574,11.6931)
\qbezier(58.4574,11.6931)(58.8194,11.6828)(59.1814,11.6725)
\qbezier(59.1814,11.6725)(59.5434,11.6625)(59.9054,11.6525)
\qbezier(59.9054,11.6525)(60.2674,11.6427)(60.6295,11.6330)
\qbezier(60.6295,11.6330)(60.9915,11.6235)(61.3535,11.6140)
\qbezier(61.3535,11.6140)(61.7155,11.6048)(62.0775,11.5955)
\qbezier(62.0775,11.5955)(62.4395,11.5865)(62.8016,11.5775)
\qbezier(62.8016,11.5775)(63.1636,11.5687)(63.5256,11.5599)
\qbezier(63.5256,11.5599)(63.8876,11.5513)(64.2496,11.5427)
\qbezier(64.2496,11.5427)(64.6116,11.5344)(64.9736,11.5260)
\qbezier(64.9736,11.5260)(65.3357,11.5179)(65.6977,11.5097)
\qbezier(65.6977,11.5097)(66.0597,11.5017)(66.4217,11.4938)
\qbezier(66.4217,11.4938)(66.7837,11.4860)(67.1457,11.4782)
\qbezier(67.1457,11.4782)(67.5078,11.4706)(67.8698,11.4630)
\qbezier(67.8698,11.4630)(68.2318,11.4556)(68.5938,11.4481)
\qbezier(68.5938,11.4481)(68.9558,11.4409)(69.3178,11.4336)
\qbezier(69.3178,11.4336)(69.6798,11.4265)(70.0419,11.4195)
\qbezier(70.0419,11.4195)(70.4039,11.4125)(70.7659,11.4056)
\qbezier(70.7659,11.4056)(71.1279,11.3988)(71.4899,11.3920)
\qbezier(71.4899,11.3920)(71.8519,11.3854)(72.2140,11.3788)
\qbezier(72.2140,11.3788)(72.5760,11.3723)(72.9380,11.3658)
\qbezier(72.9380,11.3658)(73.3000,11.3594)(73.6620,11.3531)
\qbezier(73.6620,11.3531)(74.0240,11.3469)(74.3860,11.3407)
\qbezier(74.3860,11.3407)(74.7481,11.3346)(75.1101,11.3285)
\qbezier(75.1101,11.3285)(75.4721,11.3225)(75.8341,11.3166)
\qbezier(75.8341,11.3166)(76.1961,11.3107)(76.5581,11.3049)
\qbezier(76.5581,11.3049)(76.9202,11.2992)(77.2822,11.2934)
\qbezier(77.2822,11.2934)(77.6442,11.2878)(78.0062,11.2822)
\qbezier(78.0062,11.2822)(78.3682,11.2767)(78.7302,11.2712)
\qbezier(78.7302,11.2712)(79.0922,11.2659)(79.4543,11.2605)
\qbezier(79.4543,11.2605)(79.8163,11.2552)(80.1783,11.2499)
\qbezier(80.1783,11.2499)(80.5403,11.2447)(80.9023,11.2396)
\qbezier(80.9023,11.2396)(81.2643,11.2345)(81.6264,11.2294)
\qbezier(81.6264,11.2294)(81.9884,11.2244)(82.3504,11.2194)
\qbezier(82.3504,11.2194)(82.7124,11.2146)(83.0744,11.2097)
\qbezier(83.0744,11.2097)(83.4364,11.2049)(83.7985,11.2001)
\qbezier(83.7985,11.2001)(84.1605,11.1954)(84.5225,11.1907)
\qbezier(84.5225,11.1907)(84.8845,11.1860)(85.2465,11.1814)
\qbezier(85.2465,11.1814)(85.6085,11.1769)(85.9705,11.1723)
\qbezier(85.9705,11.1723)(86.3326,11.1679)(86.6946,11.1634)
\qbezier(86.6946,11.1634)(87.0566,11.1590)(87.4186,11.1547)
\qbezier(87.4186,11.1547)(87.7806,11.1504)(88.1426,11.1461)
\qbezier(88.1426,11.1461)(88.5047,11.1418)(88.8667,11.1376)
\qbezier(88.8667,11.1376)(89.2287,11.1335)(89.5907,11.1293)
\qbezier(89.5907,11.1293)(89.9527,11.1252)(90.3147,11.1211)
\qbezier(90.3147,11.1211)(90.6767,11.1171)(91.0388,11.1131)
\qbezier(91.0388,11.1131)(91.4008,11.1092)(91.7628,11.1052)
\qbezier(91.7628,11.1052)(92.1248,11.1014)(92.4868,11.0975)
\qbezier(92.4868,11.0975)(92.8488,11.0937)(93.2109,11.0899)
\qbezier(93.2109,11.0899)(93.5729,11.0861)(93.9349,11.0824)
\qbezier(93.9349,11.0824)(94.2969,11.0787)(94.6589,11.0750)
\qbezier(94.6589,11.0750)(95.0209,11.0714)(95.3829,11.0677)
\qbezier(95.3829,11.0677)(95.7450,11.0642)(96.1070,11.0606)
\qbezier(96.1070,11.0606)(96.4690,11.0571)(96.8310,11.0536)
\qbezier(96.8310,11.0536)(97.1930,11.0501)(97.5550,11.0466)
\qbezier(97.5550,11.0466)(97.9171,11.0432)(98.2791,11.0398)
\qbezier(98.2791,11.0398)(98.6411,11.0365)(99.0031,11.0331)
\qbezier(99.0031,11.0331)(99.3651,11.0298)(99.7271,11.0265)
\qbezier(99.7271,11.0265)(100.0891,11.0233)(100.4512,11.0201)
\qbezier(100.4512,11.0201)(100.8132,11.0169)(101.1752,11.0137)
\qbezier(101.1752,11.0137)(101.5372,11.0105)(101.8992,11.0074)
\qbezier(101.8992,11.0074)(102.2612,11.0043)(102.6233,11.0012)
\qbezier(102.6233,11.0012)(102.9853,10.9981)(103.3473,10.9950)
\qbezier(103.3473,10.9950)(103.7093,10.9920)(104.0713,10.9890)
\qbezier(104.0713,10.9890)(104.4333,10.9861)(104.7953,10.9831)
\qbezier(104.7953,10.9831)(105.1574,10.9802)(105.5194,10.9773)
\qbezier(105.5194,10.9773)(105.8814,10.9744)(106.2434,10.9715)
\qbezier(106.2434,10.9715)(106.6054,10.9687)(106.9674,10.9658)
\qbezier(106.9674,10.9658)(107.3295,10.9630)(107.6915,10.9602)
\qbezier(107.6915,10.9602)(108.0535,10.9575)(108.4155,10.9547)
\qbezier(108.4155,10.9547)(108.7775,10.9520)(109.1395,10.9493)
\qbezier(109.1395,10.9493)(109.5016,10.9466)(109.8636,10.9439)
\qbezier(109.8636,10.9439)(110.2256,10.9413)(110.5876,10.9386)
\qbezier(110.5876,10.9386)(110.9496,10.9360)(111.3116,10.9334)
\qbezier(111.3116,10.9334)(111.6736,10.9308)(112.0357,10.9283)
\qbezier(112.0357,10.9283)(112.3977,10.9257)(112.7597,10.9232)
\qbezier(112.7597,10.9232)(113.1217,10.9207)(113.4837,10.9182)
\qbezier(113.4837,10.9182)(113.8457,10.9157)(114.2078,10.9132)
\qbezier(114.2078,10.9132)(114.5698,10.9108)(114.9318,10.9084)
\qbezier(114.9318,10.9084)(115.2938,10.9060)(115.6558,10.9036)
\qbezier(115.6558,10.9036)(116.0178,10.9012)(116.3798,10.8988)
\qbezier(116.3798,10.8988)(116.7419,10.8965)(117.1039,10.8941)
\qbezier(117.1039,10.8941)(117.4659,10.8918)(117.8279,10.8895)
\qbezier(117.8279,10.8895)(118.1899,10.8872)(118.5519,10.8850)
\qbezier(118.5519,10.8850)(118.9140,10.8827)(119.2760,10.8805)
\qbezier(119.2760,10.8805)(119.6380,10.8782)(120.0000,10.8760)
\qbezier(3.2781,48.0000)(3.6428,43.8561)(4.0076,39.7122)
\qbezier(4.0076,39.7122)(4.3723,36.8783)(4.7371,34.0444)
\qbezier(6.1961,26.8548)(6.5609,25.6543)(6.9256,24.4537)
\qbezier(6.9256,24.4537)(7.2904,23.4976)(7.6551,22.5415)
\qbezier(9.1142,19.7056)(9.4789,19.1686)(9.8437,18.6317)
\qbezier(9.8437,18.6317)(10.2084,18.1768)(10.5732,17.7218)
\qbezier(12.0322,16.2697)(12.3970,15.9765)(12.7617,15.6834)
\qbezier(12.7617,15.6834)(13.1265,15.4261)(13.4912,15.1689)
\qbezier(14.9503,14.3107)(15.3150,14.1303)(15.6798,13.9500)
\qbezier(15.6798,13.9500)(16.0445,13.7880)(16.4093,13.6261)
\qbezier(17.8683,13.0692)(18.2331,12.9489)(18.5978,12.8285)
\qbezier(18.5978,12.8285)(18.9626,12.7187)(19.3273,12.6088)
\qbezier(20.7864,12.2227)(21.1511,12.1375)(21.5159,12.0523)
\qbezier(21.5159,12.0523)(21.8806,11.9736)(22.2454,11.8948)
\qbezier(23.7044,11.6134)(24.0692,11.5503)(24.4339,11.4873)
\qbezier(24.4339,11.4873)(24.7987,11.4284)(25.1634,11.3696)
\qbezier(26.6225,11.1564)(26.9872,11.1080)(27.3520,11.0597)
\qbezier(27.3520,11.0597)(27.7167,11.0142)(28.0815,10.9687)
\qbezier(29.5405,10.8023)(29.9053,10.7641)(30.2700,10.7259)
\qbezier(30.2700,10.7259)(30.6348,10.6898)(30.9995,10.6537)
\qbezier(32.4586,10.5204)(32.8233,10.4896)(33.1881,10.4588)
\qbezier(33.1881,10.4588)(33.5528,10.4295)(33.9176,10.4002)
\qbezier(35.3766,10.2913)(35.7414,10.2659)(36.1061,10.2406)
\qbezier(36.1061,10.2406)(36.4709,10.2164)(36.8356,10.1922)
\qbezier(38.2947,10.1016)(38.6594,10.0804)(39.0242,10.0592)
\qbezier(39.0242,10.0592)(39.3889,10.0388)(39.7537,10.0185)
\qbezier(41.2127,9.9421)(41.5775,9.9241)(41.9422,9.9061)
\qbezier(41.9422,9.9061)(42.3070,9.8888)(42.6717,9.8715)
\qbezier(44.1307,9.8062)(44.4955,9.7908)(44.8603,9.7753)
\qbezier(44.8603,9.7753)(45.2250,9.7605)(45.5898,9.7456)
\qbezier(47.0488,9.6891)(47.4136,9.6758)(47.7783,9.6624)
\qbezier(47.7783,9.6624)(48.1431,9.6494)(48.5078,9.6365)
\qbezier(49.9668,9.5873)(50.3316,9.5756)(50.6964,9.5639)
\qbezier(50.6964,9.5639)(51.0611,9.5525)(51.4259,9.5412)
\qbezier(52.8849,9.4979)(53.2496,9.4876)(53.6144,9.4772)
\qbezier(53.6144,9.4772)(53.9792,9.4672)(54.3439,9.4572)
\qbezier(55.8029,9.4188)(56.1677,9.4097)(56.5325,9.4005)
\qbezier(56.5325,9.4005)(56.8972,9.3916)(57.2620,9.3826)
\qbezier(58.7210,9.3484)(59.0857,9.3402)(59.4505,9.3320)
\qbezier(59.4505,9.3320)(59.8153,9.3240)(60.1800,9.3160)
\qbezier(61.6390,9.2853)(62.0038,9.2779)(62.3685,9.2706)
\qbezier(62.3685,9.2706)(62.7333,9.2634)(63.0981,9.2562)
\qbezier(64.5571,9.2285)(64.9218,9.2218)(65.2866,9.2151)
\qbezier(65.2866,9.2151)(65.6514,9.2086)(66.0161,9.2021)
\qbezier(67.4751,9.1770)(67.8399,9.1709)(68.2046,9.1649)
\qbezier(68.2046,9.1649)(68.5694,9.1589)(68.9342,9.1530)
\qbezier(70.3932,9.1302)(70.7579,9.1246)(71.1227,9.1191)
\qbezier(71.1227,9.1191)(71.4874,9.1137)(71.8522,9.1083)
\qbezier(73.3112,9.0874)(73.6760,9.0823)(74.0407,9.0772)
\qbezier(74.0407,9.0772)(74.4055,9.0723)(74.7703,9.0673)
\qbezier(76.2293,9.0481)(76.5940,9.0435)(76.9588,9.0388)
\qbezier(76.9588,9.0388)(77.3235,9.0343)(77.6883,9.0297)
\qbezier(79.1473,9.0120)(79.5121,9.0077)(79.8768,9.0034)
\qbezier(79.8768,9.0034)(80.2416,8.9992)(80.6063,8.9950)
\qbezier(82.0654,8.9787)(82.4301,8.9747)(82.7949,8.9707)
\qbezier(82.7949,8.9707)(83.1596,8.9668)(83.5244,8.9629)
\qbezier(84.9834,8.9478)(85.3482,8.9441)(85.7129,8.9404)
\qbezier(85.7129,8.9404)(86.0777,8.9368)(86.4424,8.9332)
\qbezier(87.9015,8.9191)(88.2662,8.9157)(88.6310,8.9123)
\qbezier(88.6310,8.9123)(88.9957,8.9089)(89.3605,8.9055)
\qbezier(90.8195,8.8924)(91.1843,8.8892)(91.5490,8.8860)
\qbezier(91.5490,8.8860)(91.9138,8.8829)(92.2785,8.8798)
\qbezier(93.7376,8.8675)(94.1023,8.8645)(94.4671,8.8615)
\qbezier(94.4671,8.8615)(94.8318,8.8586)(95.1966,8.8557)
\qbezier(96.6556,8.8442)(97.0204,8.8414)(97.3851,8.8386)
\qbezier(97.3851,8.8386)(97.7499,8.8359)(98.1146,8.8331)
\qbezier(99.5737,8.8223)(99.9384,8.8197)(100.3032,8.8171)
\qbezier(100.3032,8.8171)(100.6679,8.8145)(101.0327,8.8119)
\qbezier(102.4917,8.8018)(102.8565,8.7994)(103.2212,8.7969)
\qbezier(103.2212,8.7969)(103.5860,8.7945)(103.9507,8.7920)
\qbezier(105.4098,8.7825)(105.7745,8.7802)(106.1393,8.7779)
\qbezier(106.1393,8.7779)(106.5040,8.7756)(106.8688,8.7733)
\qbezier(108.3278,8.7643)(108.6926,8.7621)(109.0573,8.7599)
\qbezier(109.0573,8.7599)(109.4221,8.7577)(109.7868,8.7556)
\qbezier(111.2459,8.7471)(111.6106,8.7450)(111.9754,8.7429)
\qbezier(111.9754,8.7429)(112.3401,8.7409)(112.7049,8.7389)
\qbezier(114.1639,8.7308)(114.5287,8.7289)(114.8934,8.7269)
\qbezier(114.8934,8.7269)(115.2582,8.7250)(115.6229,8.7230)
\qbezier(117.0820,8.7154)(117.4467,8.7136)(117.8115,8.7117)
\qbezier(117.8115,8.7117)(118.1762,8.7099)(118.5410,8.7080)
\put(0,10.1859){\circle*{0.35}}
\put(3,10.1859){\circle*{0.35}}
\put(6,10.1859){\circle*{0.35}}
\put(9,10.1859){\circle*{0.35}}
\put(12,10.1859){\circle*{0.35}}
\put(15,10.1859){\circle*{0.35}}
\put(18,10.1859){\circle*{0.35}}
\put(21,10.1859){\circle*{0.35}}
\put(24,10.1859){\circle*{0.35}}
\put(27,10.1859){\circle*{0.35}}
\put(30,10.1859){\circle*{0.35}}
\put(33,10.1859){\circle*{0.35}}
\put(36,10.1859){\circle*{0.35}}
\put(39,10.1859){\circle*{0.35}}
\put(42,10.1859){\circle*{0.35}}
\put(45,10.1859){\circle*{0.35}}
\put(48,10.1859){\circle*{0.35}}
\put(51,10.1859){\circle*{0.35}}
\put(54,10.1859){\circle*{0.35}}
\put(57,10.1859){\circle*{0.35}}
\put(60,10.1859){\circle*{0.35}}
\put(63,10.1859){\circle*{0.35}}
\put(66,10.1859){\circle*{0.35}}
\put(69,10.1859){\circle*{0.35}}
\put(72,10.1859){\circle*{0.35}}
\put(75,10.1859){\circle*{0.35}}
\put(78,10.1859){\circle*{0.35}}
\put(81,10.1859){\circle*{0.35}}
\put(84,10.1859){\circle*{0.35}}
\put(87,10.1859){\circle*{0.35}}
\put(90,10.1859){\circle*{0.35}}
\put(93,10.1859){\circle*{0.35}}
\put(96,10.1859){\circle*{0.35}}
\put(99,10.1859){\circle*{0.35}}
\put(102,10.1859){\circle*{0.35}}
\put(105,10.1859){\circle*{0.35}}
\put(108,10.1859){\circle*{0.35}}
\put(111,10.1859){\circle*{0.35}}
\put(114,10.1859){\circle*{0.35}}
\put(117,10.1859){\circle*{0.35}}
\put(96,9.2252){\circle*{1.4}}
\put(99,4.2252){\makebox(0,0)[l]{\scriptsize $(960,8)$}}
\put(12,40.7447){\circle*{1.4}}
\put(15,44.7447){\makebox(0,0)[l]{\scriptsize $(4240,1)$}}
\put(52,37){\small Local degradation}
\put(36,5){\small Local recovery}
\end{picture}
\caption{Analytical homogeneous recovery diagram. The solid curve is $\varphi_c=1$: small downward perturbations recover below it and grow above it. The dashed curve is $\varphi_c=0.9$, at fixed exposure, initial states with $\varphi(0)=0.9$ recover below this curve. The dotted line is $\ell_r^\varphi/\ell_f=1/\pi$. Markers identify $(\ell_r^\varphi,k_2)=(960,8)$ and $(4240,1)$ at $\sigma=1$, with $\ell_f=3330$. The curves give the homogeneous recovery boundaries as functions of scaled reaction amplitude and sensitivity to the cytotoxic phase.}
\label{fig:analytical_recovery}
\end{figure}

\subsection{Consequences for the simulations}

With $\ell_f=3330$, the saturation boundary is $\ell_f/\pi\simeq1059.97$.
For Figure~\ref{fig:finger_formation}, $\ell_r^\varphi=960$ and $k_2=8$: at fixed $\sigma=1$, homogeneous states recover if $\varphi(0)>\varphi_c(1)\simeq0.916992$, for evolving cytotoxic phase, $\varphi(0)>1/2+960\pi/(2\cdot3330)\simeq0.952842$ is sufficient regardless of the exposure history. This parameter set admits recovery even at large cytotoxic-phase values.
For Figure~\ref{fig:higher_reaction_rate}, $\ell_r^\varphi=4240$ and $k_2=1$: $\sigma_c\simeq0.414201$, whereas $\sigma=1$ gives $\varphi_c\simeq1.500026$. Fixed exposure at that level causes local degradation, exposure bounded by $\bar\sigma<\sigma_c$ permits recovery when $\varphi(0)>\varphi_c(\bar\sigma)$.

The reaction strengths of the two parameter sets also exchange their ordering with exposure. Solving
\begin{equation}
 960\arctan(8\sigma)=4240\arctan(\sigma)
\end{equation}
gives the positive crossing $\sigma\simeq0.259693$. Below this crossing, the first set has the stronger cyto reaction at the same $0<\varphi<1$, above this crossing, the second does. Their comparison involves both reaction amplitude and sensitivity to the cytotoxic phase.

\endgroup

\clearpage

\section{\texorpdfstring{Interface motion and shape perturbations}{Interface motion and shape perturbations}}
\label{sc:interface_motion}
\begingroup
\subsection{Normal velocity on a smooth angular branch}
We derive a local-equilibrium approximation to the dimensionless cyto balance. The interface is thin compared with its curvature radius, the cytotoxic phase varies little across its thickness, and curvature and reaction perturb the planar equilibrium profile. Let $z$ increase along the outward unit normal $\bs{n}$, let $\varphi=q(z)$ decrease from one to zero, and set $a(\theta)=\vartheta+\cos\theta>0$. Thus $\bs{n}=-\nabla\varphi/|\nabla\varphi|$, opposite to the profile normal used in Appendix~\ref{sc:interfacial_tension}. $\theta$ retains the principal-arctangent convention of the governing equations. With $f(q)=q^2(1-q)^2/4$ and $\epsilon(\theta)=\tilde{\epsilon}a(\theta)$, using the identities in Appendix~\ref{sc:interfacial_tension}, the planar balance and its first integral are
\begin{equation}
\ell_\varphi a^2q''=\ell_f f'(q),\qquad\frac{\ell_\varphi a^2}{2}(q')^2=\ell_f f(q),\qquad q'=-\frac{1}{a}\sqrt{\frac{\ell_f}{2\ell_\varphi}}q(1-q).
\end{equation}
Primes on $q$ denote $z$ derivatives. Integration gives $\Upsilon$, now expressed in the dimensionless energy $\ell_f f+\ell_\varphi a^2|\nabla\varphi|^2/2$ and the translation resistance:
\begin{equation}
 \Upsilon(\theta)=\frac{\sqrt{\ell_f\ell_\varphi}}{6\sqrt2}a(\theta),
 \qquad
 \int_{\bb{R}}(q')^2\di{z}=\frac{1}{6a}\sqrt{\frac{\ell_f}{2\ell_\varphi}}.
\end{equation}
Using orientation derivatives for primes on $\Upsilon$, the capillary stiffness is
\begin{equation}
 \Upsilon+\Upsilon''=\frac{\sqrt{\ell_f\ell_\varphi}}{6\sqrt2}\vartheta.
\end{equation}
This is the dimensionless form of the tension correspondence in Appendix~\ref{sc:interfacial_tension}. Although the stiffness is constant on a smooth branch, the inverse translation resistance depends on orientation.

Let $V_n$ be the outward normal velocity, $\kappa=\Divs\bs{n}$ (positive for a convex cell), and $\sigma_\Gamma$ the cytotoxic phase at the interface. Projection of the cyto balance onto $q'$ gives
\begin{equation}
 -V_n\int_{\bb{R}}(q')^2\di{z}
 = (\Upsilon+\Upsilon'')\kappa
 -\ell_r^\varphi\arctan(k_2\sigma_\Gamma)
 \int_{\bb{R}}q(1-q)q'\di{z}.
\end{equation}
Since $\int_{\bb{R}}q(1-q)q'\di{z}=-\int_0^1q(1-q)\di{q}=-1/6$, the leading velocity is
\begin{equation}\label{equ:interface_velocity}
 V_n=-\ell_\varphi a(\theta)\vartheta\kappa
 -a(\theta)\sqrt{\frac{2\ell_\varphi}{\ell_f}}
 \ell_r^\varphi\arctan(k_2\sigma_\Gamma).
\end{equation}
Both curvature and nonnegative cytotoxic-phase values drive a convex boundary inward. The angular factor changes the speed.

The principal-arctangent convention also requires care. Away from zero gradients and angular branch junctions,
\begin{equation}
 \cos\!\left(\arctan\frac{\partial\varphi/\partial y}{\partial\varphi/\partial x}\right)
 =\frac{|\partial\varphi/\partial x|}{|\nabla\varphi|}=|n_x|.
\end{equation}
Its continuous extension gives $a=\vartheta+|n_x|$, where $n_x$ is the first component of $\bs{n}$. The resulting tension has angular cusps at $n_x=0$. The velocity formula applies locally on the smooth angular branches, where the stiffness and translation resistance are differentiable.

\newpage
\subsection{A shrinking circular reference interface}
To separate curvature smoothing from spatially varying exposure and angular branch junctions, consider an isotropic reference with constant $a=a_0>0$ and the cytotoxic phase prescribed uniformly along the interface. Define
\begin{equation}
 D:=\ell_\varphi a_0^2,\qquad
 v(\bar t):=\sqrt{\frac{2D}{\ell_f}}\ell_r^\varphi
 \arctan(k_2\sigma_\Gamma(\bar t)).
\end{equation}
The same projection gives $V_n=-D\kappa-v(\bar t)$, using $\kappa=1/R$ so a circle of radius $R$ obeys $\dot R=-D/R-v$. Dots denote dimensionless-time derivatives. Even with prescribed $\sigma_\Gamma=0$, thus $v=0$, $R^2=R(0)^2-2D\bar t$ while the interface remains thin. This prescribed-exposure reference isolates the contribution of curvature to boundary motion.

Let $\chi$ be polar angle and perturb the circle by $r(\chi,\bar t)=R(\bar t)+u_m(\bar t)\cos(m\chi)$, with integer $m\ge2$ and $|u_m|\ll R$. For fixed $m$, expanding the curvature gives
\begin{equation}
 \kappa=\frac1R+\frac{m^2-1}{R^2}u_m\cos(m\chi)
 +O\!\left(\frac{u_m^2}{R^3}\right).
\end{equation}
With $v(\bar t)$ unchanged by the perturbation, subtracting the circular velocity yields
\begin{equation}\label{equ:circle_modes}
 \dot u_m=-\frac{D(m^2-1)}{R^2}u_m,
 \qquad
 \frac{\di{}}{\di{\bar t}}\left(\frac{u_m}{R}\right)
 =\left[\frac{v}{R}-\frac{D(m^2-2)}{R^2}\right]\frac{u_m}{R}.
\end{equation}
All shape modes decay in absolute amplitude, but their amplitude relative to the shrinking radius can increase. The distinction between absolute and relative amplitudes separates curvature smoothing from the geometric effect of shrinkage.

In the coupled spatial problem, the cytotoxic phase sampled at the perturbed interface changes. Its first-order contribution to the reference velocity includes
\begin{equation}
 -\sqrt{\frac{2D}{\ell_f}}
 \frac{\ell_r^\varphi k_2}{1+(k_2\sigma_\Gamma)^2}\,\delta\sigma_\Gamma,
\end{equation}
where $\delta\sigma_\Gamma$ includes both field variation and interface displacement. This term couples shape evolution to the cytotoxic phase. The circular reference quantifies curvature smoothing and relative amplification during shrinkage, while the interfacial response to the cytotoxic phase supplies an additional contribution to the shape dynamics. Internal cavities, their opening onto the exterior, and detached fragments are distinct topological changes. $u_m$ describes perturbations of the connected outer boundary.
\endgroup

\clearpage

\section{\texorpdfstring{Controlled numerical comparisons}{Controlled numerical comparisons}}
\label{sc:controlled_simulations}
\begingroup
\subsection{Parameters and measurements}
The analytical diagram in Appendix~\ref{sc:local_threshold} describes homogeneous recovery boundaries. Here, four simulations cross two reaction-coefficient pairs with two sensitivities, and two isotropic controls assess the interfacial law (Table~\ref{tab:controlled_runs}). All cases use the same pair of initial field arrays: a unit field value inside a disk of radius $3$, with independent Gaussian perturbations of standard deviation $0.1$ inside the disk, clipped to $[0,1]$, and zero outside. The PCG64 generator with seed $20260927$ produces the common arrays, which are represented in the Fourier basis. The independence is between the two fields and their realisations are identical across the six cases.

\begin{table}[htbp]
\centering

\caption{Controlled simulation parameters. S1/S2 and S3/S4 isolate sensitivity. S1/S3 and S2/S4 compare the two reaction-coefficient pairs. S2/S5 and S3/S6 isolate the interfacial law.}
\label{tab:controlled_runs}
\begin{tabular}{ccccc}
\hline
Case & $\ell_r^\varphi$ & $\ell_r^\sigma$ & $k_2$ & Interfacial law\\
\hline
S1 & 960 & 0.28 & 1 & Anisotropic\\
S2 & 960 & 0.28 & 8 & Anisotropic\\
S3 & 4240 & 1.27 & 1 & Anisotropic\\
S4 & 4240 & 1.27 & 8 & Anisotropic\\
S5 & 960 & 0.28 & 8 & Isotropic\\
S6 & 4240 & 1.27 & 1 & Isotropic\\
\hline
\end{tabular}
\end{table}

The shared coefficients are $\ell_f=3330$, $\ell_\varphi=1.333\times10^{-4}$, $\ell_1=1$, $\ell_2=1.5$ and $\vartheta=25$. We use $512\times512$ Fourier modes on the periodic domain $[-4,4)^2$, $3/2$ dealiasing, the RK443 scheme and $\Delta\bar t=5\times10^{-5}$ up to $\bar t=2$. Fields are saved every $0.05$.

For the implemented principal-arctangent law, $a=\vartheta+|n_x|$, where $n_x$ is the first component of the interface unit normal. The isotropic controls use
\[
 a_{\rm iso}=\sqrt{\vartheta^2+\frac{4\vartheta}{\pi}+\frac12},
 \qquad \ell_\varphi a_{\rm iso}^2\Delta\varphi
\]
as the cyto interfacial operator. This matches the orientation average $a_{\rm iso}^2=\langle a^2\rangle$ while retaining $\ell_\varphi$.

The effective area fraction is $A=|\Omega|^{-1}\int_\Omega\varphi\di{\bs{x}}$ and we report $A(\bar t)/A(0)$. Geometry uses $\varphi\ge1/2$, periodic eight-connected foreground and four-connected background. Cavities are holes enclosed by the largest foreground component after other components are removed for this measurement. Their total area is normalised by that component's initial filled-envelope area $A_{\rm filled}(0)$. The detached-area fraction is $A_{\rm detached}/A_{\rm foreground}$, the area in retained components other than the largest divided by all foreground area. Reported cavities and detached components contain at least $16$ mesh cells (area $0.00390625$).

The outer-envelope circularity is $4\pi A_{\rm filled}/P_{\rm outer}^2$, where $A_{\rm filled}$ is the enclosed area and $P_{\rm outer}$ is the perimeter obtained from the interpolated outer $\varphi=1/2$ contour. Holes do not contribute to this perimeter. After fragmentation, this measure describes the largest fragment at each time. A change in its identity can produce a discontinuity. To measure the entire cell or fragment cloud, we also take the convex hull of all retained components' interpolated contours. Its area $A_{\rm hull}$ and perimeter $P_{\rm hull}$ give the same circularity measure, $4\pi A_{\rm hull}/P_{\rm hull}^2$. All retained components lie inside the displayed periodic cell, so this hull requires no unwrapping. The hull fills cavities and boundary indentations and therefore measures the gross envelope separately from local protrusions and connectivity.

\clearpage
\subsection{Reaction dependence}
\begin{figure}[htbp]
\centering
\includegraphics[width=\textwidth]{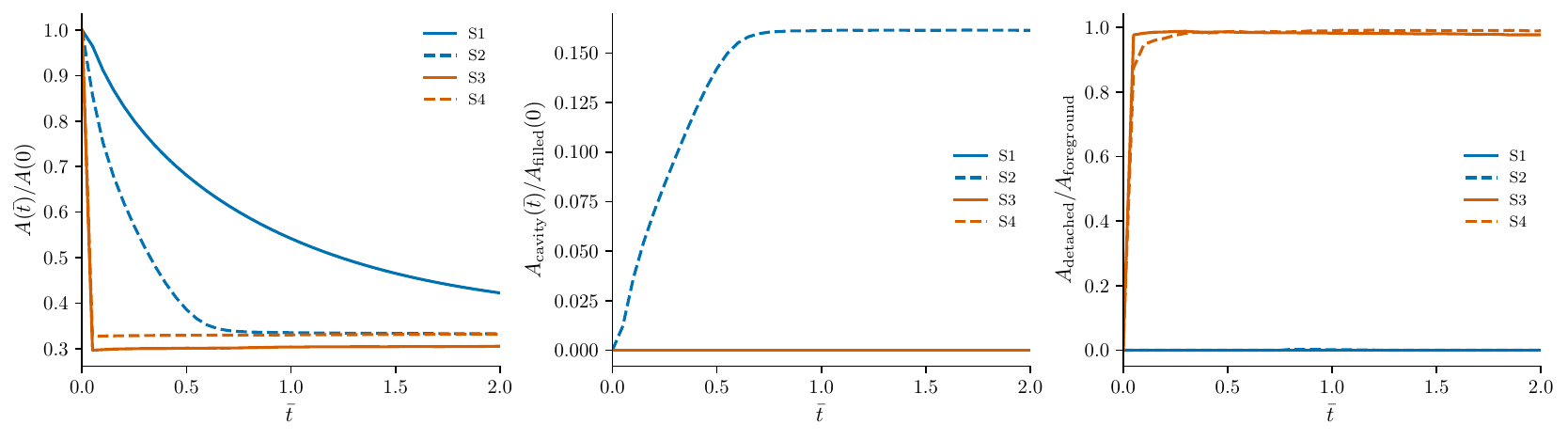}

\caption{Reaction comparison, S1--S4: normalised effective area, enclosed cavity area and detached-area fraction. Blue and orange distinguish the two reaction-coefficient pairs. Solid and dashed curves distinguish $k_2=1$ and $k_2=8$.}
\label{fig:controlled_reaction}
\end{figure}
As can be seen in Figure~\ref{fig:controlled_snapshots}, at the lower reaction amplitude, S1 mainly shrinks, whereas the higher sensitivity in S2 produces a pronounced cavity and an irregular outer boundary. The higher-amplitude cases S3 and S4 have already fragmented by the first saved positive time, $\bar t=0.05$. Increasing $k_2$ within this pair changes fragment size and the retained cyto-phase fraction. In particular, the final effective area is larger in S4 than in S3, despite the larger instantaneous reaction magnitude at equal cyto phase and positive cytotoxic-phase values. The coupled trajectories therefore cannot be ordered by sensitivity alone.

As shown in Figure~\ref{fig:controlled_reaction}, at $\bar t=2$, the effective-area ratios for S1--S4 are approximately $0.4225$, $0.3325$, $0.3053$ and $0.3318$. Together with the snapshots in Figure~\ref{fig:controlled_snapshots}, the area and topology measures distinguish recession, cavity formation and fragmentation. The recovery criterion for homogeneous states is established analytically in Appendix~\ref{sc:local_threshold}.

\clearpage
\begin{figure}[p]
\centering
\includegraphics[width=\textwidth,height=.85\textheight,keepaspectratio]{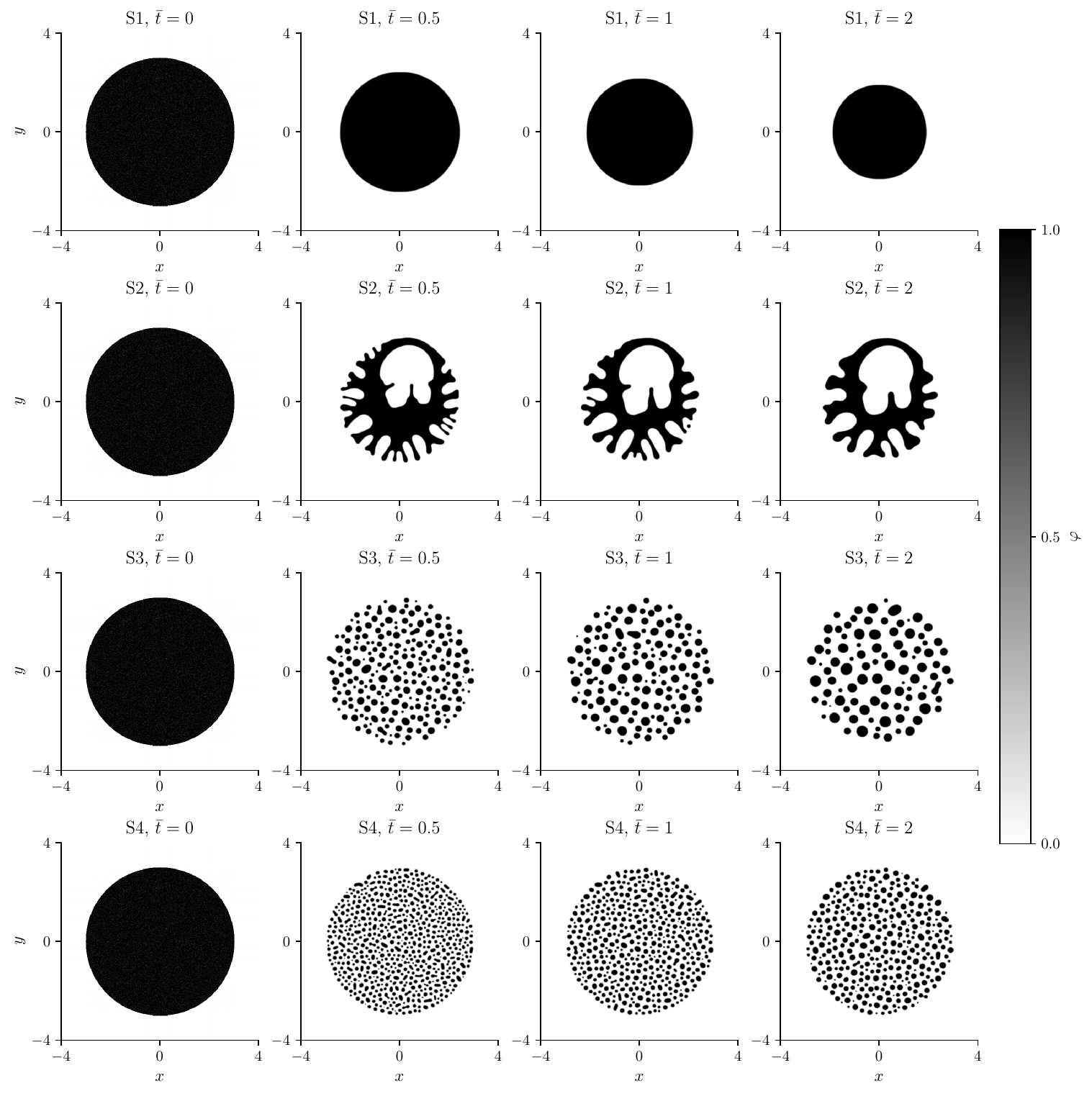}

\caption{Cyto-phase evolution for S1--S4 at the common times $\bar t=0,0.5,1,2$. White denotes $\varphi=0$ and black $\varphi=1$. All panels have identical spatial and colour limits. The continuous field is displayed with colour limits $[0,1]$, with values outside the colour limits rendered at the corresponding endpoint colours.}
\label{fig:controlled_snapshots}
\end{figure}

\clearpage
\subsection{Role of anisotropy}
\begin{figure}[htbp]
\centering
\includegraphics[width=\textwidth]{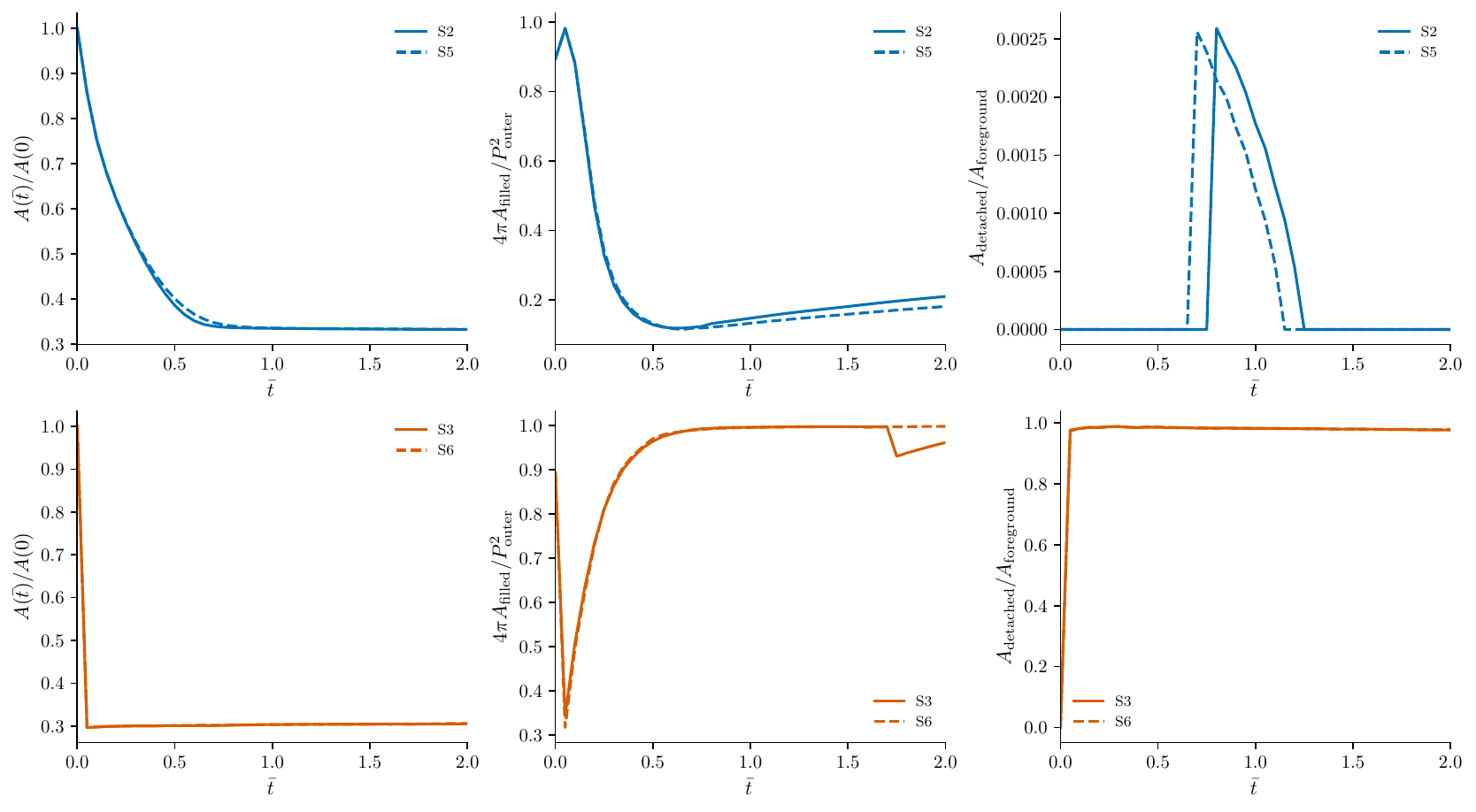}

\caption{Matched anisotropic (solid) and isotropic (dashed) cases: S2/S5 above and S3/S6 below. Columns show normalised effective area, largest-component outer-envelope circularity and detached-area fraction. In S3/S6, post-fragmentation circularity measures the largest fragment.}
\label{fig:controlled_anisotropy}
\end{figure}
The isotropic controls retain the principal features of their anisotropic counterparts: S5 develops a cavity and boundary protrusions, and S6 fragments, as observed in Figure~\ref{fig:controlled_anisotropy}. Their final effective-area ratios, $0.3328$ and $0.3058$, are close to those of S2 and S3. Anisotropy changes details of the boundary shape and the timing of small detachment events in S2/S5, while the main morphological distinction persists under the matched isotropic law. Thus, anisotropy is not required for these features in the tested cases.

\clearpage
\begin{figure}[htbp]
\centering
\includegraphics[width=\textwidth]{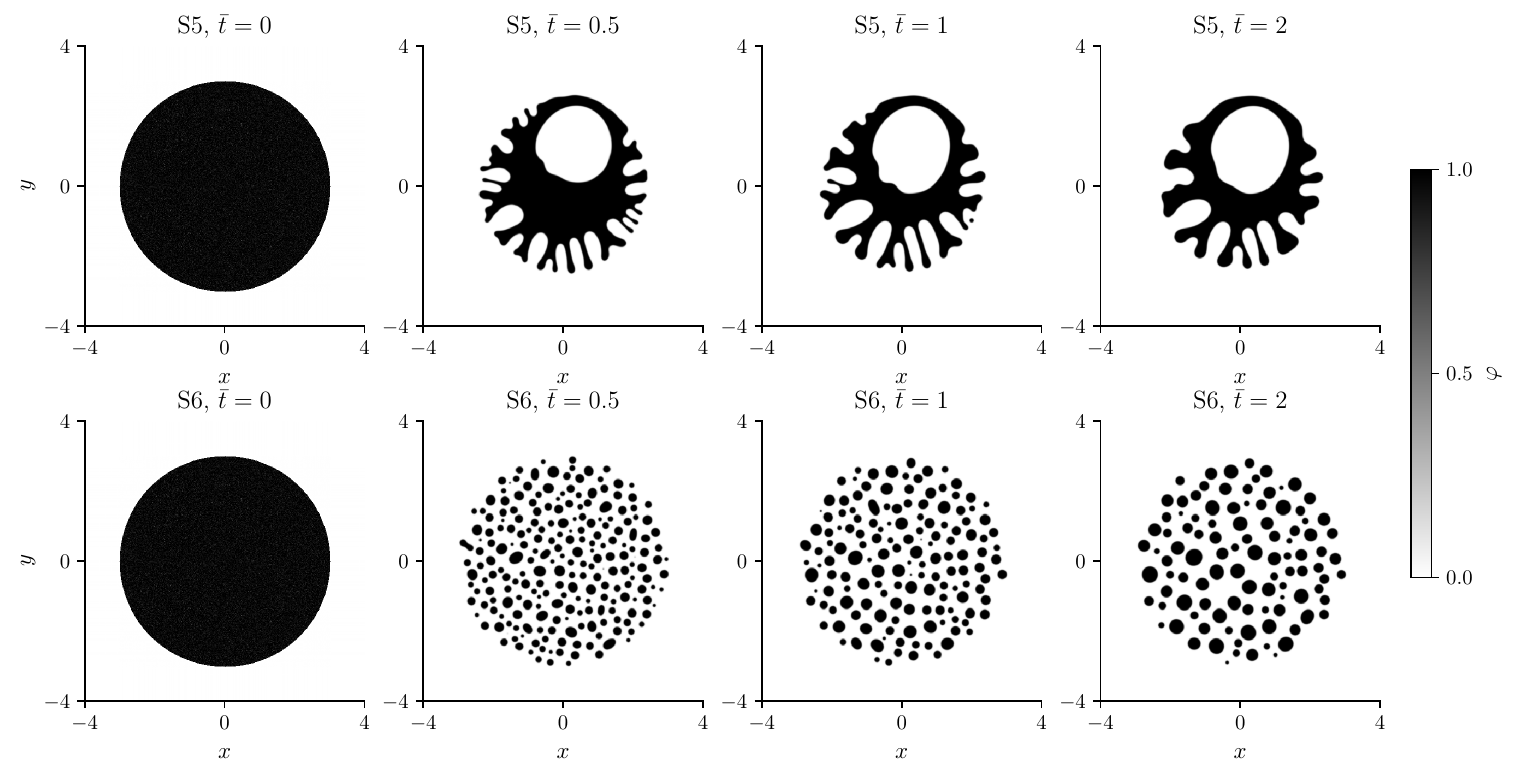}

\caption{Isotropic controls S5 and S6 at $\bar t=0,0.5,1,2$, with the same grayscale and spatial limits as the anisotropic snapshots. S5 retains the cavity and protrusion morphology of S2. S6 retains the fragmented morphology of S3.}
\label{fig:controlled_isotropic_snapshots}
\end{figure}

\subsection{Absolute deformation and relative prominence}

For a single-valued outer radius $r(\chi,\bar t)$ about the filled-area centroid, with polar angle $\chi$, define
\begin{equation}
R=\frac{1}{2\pi}\int_0^{2\pi}r\di{\chi},\qquad U_m=\left|\frac{1}{\pi}\int_0^{2\pi}(r-R)e^{-\mathrm i m\chi}\di{\chi}\right|.
\end{equation}
For the single cosine perturbation in Appendix~\ref{sc:interface_motion}, this definition gives $U_m=|u_m|$. We sample $512$ equally spaced rays and retain modes with wavelength $2\pi R/m$ at least eight mesh spacings. Figure~\ref{fig:controlled_modes} shows the fixed low modes $m=2,3,4$ for S2/S5. Both $U_m$ and $U_m/R$ show net growth over the resolved early interval. The boundary deformation therefore grows in absolute magnitude, as well as becoming more prominent relative to the shrinking radius. This occurs for both interfacial laws. It differs from the decay of absolute modes in the uniformly forced circular reference of Appendix~\ref{sc:interface_motion}, which isolates curvature smoothing. The coupled simulations also include a spatially varying cytotoxic phase and heterogeneous initial fields.

\clearpage
\begin{figure}[htbp]
\centering
\includegraphics[width=\textwidth]{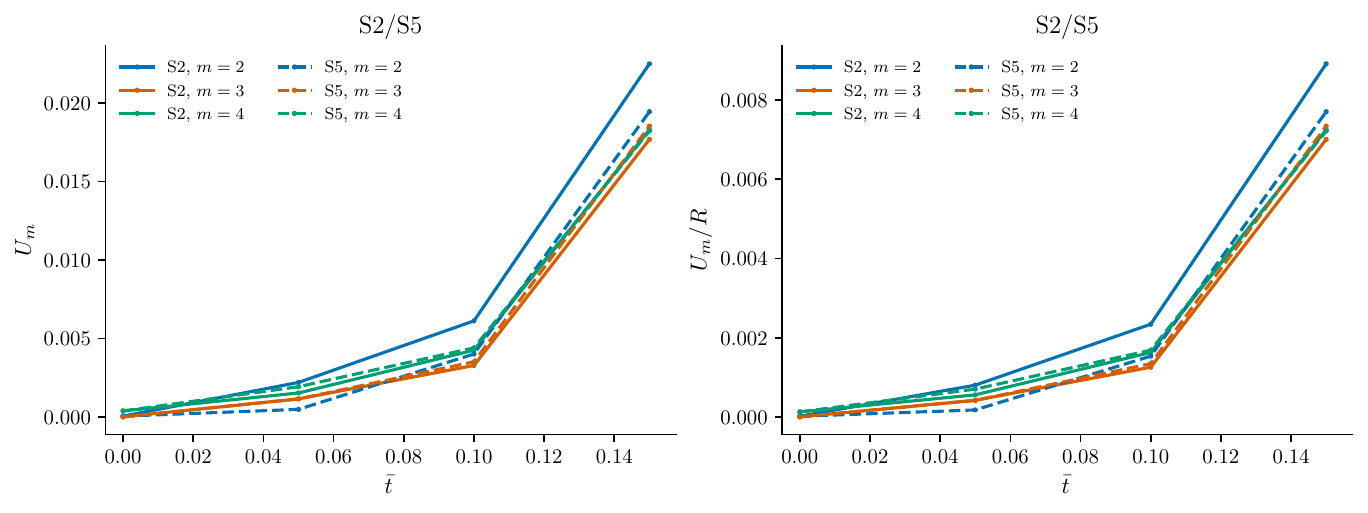}

\caption{Absolute and relative boundary-mode amplitudes for S2/S5. Curves join the saved times $\bar t=0,0.05,0.10,0.15$. At the next saved time, the outer contour is no longer a single-valued radial graph about the filled-area centroid. S3/S6 have already fragmented at their first positive output time and therefore do not provide a resolved pre-fragmentation mode trajectory.}
\label{fig:controlled_modes}
\end{figure}

\subsection{Envelope of the fragment cloud}
\begin{figure}[htbp]
\centering
\includegraphics[width=\textwidth]{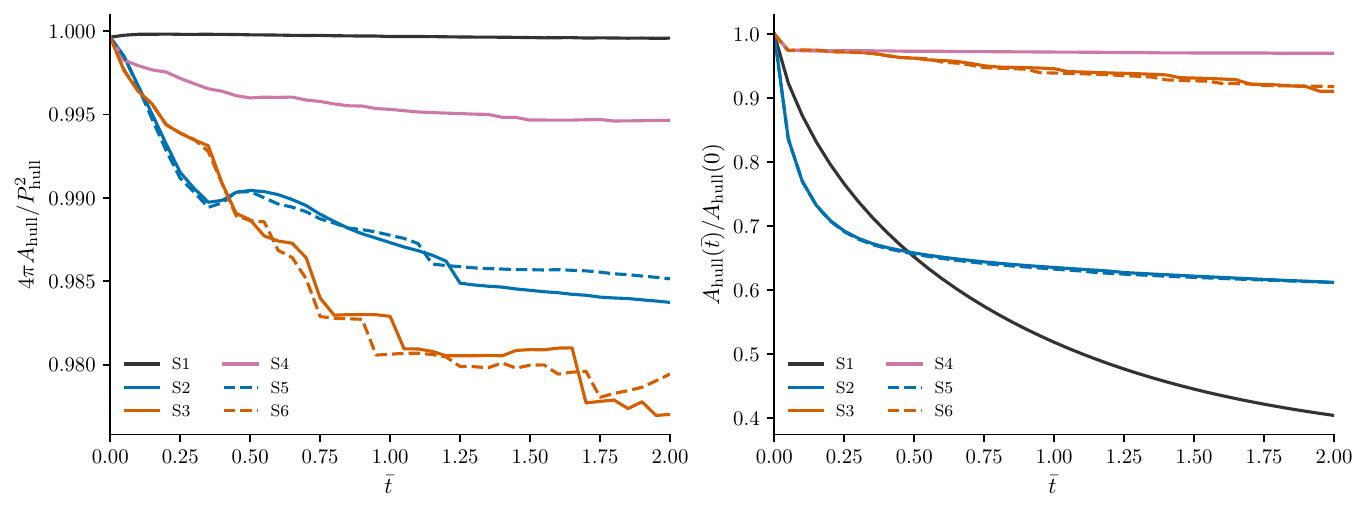}

\caption{Circularity and normalised area of the convex hull of the complete retained cyto-phase set. The hull includes gaps between fragments. The area is an envelope and does not represent the amount of cyto material.}
\label{fig:cloud_envelope}
\end{figure}
In Figure~\ref{fig:cloud_envelope}, the hull circularity remains above $0.976$ in all six cases. For the fragmented cases S3, S4 and S6, the hull retains approximately $91.1\%$, $97.0\%$ and $91.8\%$ of its initial area at $\bar t=2$, while the effective cyto-phase area retains only $30.5\%$, $33.2\%$ and $30.6\%$. Thus, extensive breakup occurs within a nearly circular spatial envelope. This separates actual detachment from both whole-envelope contraction and the circularity of individual fragments.

\endgroup


\newpage
\footnotesize

\bibliographystyle{abbrvnat}
\bibliography{bib}

\end{document}